\documentclass[a4paper]{jpconf}
\usepackage{graphicx}
\usepackage{color}
\usepackage{amsmath,amssymb,amsxtra}
\usepackage{wrapfig}

\def\gr{$\gamma$-ray}

\begin{document}
\title{Introduction to multi-messenger astronomy}

\author{Andrii Neronov}

\address{APC, Paris \& University of Geneva}


\begin{abstract}
The new field of multi-messenger astronomy aims at the study of astronomical sources using different types of "messenger" particles: photons, neutrinos, cosmic rays and gravitational waves. These lectures provide an introductory overview of the observational techniques used for each type of astronomical messenger, of different types of astronomical sources observed through different messenger channels and of the main physical processes involved in production of the messenger particles and their propagation through the Universe.
\end{abstract}

\section{Introduction}

Our knowledge of the Universe around us was acquired throughout centuries via detection of electromagnetic signals from different types of astronomical sources. This has changed in 2013, with the discovery of new astronomical messengers transmitting signals from distant sources outside the Solar system: the high-energy neutrinos \cite{icecube_science}. Further breakthrough has been achieved in 2015 with addition of one more ``messenger'': gravitational waves \cite{gw}. In this way new field of ``multi-messenger'' astronomy has been born. The range of astronomical observation tools has been extended over the last decade not only through the inclusion of new types of astronomical messengers but also via dramatic extension of the energy window through which the Universe is observed. Optical telescopes detect photons with energies about 1 eV. Astrophysical neutrinos discovered by IceCube have energies which are fifteen orders of magnitude higher, largely exceeding the energies of particles accelerated at the Large Hadron Collider at CERN. 

\begin{figure}[h!]
\begin{center}
\includegraphics[width=0.75\linewidth]{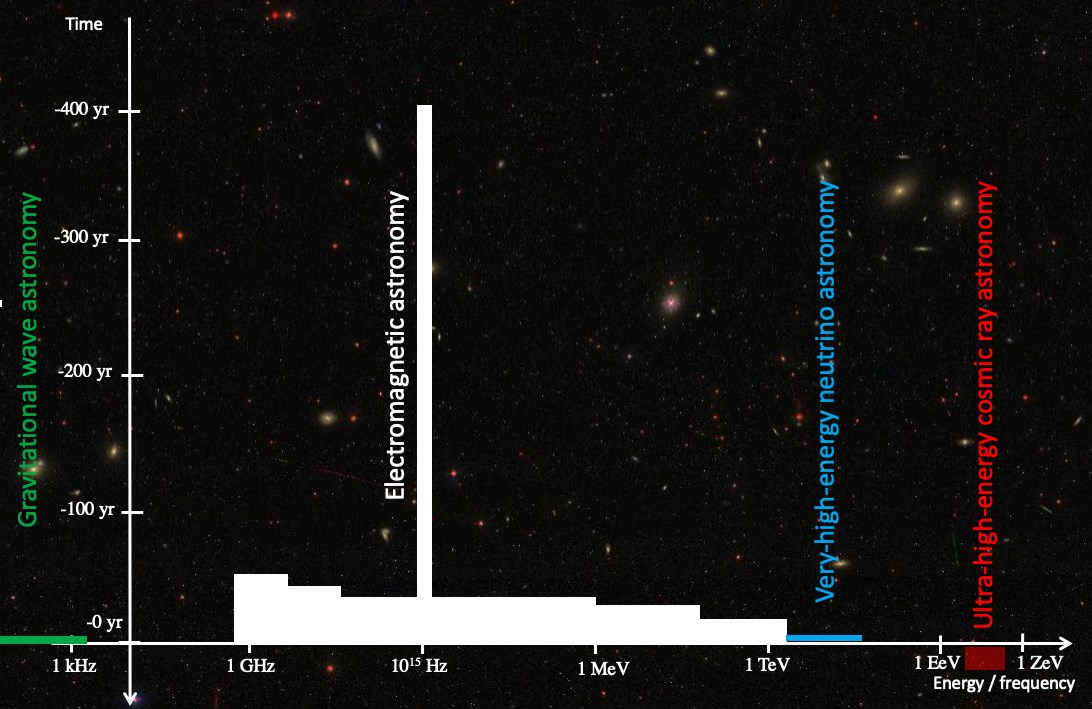}
\caption{Timeline of multi-messenger astronomy}
\label{fig:timeline}
\end{center}
\end{figure}

These introductory lectures given at 2018 Baikal-ISAPP summer school "Exploring the Universe through multiple messengers"  provide an overview of the multi-messenger astronomy observational tools (section \ref{sec:instruments}), of the relevant physical processes (section \ref{sec:processes})  in multi-messenger astronomical sources and of the types of sources detected in various messenger channels (section \ref{sec:sources}). 

\section{Multi-messenger astronomy tools} 
\label{sec:instruments}

\subsection{Astronomy with optical telescopes} 

Over the last five hundred years since the invention of telescope by Galileo, our knowledge of the Universe around us was based on the information collected through the visible light photons. The energy of photons of wavelength $\lambda$ is  $E_\gamma=hc/\lambda\simeq 3\left[\lambda/400 \mbox{ nm}\right]^{-1}\mbox{ eV}$
where $c$ is the speed of light and $h=2\pi \hbar$ is the Planck constant\footnote{In what follows the Natural system of units is used in which $\hbar=c=1$}. Such photons are produced e.g. by objects heated up to the temperature at which the typical energy or blackbody photons $E_{bb}=3k_BT\simeq 3\left[T/10^4 \mbox{ K}\right]\mbox{ eV}$
falls into the visible band. This is the case for the thermal emission from  stars. Universe known to the mankind the centuries and millennia was the Universe of individual stars and galaxies (collections of stars). 

\begin{figure}[ht]
\begin{center}
\includegraphics[width=0.85\linewidth]{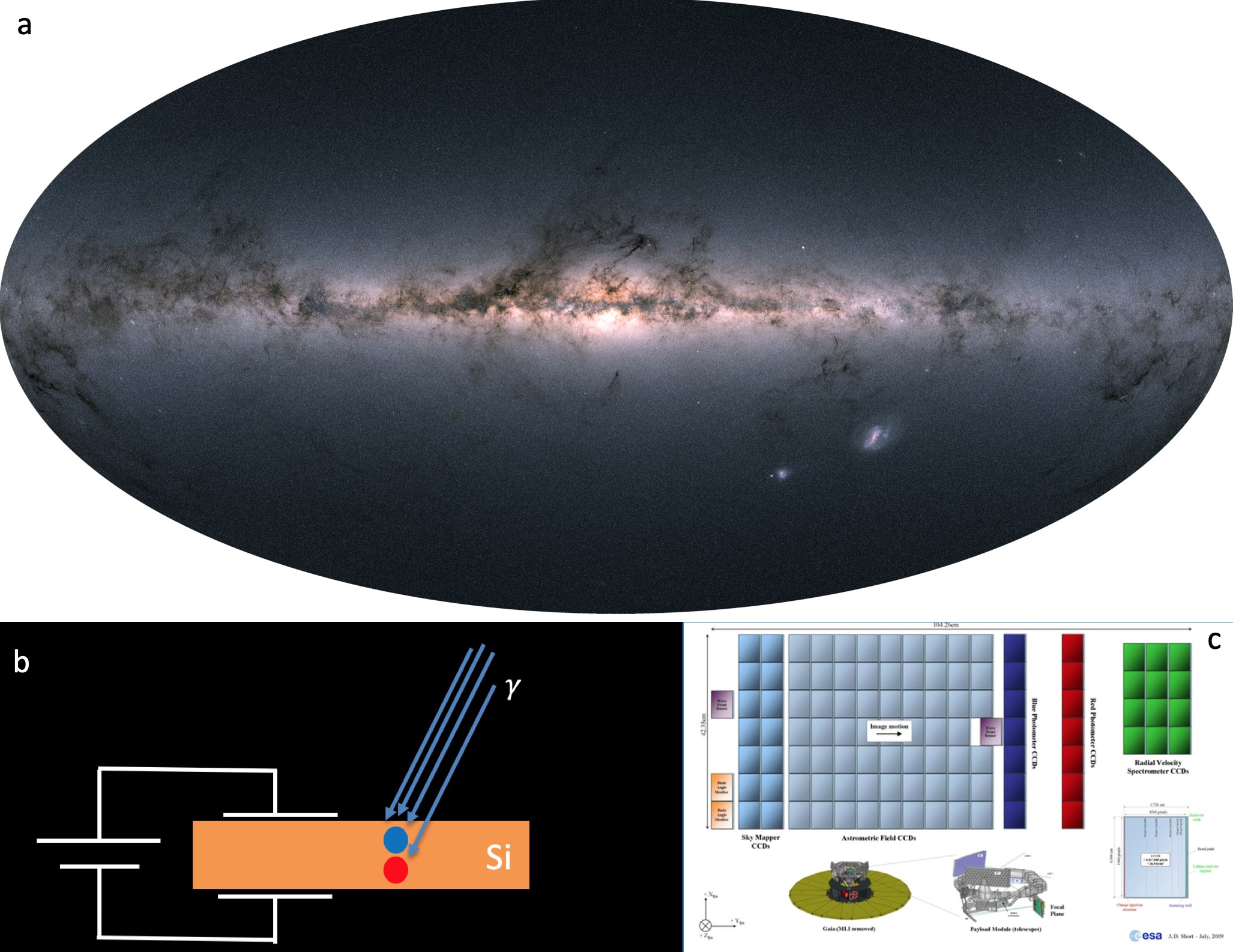}
\caption{Top: all-sky  visible band image based on the second data release of GAIA telescope \cite{gaia}. Bottom left: the principle of photon detection by CCD, right: the giga-pixel CCD of GAIA telescope. }
\label{fig:gaia}
\end{center}
\end{figure}

Modern visible band astronomy builds upon Galileo's invention, but employs telescopes with larger aperture (up to $\sim 10$~m in diameter, e.g. the Very Large Telescope VLT in Chile\footnote{https://www.eso.org/public/teles-instr/paranal-observatory/vlt/}) and uses photodetectors different from the human eye in their focal plane. Optical astronomy has been revolutionised by the Charged-Coupled Device (CCD) photodetectors, of the type similar to those used e.g. in the smartphones. These detectors allow to directly record sky images in the digital form and enable the technique of massive digitized sky surveys, pioneered by the Sloan Digital Sky Survey (SDSS)\footnote{https://www.sdss.org}. One most recent example of massive sky survey which provides precision measurements of positions of stars on the sky (astrometry) is given by GAIA telescope\footnote{http://sci.esa.int/gaia/} shown in the bottom right panel of Fig.~\ref{fig:gaia}. Its focal plane which contains a "Gigapixel" CCD with overall area $\sim 0.3$~m$^2$ \cite{gaia1}. 

The epoch of next-generation massive sky surveys is about to start, with two "flagship" projects: space-based telescope EUCLID\footnote{http://sci.esa.int/euclid/} and ground-based facility Large Synoptic SurveyTelescope (LSST)\footnote{https://www.lsst.org}. The automatised sky survey approach allows to simultaneously pursue different types of scientific research programs: survey sky in the search of "bursting" transient sources like supernovae and related phenomena, measure shapes of galaxies for the study of gravitational lensing by dark matter structures etc.  

Large aperture of modern optical telescopes allows to collect larger number of photons from weaker sources and thus reach higher sensitivity level (minimal detectable source flux). This principle will be pushed to a limit with the next-generation projects like Thirty Meter Telescope\footnote{https://www.tmt.org} (TMT) and European Extremely Large Telescope\footnote{https://www.eso.org/sci/facilities/eelt/} (ELT) which plan to use optical systems with apertures up to 40 m in size. 

The angular resolution of ground-based telescopes is limited by the distortions of the wavefront of optical light by the fluctuations of parameters of the atmosphere. This limitation is relaxed via the use of adaptive optics. In such approach optical elements of the telescope are continuously adjusted in real time to compensate for the variable waveform distortions by the atmosphere.

\begin{figure}[ht]
\begin{center}
\includegraphics[width=0.85\linewidth]{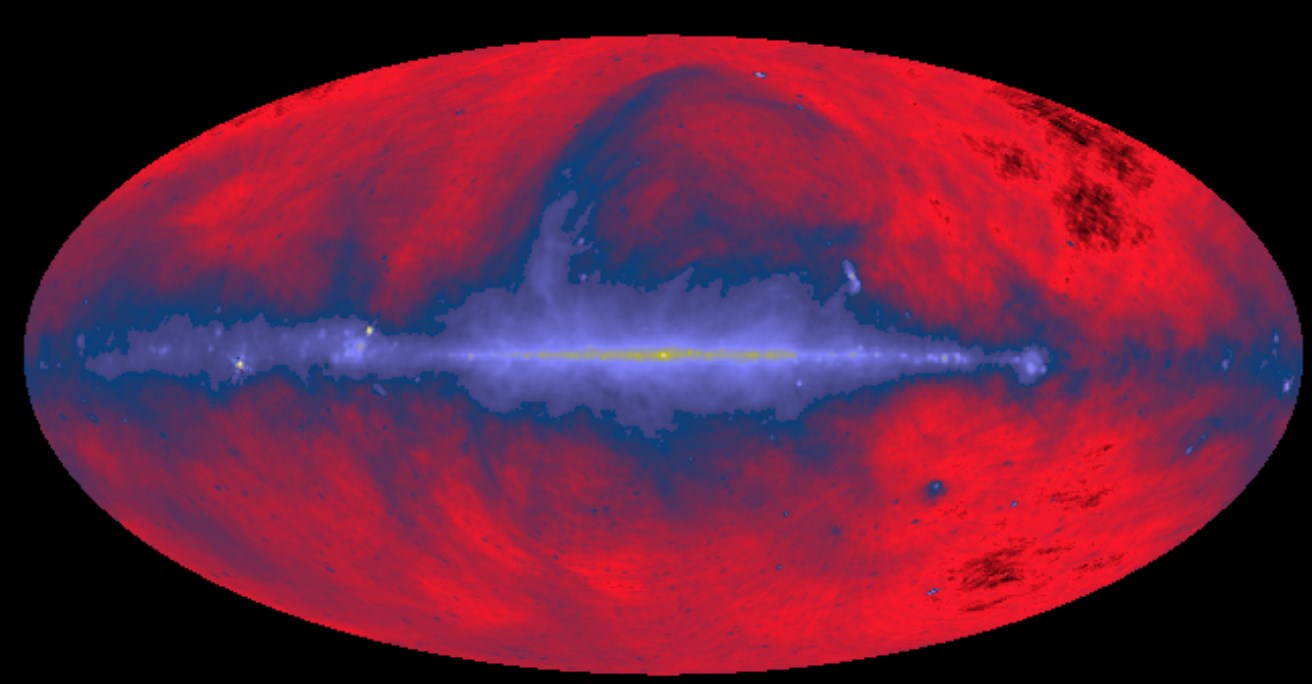}\\[2mm]
\includegraphics[height=3.5cm]{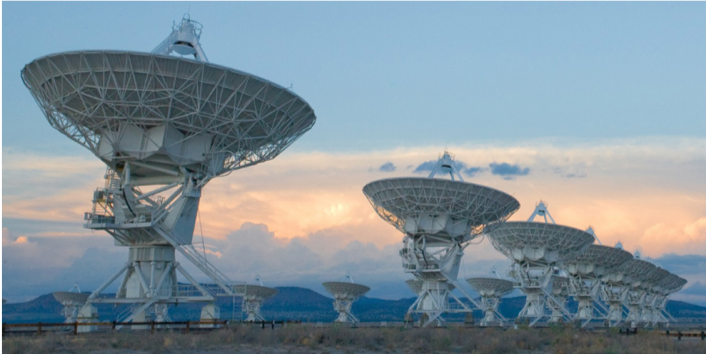}\hspace{1mm}
\includegraphics[height=3.5cm]{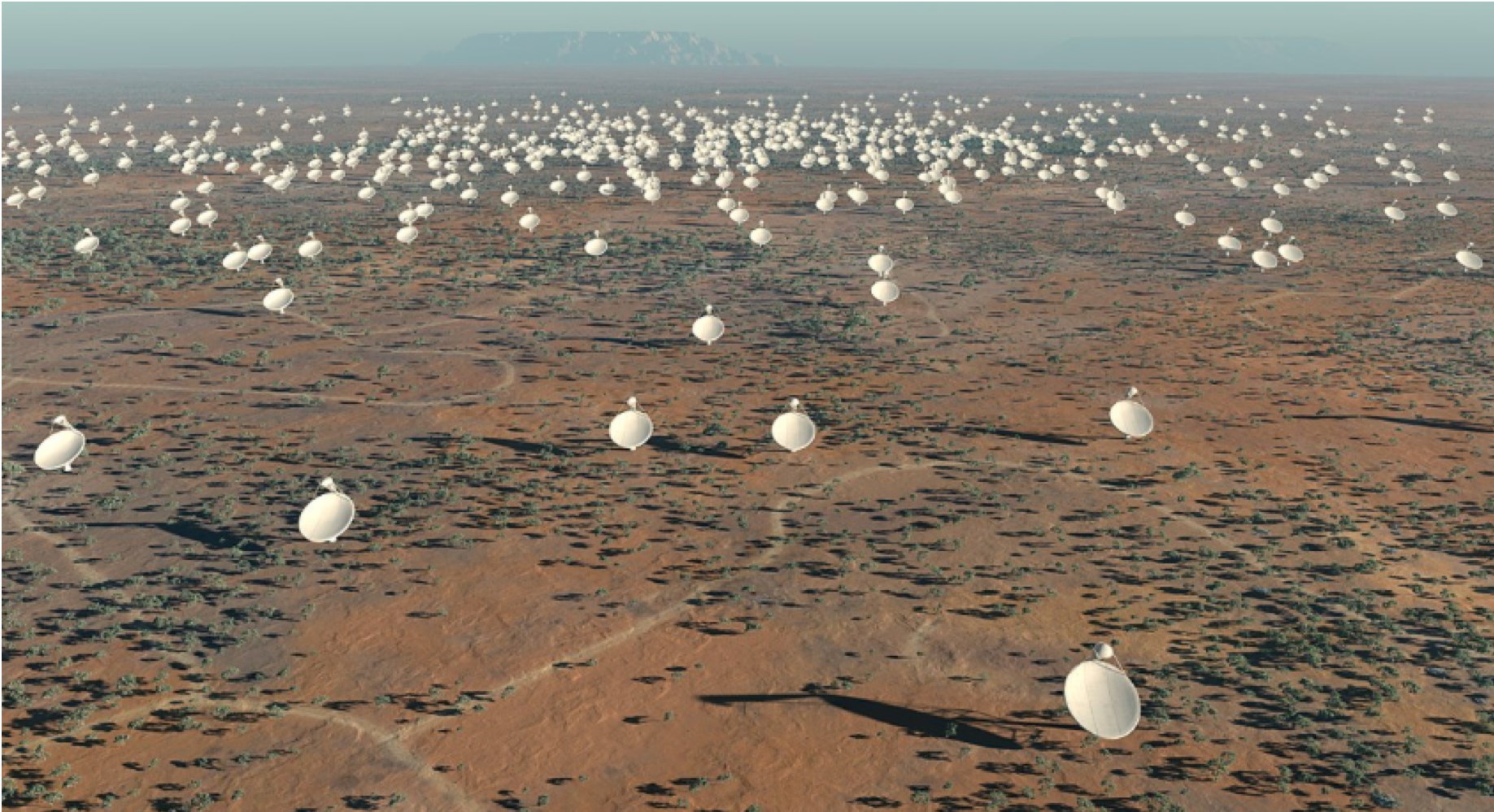}
\caption{Top: all-sky image at 408 MHz frequency \cite{408mhz}. Bottom the VLA radio telescopes. Right: layout of the next generation observatory SKA.}
\label{fig:radio}
\end{center}
\end{figure}

The energy of the visible light photons is about 1~eV, i.e. comparable to the energy gap width in the CCDs semiconductor. This does not allow precision measurement of the photon's energy on photon-by-photon basis. Besides, the number of photons incident on the CCD is large, so that the CCD measures the intensity of light, rather than counts photons. Measurement of the energy dependence of the photon flux in the optical band (i.e. spectroscopy) is done by deviating the light from selected sources, e.g. with a prism. In this way, photons of different energy (wavelength) are focused at different locations at the CCD. Sampling these locations provides a possibility to measure the flux as a function of energy, with high spectral resolution $\Delta E_\gamma/E_\gamma\ll 1$. This is particularly interesting in the visible band, in which the emission spectra of astronomical objects have atomic emission and absorption lines. High spectral resolution allows to measure the line shapes (which is influenced 	by e.g. Doppler broadening induced by random motions of line-emitting material) and position (e.g. Doppler shifts induced by the overall movement of the emitting object).

\subsection{Radio astronomy} 

The observational window on the Universe has started to extend  in the course of 20th century with the invention of radio telescopes. Astronomical observations in the frequency $\nu\sim 0.1-10$ GHz frequency range  have revealed new classes of astronomical objects different from stars and emitting radiation at much longer wavelength  $\lambda\simeq 21\left[\nu/1.4\mbox{ GHz}\right]^{-1}\mbox{ cm}$ with spectra  that are not following the blackbody spectrum. Instead, they are generically of "power law" type: $dN_\gamma/dE_\gamma\propto E_\gamma^{-\Gamma}$
where $\Gamma$ is the powerlaw slope (or "spectral index").  

Modern radio telescopes are networks of different types of radio antennae, including "dish" and "pole" types. An example of existing Very-Large Array (VLA) radio observatory\footnote{https://public.nrao.edu/telescopes/vla/} and future Square Kilometer Array (SKA)\footnote{https://www.skatelescope.org} observatories are shown in Fig.~\ref{fig:radio}. Each dish antenna detects radio waves from particular sky  direction. Larger-size antennae of diameter $D$ observing at the wavelength $\lambda$ could achieve better angular resolution which is limited by the diffraction limit: $\Theta_{min}=\lambda/D\simeq1^\circ\left[\lambda/21\mbox{ cm} \right]\left[D/10\mbox{ m}\right]^{-1}$
Organizing antennae into a network in which each antenna is able to record the waveform (amplitude and phase) of the signal form given sky direction, allows to obtain a much better angular resolution using the "interferometry" technique, i.e. combining the waveform measurements from all antennae. This efficiently shifts the diffraction limit down to the milli-arcsecond range
$\Theta_{min}\simeq 20 \left[\lambda/1\mbox{ mm} \right]\left[D/1\mbox{ km}\right]^{-1}\mbox{ mas}$
for observations at millimeter wavelength with an array of antennae spread across kilometer-scale area. This is the configuration of ALMA\footnote{https://www.almaobservatory.org/en/home/} observatory in Chile.  The separation of antennae is called "baseline", and the techniques is called Very Long Baseline Interferometry (VLBI). The largest baseline is achieved by organising telescopes scattered across the Earth surface to work in interferometry mode, thus reaching the separations up to $10^4$ km. In this case the angular resolution reaches $\sim 10\ \mu$as range. This resolution is achieved with a network called Event Horizon Telescope\footnote{https://eventhorizontelescope.org}, which is aimed at resolving the event horizons of supermassive black holes in the Milky Way and in the nearby  elliptical galaxy M87.  Still larger baseline could be achieved via deployment of an antenna in space, an idea implemented for the first time by the RadioASTRON\footnote{http://www.asc.rssi.ru/radioastron/} project which uses an antenna in an elliptical orbit around the Earth with the apogee reaching the altitude about the Earth-Moon distance.

The non-thermal sources  observed by radio telescopes  include sources in the Milky Way galaxy: different supernova-related objects, like pulsars, pulsar wind nebulae and supernova remnants. Among extragalactic sources the dominant class is radio galaxies, which are "radio-loud" Active Galactic Nuclei (AGN) powered by activity of supermassive black holes in the centres of galaxies. 

\subsection{X-ray astronomy} 

Further extension of the astronomical observational window toward higher energies (X-rays)  became possible with the start of the space age at the end of 60th of the last century. Contrary to the visible light or radio waves, the X-ray photons are interacting in the Earth atmosphere and are not able to directly reach the ground level. Thus, telescopes observing in the X-ray band have to be deployed outside the atmosphere, in orbit around the Earth.

\begin{figure}[h!]
\begin{center}
\includegraphics[width=0.9\linewidth]{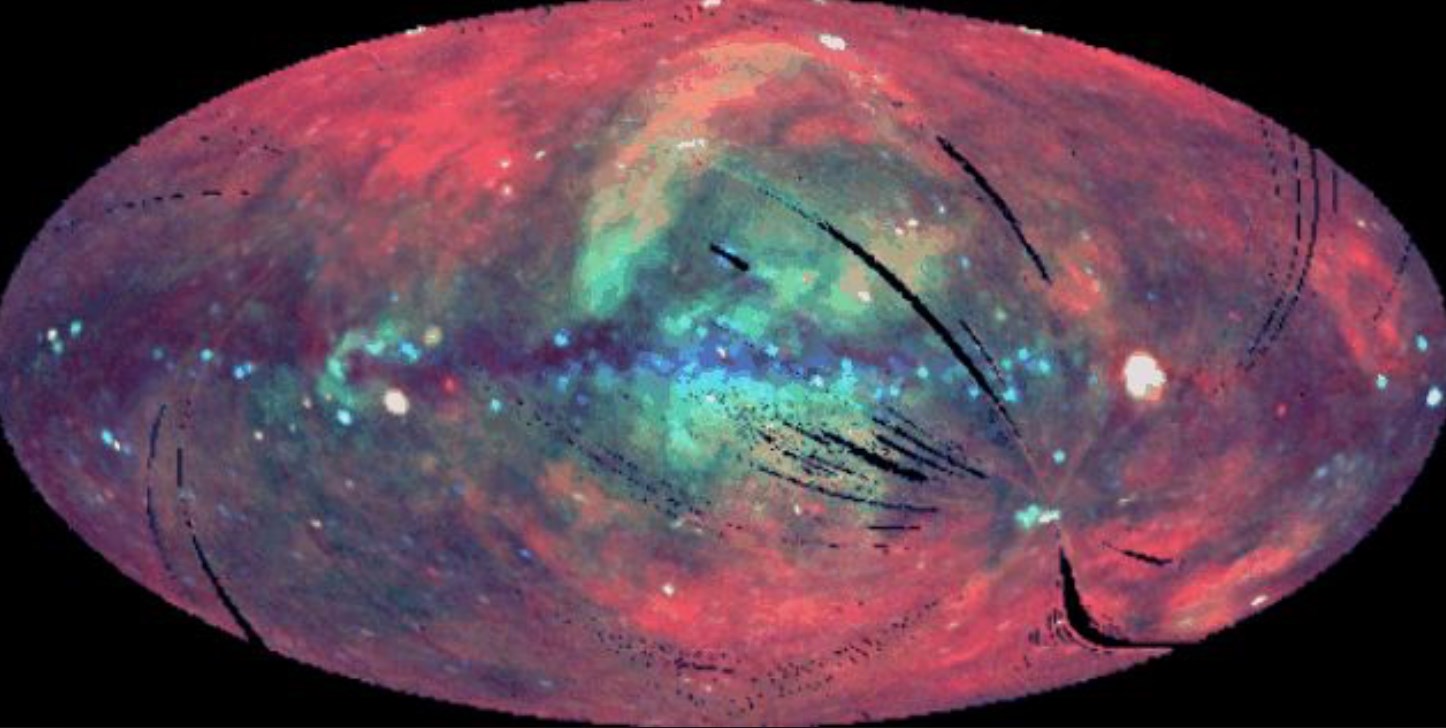}\\[2mm]
\includegraphics[height=4.4cm]{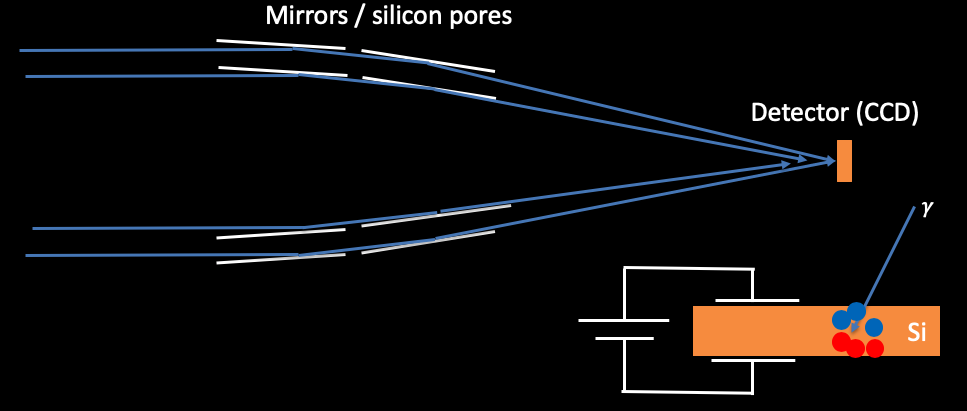}\hspace{1mm}
\includegraphics[height=4.4cm]{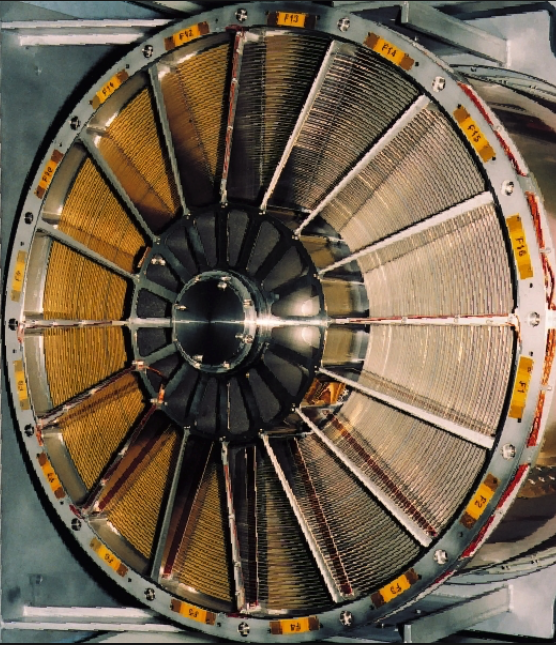}
\caption{Top: X-ray map of the sky produced by ROSAT all-sky survey \cite{rass}. Bottom left: grazing incidence optics principle of X-ray telescopes. Also shown is the principle of spectroscopic measurements with CCD in the focal surface. Bottom right right: mirror stack of the XMM-Newton X-ray telescope.}
\label{fig:X_ray_optics}
\end{center}
\end{figure}

The wavelength of the X-ray photons is  $\lambda=E_\gamma^{-1}\simeq 10^{-7}\left[E_\gamma/1\mbox{ keV}\right]\mbox{ cm}$, i.e. 
about the size of atoms. Because of this, X-ray photons could not be focused using refractive optics. Instead of being deviated by collective interactions with atoms in materials, they interact with individual atoms ionising them. To focus X-rays, the X-ray telescopes use the effect of total reflection to focus X-rays, the technique known as "grazing incidence" optics (Fig.~\ref{fig:X_ray_optics}). The optical system of X-ray telescope is a stack of concentric mirrors (shown in cross-section in Fig.~\ref{fig:X_ray_optics}). Each mirror is oriented almost parallel to the direction of arrival of X-ray photons, so that X-rays incident on the mirror are totally reflected and their direction is slightly deflected. Further set of mirrors deflected the X-ray by still further angle so that the X-ray signal from a particular sky direction arrives at one point in the focal plane. Current generation X-ray telescopes (Chandra\footnote{http://chandra.harvard.edu}, XMM-Newton\footnote{https://www.cosmos.esa.int/web/xmm-newton}) use metallic mirror stacks, the right panel of Fig.~\ref{fig:X_ray_optics} shows an example of the gold-coated mirror stack of XMM-Newton telescope. Next generation X-ray observatory Athena\footnote{https://www.the-athena-x-ray-observatory.eu} will use a different technology, in which the X-rays are guided (by the same total reflection effect) through the pores in silicon (silicon pore optics). This allows to achieve modular design of the optical system, enabling large aperture optics (the Athena aperture will be about 1.5 m$^2$).

 The flux of radiation in the X-ray band is much lower than that in the visible band. This could be understood by considering an astronomical source which has certain power, or luminosity $L$, at comparable levels in the visible and X-ray bands. The photon flux from the source is  $F=L/(4\pi D^2 E_\gamma)$
where $D$ is the distance to the source. The energy $E_\gamma$ of X-ray photons is three orders of magnitude higher than that of the visible light photons. Thus the X-ray photon flux is by three orders of magnitude lower. 

The focal surface detectors of X-ray telescopes are similar to those of the optical telescopes: both use CCDs. However, in the case of X-rays, the CCD is used in a different way. Instead of measuring the intensity of light, the CCD detects individual X-ray photons. In this respect, lower rate and higher energy  of X-ray photons is beneficial for the spectroscopic measurements with CCD detector. The X-ray energy is by a factor $\sim 10^3$ higher than the energy gap in the silicon, so that one X-ray photon  interacting in a CCD pixel produces about $N_{e-h}\sim 10^3$ electron-hole pairs (Fig.~\ref{fig:X_ray_optics}). The number of electron-hole pairs and, as a result, the current pulse produced by the X-ray hit is directly proportional to the energy of the X-ray. This allows to measure the energy of each incident X-ray directly, on hit-by-hit basis, with precision $\Delta E_\gamma/E_\gamma\sim 1/\sqrt{N_{e-h}}\sim 3\%\left[N_{e-h}/10^3\right]^{-1/2}$
Thus, there is no need to deviate photons with the prism (as it is done in the optical spectroscopy). Instead, X-ray observations provide a possibility to perform "field spectroscopy", i.e. measure spectra of all observed sources simultaneously. Still, both Chandra and XMM-Newton telescopes use the technique of deviating the photons with diffactive grating (rather than prism) to perform high-resolution spectroscopy (with $\Delta E_\gamma/E_\gamma\ll 1\%$). 

Alternatively, new technologies for the high-resolution X-ray spectroscopy are developing, such as that of "transition edge sensors". In this case the camera pixels are made from superconducting material which is kept at the temperature just below the normal-superconducting phase transition. Each time an X-ray hits the pixel, it "reheats" the material and produces a small temperature increase. This leads to the loss of superconductivity, which is readily detectable as a jump in resistivity. This technology was used for the first time by HITOMI telescope \cite{hitomi} which was briefly operating in space in 2016, before mission termination due to an unrelated failure. The same type of detector will be implemented in the successor of HITOMI, XRISM\footnote{https://heasarc.gsfc.nasa.gov/docs/xrism/} and in the X-ray Integral Field Unit (X-IFU) instrument on board of the next-generation X-ray observatory Athena\footnote{http://x-ifu.irap.omp.eu}. The interest in high  resolution X-ray spectroscopy stems from the fact that spectra of objects emitting in the X-ray band, have emission and absorption lines, which are broadened by the random motions and red / blue-shifted by the bulk motions of matter in the astronomical objects. 

\begin{wrapfigure}{L}{0.6\linewidth}
\includegraphics[width=\linewidth]{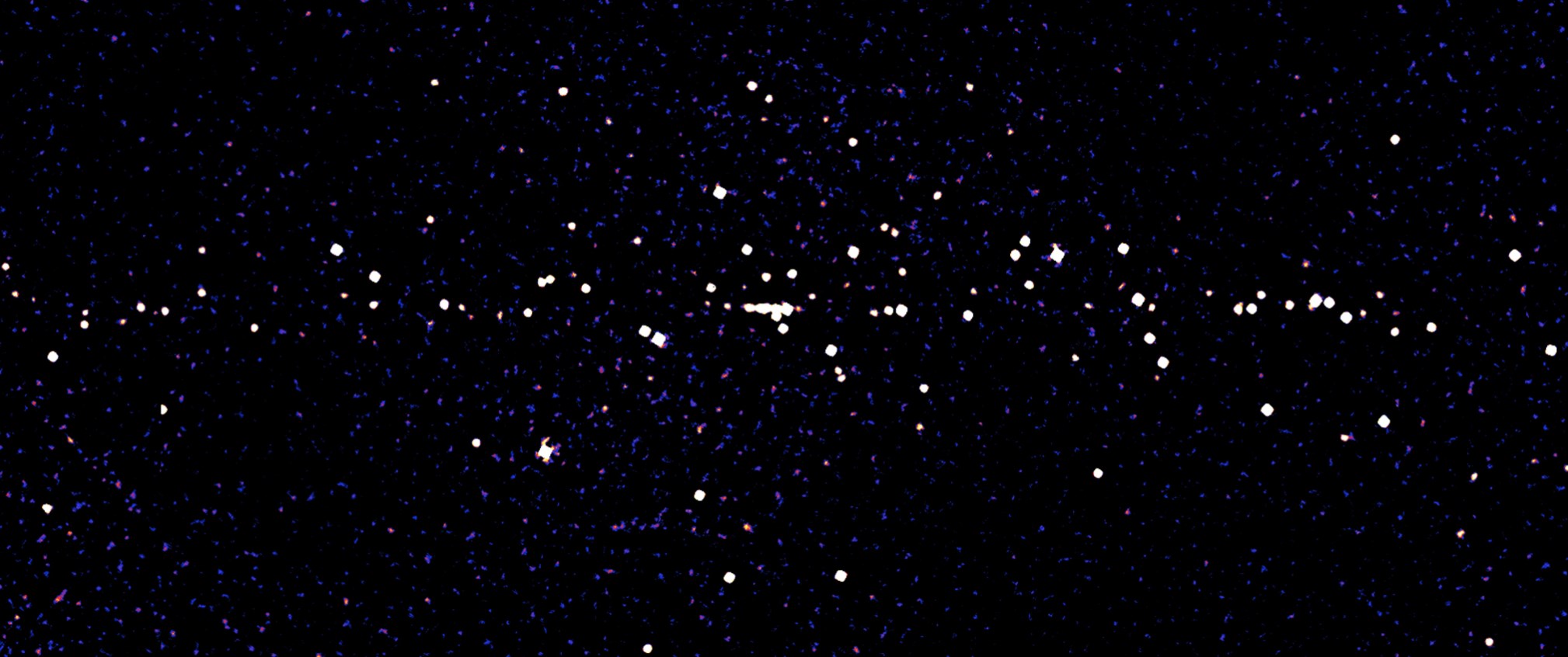}
\includegraphics[width=0.3\linewidth]{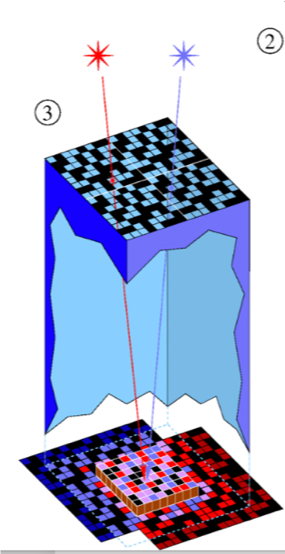}
\includegraphics[width=0.59\linewidth]{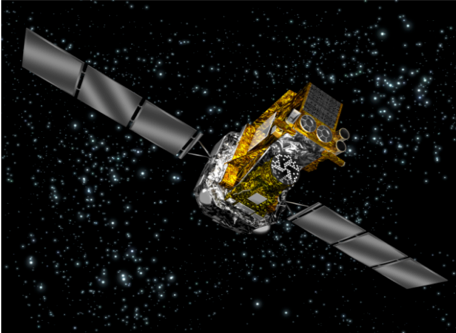}
\caption{Top: mosaic of the inner Galaxy sky region (in Galactic coordinates, Galactic Plane goes through the middle of the image) observed by INTEGRAL telescope in the energy band above 20 keV. Most of the sources in the map are X-ray binary systems. Bottom left: coded mask telescope principle. Bottom right: INTEGRAL telescope based on the coded mask technique.}
\label{fig:coded_mask}
\end{wrapfigure}

The Universe observed by X-ray telescopes includes both thermal and non-thermal sources. The thermal sources visible in the X-ray band have temperature about $T\sim 10^7$~K, so that the characteristic energy of the blackbody radiation photons is  in the keV range. Such temperatures are encountered in stellar coronae, including the corona of our own Sun, but the range of sources heated to ten-million degree temperatures extends much beyond the conventional Sun-like stars. These temperatures are reached in compact stars, like the neutron stars of white dwarfs, in supernova remnants, in galaxy clusters. The non-thermal sources visible in the X-ray band include the same source classes as observed by the radio telescopes: AGN, pulsars, supernovae. 

\subsection{Hard X-ray / soft \gr\  telescopes} 
\label{sec:compton}

Focusing photons of energies much higher than that of the  X-rays is not possible and telescopes operating in the "hard X-ray" (100 keV) band use different approach for imaging. This approach is close to that of the "pinhole camera". Its principle is illustrated  in Fig.~\ref{fig:coded_mask}. Hard X-rays enter the telescope through a plate, called "coded mask" which blocks part of the photon flux following a certain geometrical pattern. As a result, this pattern re-appears as a shadow of specific shape on the detector plane. "Shadowgrams" produced by the sources in different directions on the sky are displaced with respect to each other. The shape of the shadowgram pattern is chosen in such a way that it is most easily recognisable during the data analysis. With such an approach, signals from different sources are "mixed" on the detector plane, so that statistical fluctuations of the signal of brighter sources prevent detection of weaker sources (contrary to the focusing optics where each source produces a localised signal on the detector). The coded mask technique is implemented in the hard  X-ray telescopes of INTEGRAL sattelite\footnote{http://sci.esa.int/integral/} shown in the second panel of Fig.~\ref{fig:coded_mask}  and in the SWIFT/BAT telescope\footnote{https://swift.gsfc.nasa.gov/about\_swift/bat\_desc.html}. 

Detection of higher and higher energy photons poses a challenge for the focal surface instrumentation of telescopes.  Visible light and X-ray photons interact with the silicon material of CCD by ionising and /or exciting the atoms and molecules. The cross-section of such interactions decreases with energy so that the higher energy photons  are penetrating deeper and deeper into the material. They would finally penetrate through the hole detector material without interacting once. This means that the "efficiency" of detector drops with energy. The only way to intercept the higher energy photons is to put more material on their way. This implies larger and larger volume and mass of the detector material. The mass-vs-efficiency optimisation results in the use of specific materials in the hard X-ray telescope detectors doped with heavy nuclei (like e.g. the Cadmium-Telluride CdT) detector of INTEGRAL/ISGRI telescopes.

\begin{wrapfigure}{R}{0.6\linewidth}
\includegraphics[width=\linewidth]{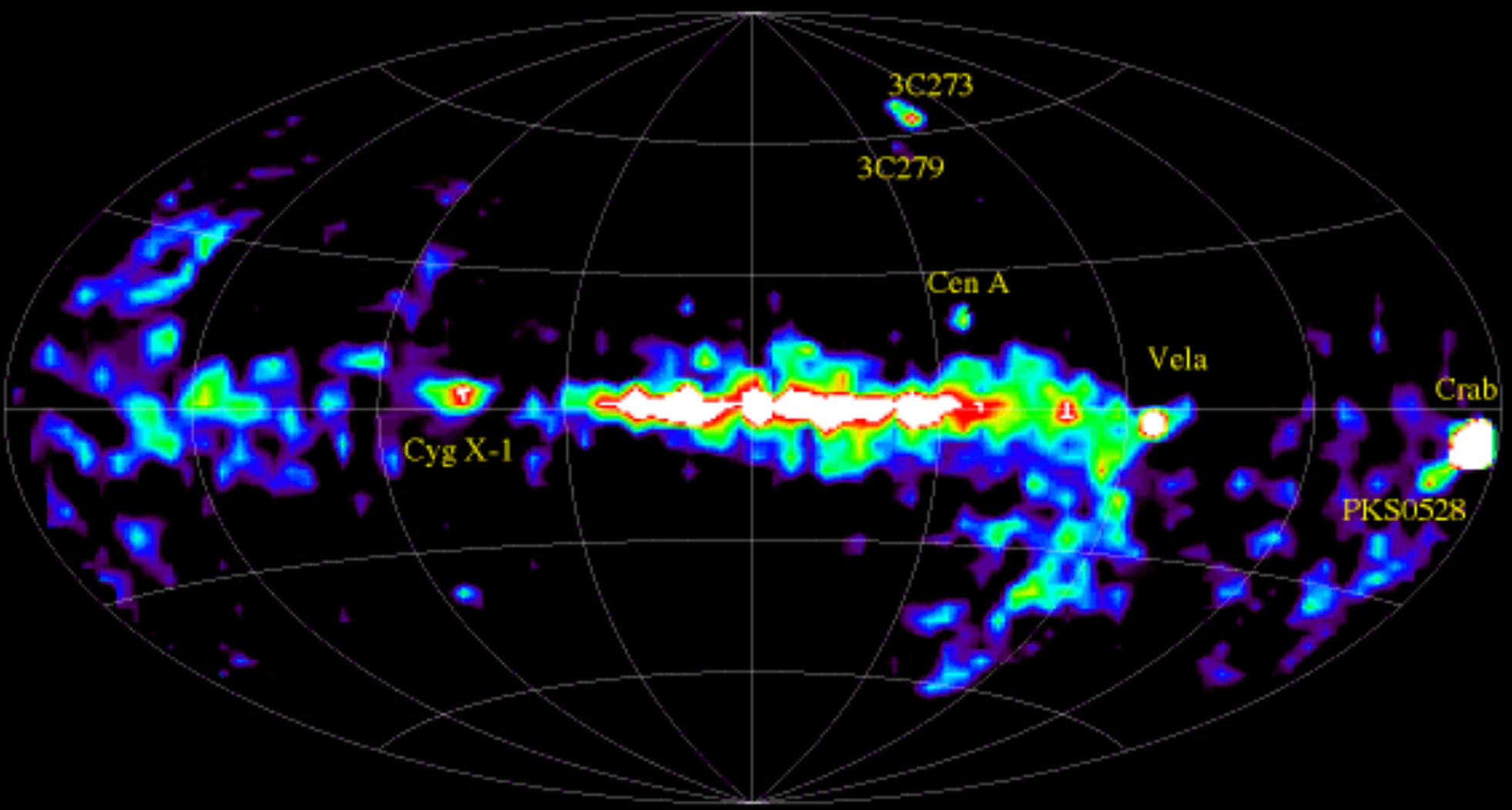}
\includegraphics[width=0.45\linewidth]{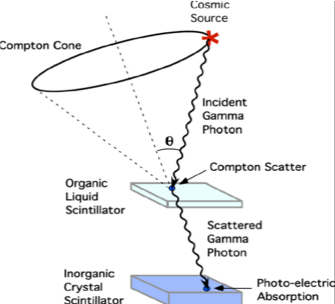}
\includegraphics[width=0.5\linewidth]{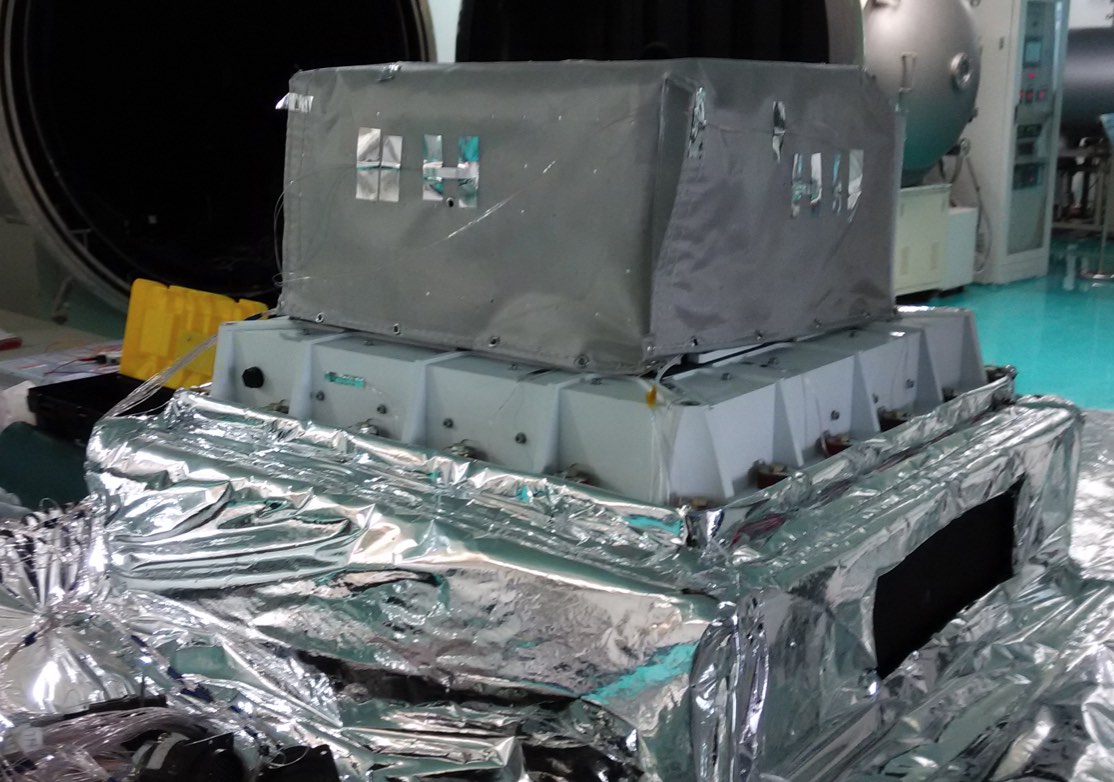}
\caption{Top map of the sky produced from COMPTEL telescope observations \cite{comptel}. Bottom left: Compton telescope principle. Bottom right: POLAR gamma-ray burst Compton telescope which operated on Chinese space station in 2017 \cite{polar} }
\label{fig:compton}
\end{wrapfigure}

The photoelectric interaction cross-section drops below the Compton scattering cross-section in most materials above the photon energy $E_\gamma\sim 100$~keV. This further reduces the efficiency of telescopes and detectors relying on the photoelectric effect, but it opens a possibility for a completely new observational technique of "Compton telescopes". 

The Compton telescope technique uses the measurement of parameters of Compton scattering interactions to measure the energy and arrival direction of every incoming photon. The measurement principle is illustrated in the first bottom panel of Fig.~\ref{fig:compton}. A \gr\ of energy $E_\gamma$ scattered off an electron at rest bounces off with an energy 
$E_\gamma'=E_\gamma/\left(1+E_\gamma (1-\cos\theta)/m_e\right)$,
where $\theta$ is the scattering angle and $m_e$ is electron mass. 
The energy transferred  to the electron $E_e=E_\gamma-E_\gamma'$ could be measured in the detector, using techniques similar to those described above for photon detection (e.g. in the semiconductor detectors, or, more conventionally, in scintillators where the amount of scintillation light produced by high-energy electron is directly proportional to the electron energy). The lower energy scattered photon could interact second time in the detector volume via photoelectric effect in which the photon is absorbed. The energy of the absorbed photon, $E_\gamma'$ could be measured in the same way: the current pulse strength  in semiconductor detector, the amount of scintillation photon signal. In this way, two parameters of the Compton scattering are readily available: $E_\gamma', E_\gamma$. This allows to reconstruct the third parameter: the angle $\theta$. If the detection volume is segmented in pixels, the reference direction from the first to second interaction point could be recorded. Knowing the scattering angle $\theta$ one could reconstruct the arrival direction of the primary \gr\ up to an azimuthal angle around the reference direction. Thus, the Compton telescope technique does not provide full reconstruction fo the photon arrival direction, it only constrains it to be within a "circle on the sky", as shown in the first bottom panel of Fig.~\ref{fig:compton}.

Limited sensitivity of the coded mask and Compton telescope techniques explains smaller and smaller number of sources detected on higher energy sky. The "state-fo-art" Compton telescope COMPTEL which operated in 90th of the last century on board of CGRO mission has detected only a handful of isolated sources and the diffuse emission from the Milky Way in the $E_\gamma\sim 1$~MeV energy band during several years of operation (see Fig.~\ref{fig:compton}, top panel) \cite{comptel}. The sparsity of our knowledge of the soft \gr\ sky is known in astronomy as the "MeV sensitivity gap". Compton telescope technique has been recently re-used in a COSI high-altitude balloon-based telescope\footnote{http://cosi.ssl.berkeley.edu} and in the POLAR gamma-ray burst detector  \cite{polar} which operated on the Chinese space station, see Fig.~\ref{fig:compton} (POLAR-2, an upgrade of POLAR Compton telescope, is planned for launch in 2024). 

The MeV energy range hosts a peculiar signal  from the inner Milky Way in the form of a spectral line (almost mono-energetic emission) at the energy $E_\gamma=511$~keV, i.e. at the rest energy of electron \cite{511_discovery,511line}. This emission is produced by "positronium" atoms which could form for short periods of time if free positrons are available in the medium (in positronium an electron and positron form a bound state). This emission from the inner Galaxy direction(about 10 degrees around the Galactic center) was studied at low spectral resolution by COMPTEL telescope \cite{511kev_comptel}  and further with with very high energy resolution by SPI instrument on board of INTEGRAL \cite{511kev_integral}. In principle, positrons are produced in interactions of high-energy particles in the interstellar medium and in the sources operating particle accelerators. Thus, it is not surprising to find positrons in the inner Galaxy. What is surprising is that these positrons are able to form positronium atoms. This implies that they are injected with energies close to their rest energy (more precisely, below 2 MeV \cite{511_beacom}), a fact which is difficult to explain with conventional models of positron population in the Galaxy. 

\subsection{Gamma-ray telescopes} 

\begin{figure}[ht]
\begin{center}
\includegraphics[width=0.9\linewidth]{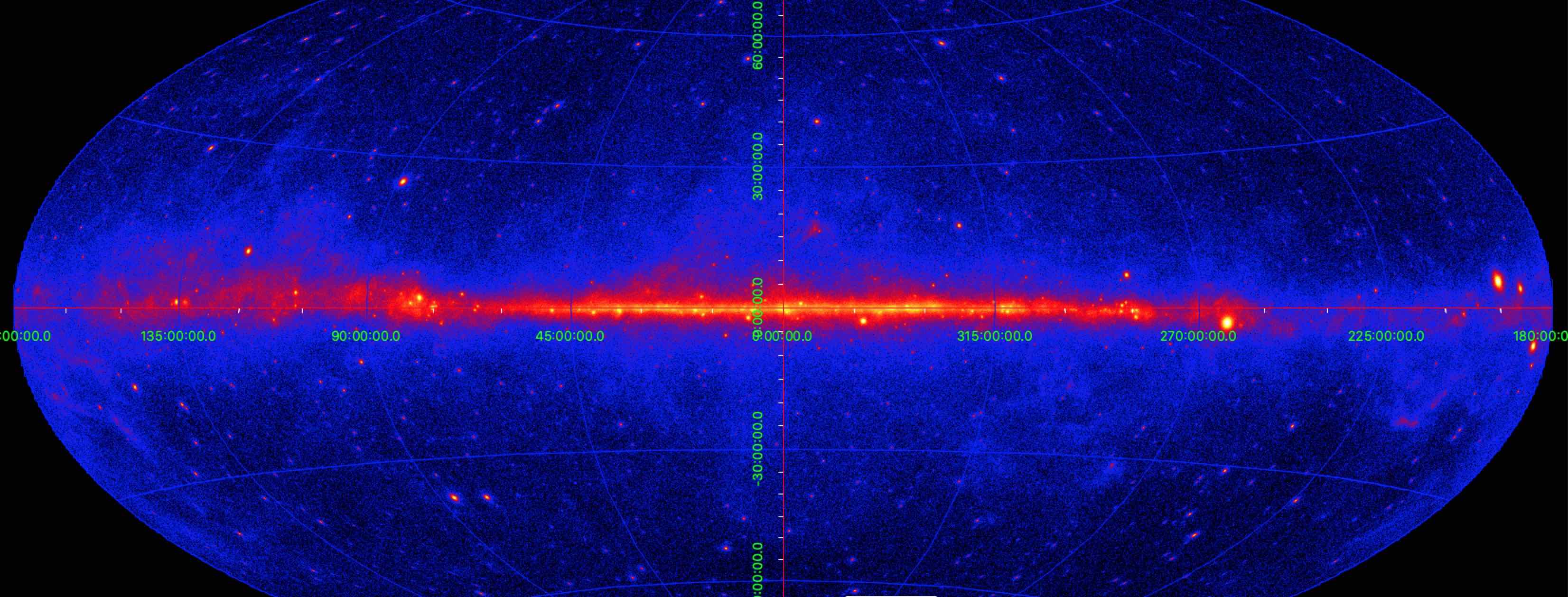}\\[2mm]
\includegraphics[height=4.5cm]{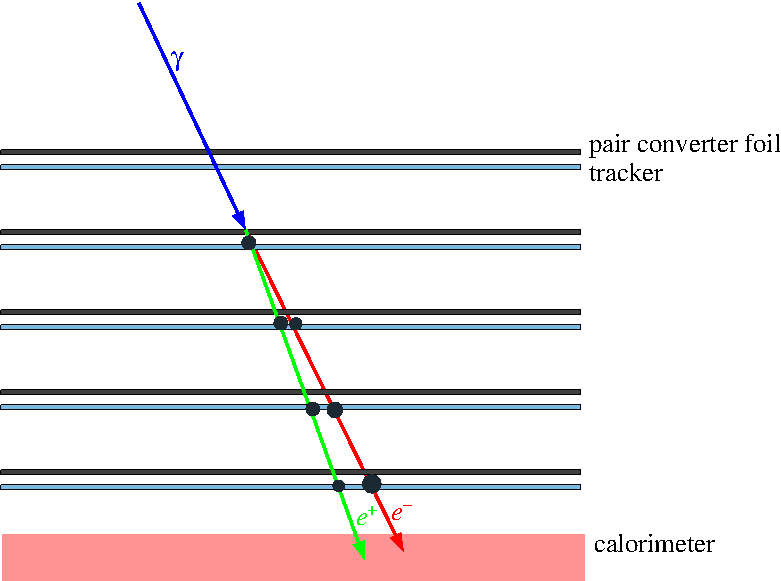}\hspace{10mm}
\includegraphics[height=4.5cm]{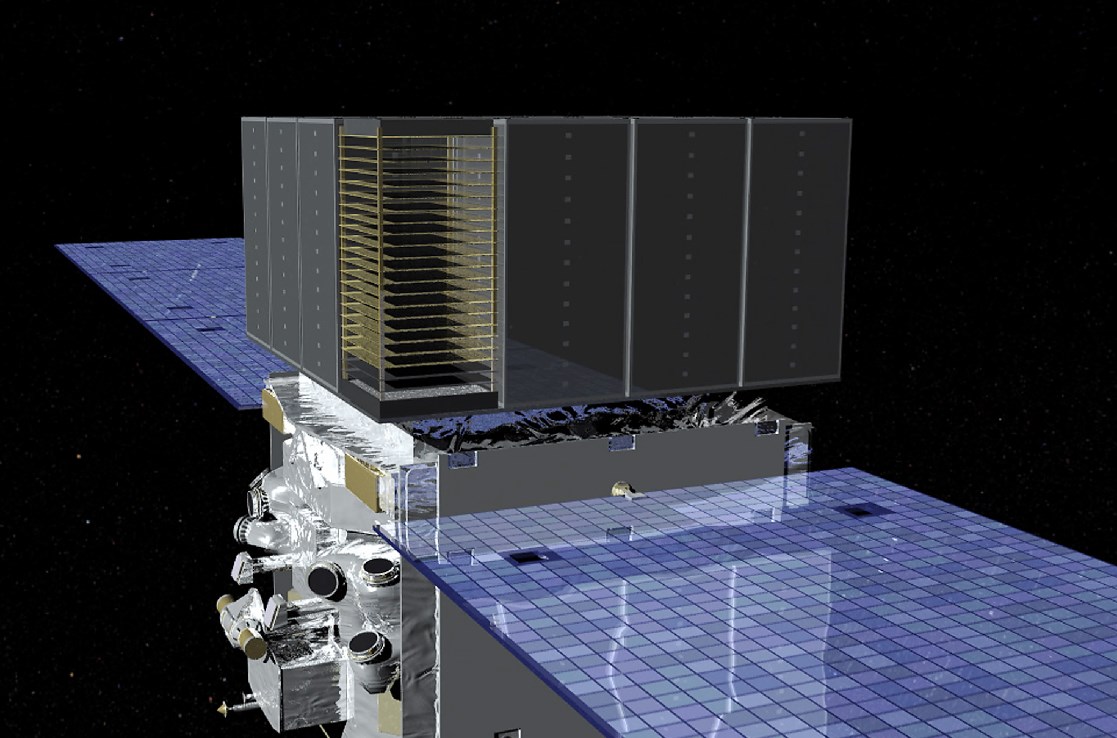}
\caption{Top: sky map in the energy range $E_\gamma>1$~GeV observed by Fermi/LAT telescope. Bottom left: the principle of \gr\ detection by pair conversion telescope. Bottom right: Fermi Large Area Telescope: a pair conversion telescope currently operating in orbit.}
\label{fig:fermi}
\end{center}
\end{figure}

At the energies much higher than MeV, the Compton scattering cross section diminishes and becomes smaller than the cross-section of electron-positron pair production. The pair production interaction channel provides a new possibility for photon detection and measurement of its energy and arrival direction. This possibility is realised in "pair conversion" telescopes, such as  Fermi space telescope shown in Fig.~\ref{fig:fermi} \cite{fermi_atwood}. Gamma-rays entering the telescope volume are converted into electron-positron pairs. Tracking the direction of motion of the two particles and measuring their energy in a "calorimeter" provides information about the direction of initial photons and its energy.  The tracker of Fermi/LAT is a "tower" of layers, each layer is a sandwich of high atomic number material foil, below which thin strips of silicon a layered in two perpendicular directions. Similarly to the CCD device, electrons / positrons passing through silicon produce electron-hole pairs which which produce a pulse of current, because of the voltage applied to each strip. The two perpendicular strips which register the current pulse determine the $x,y$ coordinates of the electron / positron passage through each layer of the tracker. The calorimeter is made of scintillator material in which the amount of light produced by electrons/positrons is proportional to their energy. 

The sky observed in  \gr s with energies larger than the rest energy of proton, $m_p\simeq 1$~GeV, is by definition, dominated by emission from sources which host particle accelerators. Photons (electromagnetic waves)  are produced by charged particles, and to generate photons with energy above 1 GeV, the particle should have energy higher than that. High-energy particles ejected by cosmic particle accelerators typically do not form thermal distributions. The high-energy \gr\ sources are all "non-thermal". Among those sources one finds the same types of non-thermal objects as in the radio band: pulsars, supernovae, AGN. Not only individual sources operating particle accelerators contribute to the \gr\ sky flux. High-energy particles injected by those sources in the interstellar medium of the Milky Way interact with matter and radiation present in this medium. This results in production of diffuse \gr\ emission from the Milky Way disk, clearly seen in the top panel of Fig.~\ref{fig:fermi}. 

\subsection{Very-high energy gamma-ray telescopes} 

The flux of \gr s from individual sources luminous in the very-high-energy band (above 100 GeV) is typically too low to be detectable with sufficiently high statistics by space-based telescope like Fermi/LAT. The brightest sources: the Crab pulsar wind nebula in the Milky Way and an AGN Mrk 421 outside the Galaxy generate photon flux at the level $F_{crab}\sim 3/\mbox{ m}^2\mbox{ yr}$
  in the TeV range. In such conditions, accumulation of signal statistics with the space-based telescope, which is limited to have geometrical area about 1 m$^2$ is very slow. Larger signal statistics could be achieved with ground-based telescopes which use the Earth atmosphere as "tracker" and "calorimeter", similar to those of Fermi/LAT telescope. 
  
Very-high-energy \gr s entering the atmosphere interact, similarly to the \gr s entering the tracker of Fermi/LAT telescope,  via the Bethe-Heitler pair production. Contrary to the space-based detector, it is not possible to track the electron and positron directly, because they interact themselves via production of Bremsstrahlung photons, which are themselves \gr s able to produce further electron-positron pairs. The result of the multiple interactions of electrons, positrons and \gr s in the air is an electromagnetic cascade, or an "Extensive Air Shower" (EAS) which develops along the direction of the primary \gr s. 

\begin{figure}[ht]
\begin{center}
\includegraphics[width=0.9\linewidth]{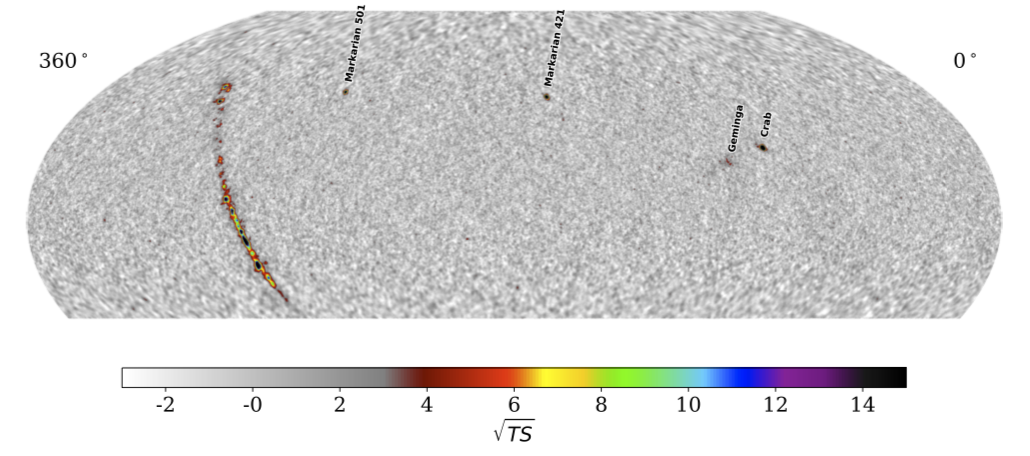}\\[1mm]
\includegraphics[height=2.2cm]{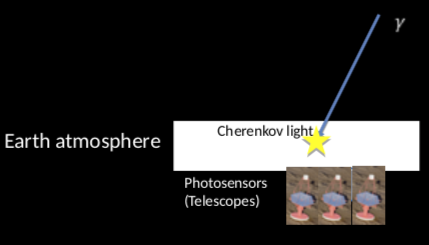}
\includegraphics[height=2.2cm]{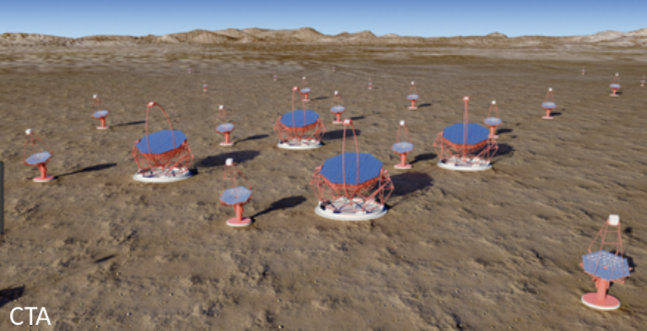}
\includegraphics[height=2.2cm]{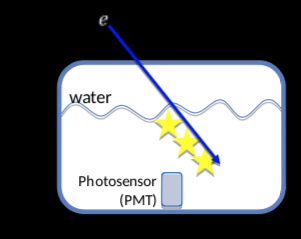}
\includegraphics[height=2.2cm]{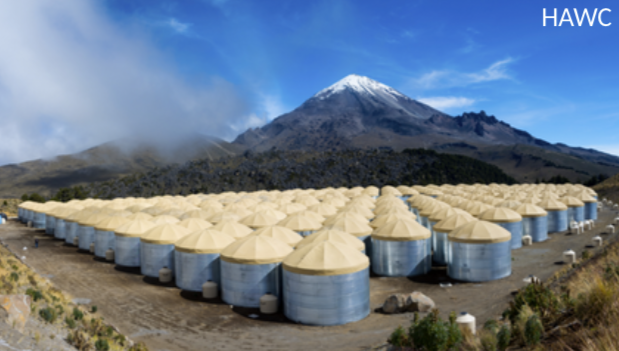}
\caption{Top: significance map of the sky observed by HAWC telescope \cite{hawc_survey}. Bottom left: the principle of \gr\ detection by ground-based Cherenkov telescopes. Bottom 2nd panel: Layout of CTA IACT network. Bottom third panel: principle of operation of water Cherenkov detector. Bottom right: HAWC water water Cherenkov detector array.}
\label{fig:cta}
\end{center}
\end{figure}

The number of particles in the EAS roughly doubles with each next generation of electromagnetic cascade (after each interaction). If the energy of the primary particle which initiated the EAS is $E_0$, the number of particles after $n$ interactions is $N\sim 2^n$ and the characteristic energy of the particles is $E=E_0/N$. Once the energy scale of the EAS drops below a critical energy at which the energy loss to ionisation becomes more important than the energy loss through the Bremsstrahlung (this happens at the energy $E_*\sim 100$~MeV), the cascade development is stopped. The maximal number of particles in the EAS is then  $N_{max}=2^{n_{max}}\sim E_0/E_*$.
It is roughly proportional to the primary particle energy. Electrons with energy about $E_*$ have gamma factors $\gamma_e\sim 10^2$ and their trajectories are aligned with the EAS axis to within an angle 
$\theta_*\sim 1/\gamma_e\lesssim 1^\circ$.
By the same reason the EAS axis is aligned with the direction of motion of the primary \gr.  Measuring the direction of development of the EAS and the amount of particles in it one obtains a way to infer the energy and direction of the primary \gr\ and hence use the atmosphere as part of a giant pair conversion telescope, see the left bottom panel of Fig.~\ref{fig:cta}). 

Several techniques are used for the measurement of the energies and arrival directions of the EAS. One possibility is to track the ultraviolet Cherenkov emission produced by electrons and positrons moving with the speed faster than the speed of light in the air (which is  $c_{air}=c/n_{air}$
with $n_{air}\simeq 1+3\times 10^{-4}$). The weak UV glow of the EAS is usually sampled with Imaging Atmospheric Cherenkov Telescopes (IACT), which are large (up to 25 m in diameter) deflectors, like those shown in the second bottom row panel of Fig.~\ref{fig:cta} where the layout of the next generation Cherenkov Telescope Array (CTA)\footnote{https://www.cta-observatory.org} is shown.  The Cherenkov light is beamed in the forward direction to within the angle $\cos\theta_{Ch}=1/n_{air}$, $\theta\simeq 1^\circ$. Thus, IACTs observe EAS which have axes nearly aligned with the direction toward the telescope. This alignment produces strong Doppler effect. The EAS crosses the atmosphere scale height $H\sim 10$~km on the time scale $T=H/c\sim 30\ \mu$s. However, the Doppler effect shortens the signal time scale to $\sim 10$~ns. Because of this, the cameras of IACTs are equipped with ultra-fast readout electronics able to take images on the 10~ns time scales (and even short movies of the observed region in the atmosphere with sub-nanosecond time resolution). In these movies, the EAS appears moving through the field of view, similarly to a shooting star. Measurement of the direction of motion of the EAS through the field-of-view provides information on the arrival direction of the primary \gr. The amount of Cherenkov light from the EAS is proportional to the number of particles and hence to the energy of the primary \gr. 

An alternative observational technique is to use a network of detectors on the ground to measure the characteristics of the EAS front. One of the simplest particle detectors is a tank of water equipped with a photosensor, like the detectors of HAWC\footnote{https://www.hawc-observatory.org/ } array shown in the right panel of Fig.~\ref{fig:cta}. Particles moving with the speed faster than the speed of light in water (the refraction index of water is $n_{water}=1.33$) generate ultraviolet Cherenkov light which has wide angular distribution ($\cos\theta_{Ch}=1/n_{water}$) and could be detected with e.g. photomultipliers. 

The front of an EAS incident on the surface detector array of the size $R_{array}$ at zenith angle $\Theta_z$ arrives earlier at one side of array, compared to an opposite side, with the 
$\Delta t=R_{array}\sin\Theta_z$. Measuring the time difference between the earliest and the latest signal and the direction from the earliest to the latest triggered detectors provides a measurement of the zenith and azimuth angle of the EAS, i.e. the information on the arrival direction of the primary \gr. The number of particles in the EAS surviving to the ground level scales proportionally to the overall number of particles in the EAS, so that measuring this number one gets a handle on the energy of the \gr.

\begin{figure}[ht]
\begin{center}
\includegraphics[width=0.9\linewidth]{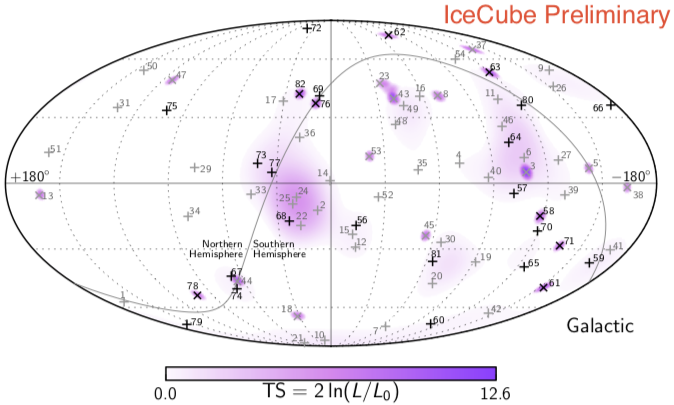}\\[2mm]
\includegraphics[height=5cm]{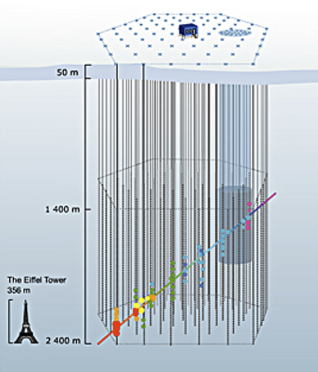}\hspace{1mm}
\includegraphics[height=5cm]{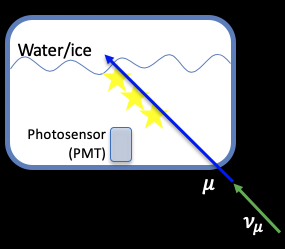}\hspace{1mm}
\includegraphics[height=5cm]{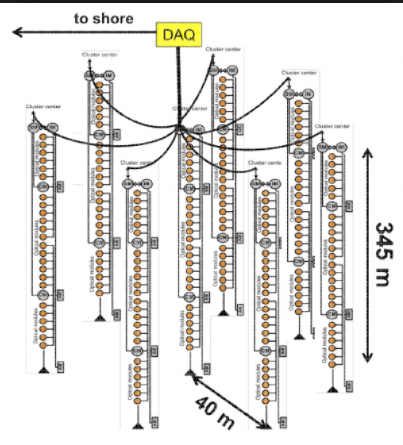}
\caption{Top: sky map produced based on the IceCube data collected in the HESE mode \cite{icecube_icrc2017}. Bottom left: Layout of IceCube neutrino telescope. Middle: the principle of neutrino detection with water / ice Cherenkov detectors. Right: layout of Baikal GVD detector \cite{baikal}. }
\label{fig:neutrino}
\end{center}
\end{figure}

The main challenge of astronomical source observations with ground-based \gr\ telescopes is to distinguish the EAS produced by \gr s from those produced by much more numerous charged cosmic rays. One obvious method to suppress the cosmic ray background is to collect only the EAS events coming from the direction of an isolated source of interest. Charged cosmic rays arrive from random sky directions and would not exhibit and excess from the direction of a particular astronomical source. This method does not work for detection of diffuse emission like e.g. the diffuse \gr\ flux from the Milky Way galaxy. An alternative way of suppression of the part of the cosmic ray background produced by the EAS initiated by atomic nuclei is to catch the differences in morphologies of nuclear and \gr\ induced showers. EAS initiated by the primary \gr\ and proton with the same energy $E_0$ would have different morphology because the gamma-factor of proton $\gamma_p=E_0/(m_pc^2)$ is three orders of magnitude lower than that of the first electron or positron produced in Bethe-Heitler pair production. This difference persists all over the EAS path, because the proton-induced showers contain also pions, particles of the masses $m_\pi\sim 10^2$~MeV, much heavier than electrons. Because of this, particles in the proton and nuclei induced showers have wider angular spread, a difference which could be seen both in the EAS images in the IACTs and in the lateral profiles of particle density on the ground, measured by the surface detectors. This difference can not be spotted on the EAS-by-EAS basis. Instead a statistical study based on Monte-Carlo simulations of proton / nuclei and \gr\ induced showers has to be done. 

The statistical studies performed for cosmic ray background suppression in ground-based \gr\ telescopes allow to reduce the cosmic ray background by about two orders of magnitude. This is still not enough for detection of large-scale diffuse \gr\ sky emission in multi-TeV energy range, except for the innermost part of the Galactic Plane \cite{hess_diffuse_gp}. On top of these diffuse emission, numerous sources in the Galactic Plane and in the extragalactic sky were discovered by the HESS \cite{hess_survey} and HAWC \cite{hawc_survey} surveys. Large fraction of those sources are "unidentified", i.e. it is not possible to unambiguously associate them to the known astronomical source. 

\subsection{Neutrino astronomy} 

Surprisingly, this emission is detectable at still much higher energy, above 30 TeV, with neutrino telescopes (Fig.~\ref{fig:neutrino}) \cite{icecube_science,icecube_icrc2017}. These telescopes use detection technique similar to that of HAWC, but for detection of charged particles produced in neutrino interactions directly in the detector volume (so-called "High-Energy Starting Events, HESE") or outside it, in the Earth's interior. Charged particles propagating through the detector volume produce Cherenkov light which is sampled by 3D network of photomultipliers, as shown in the bottom panels of Fig.~\ref{fig:neutrino}. Particles with energies in the 10 TeV band propagating through the detector volume produce phenomenon similar to the EAS, but in water or ice medium of the detector. Timing of the signal in different PMT modules allows to reconstruct the direction of the particle cascade developing preferentially in the direction of the primary particle track.  The precision of such reconstruction is better for muons which loose energy on distance scales larger than the size of detectors (kilometer-scale). The cascades initiated by electrons or tau leptons develop on much shorted distance and are typically contained inside the detector volume. The short cascade development distance diminishes the precision of direction reconstruction for electron- or tau-lepton-induced events. To the contrary, the containment effect enables the measurement of the primary particle energy, because the amount of UV light registered by the photomultipliers is proportional to the number of particles in the cascade which is in turn proportional to the primary particle energy. In the case of muon-induced events, one could measure, to some extent, the energy of the muon passing through the detector volume, because its energy loss rate is approximately proportional to the muon energy. However, it is not possible to measure the energy of muon as it was at the moment of production, before it entered the detector volume. Because of this, neutrino interaction events detected through the "throughgoing muon" channel have very limited energy resolution.  

Apart from the rare particles originating from astrophysical neutrino interactions, these detectors also suffer from backgrounds produced by the particles of EAS initiated by charged cosmic rays and \gr s. This type of background could be efficiently rejected by accepting only events which start inside the detector volume (in the case of HESE event selection) or only events coming from the directions toward interior of the Earth \cite{icecube_icrc2017}. In the latter case, charged EAS particles are absorbed by the rock and could not reach the detector volume. This allows to suppress the cosmic ray background much more efficiently, by a factor $\sim 10^6$ is the detector is situated deep underwater (in the case of ANTARES \cite{antares}, Baikal-GVD \cite{baikal} and, in the future, km3net \cite{km3net} neutrino detectors) or under a deep layer of ice  (in the case of IceCube detector in Antarctica) \cite{icecube_icrc2017}. 

Apart from the charged cosmic ray background, the neutrino telescopes have irreducible ``atmospheric neutrino'' background composed of neutrinos originating from the EAS \cite{icecube_icrc2017}. Contrary to the charged particles from the same EAS, the EAS neutrino flux is not attenuated after propagation through the Earth volume (below PeV energies), so that the atmospheric neutrino arrive from all sky directions. Being essentially the same particles as the astrophysical neutrinos, the atmospheric neutrinos could not be rejected based on the differences of signal appearance in the detector. Still, the atmospheric neutrino background could be suppressed to some extent for events coming into detector from above, because those atmospheric neutrinos often come simultaneously with the muons from the same EAS. Imposing a veto on the time intervals of muon passage through the detector allows to reject the atmospheric neutrino background. 

This additional suppression of the atmospheric background in the HESE event sample has allowed IceCube collaboration to discover the astrophysical neutrino signal in the energy range above 30 TeV \cite{icecube_science,icecube_icrc2017}. As mentioned above, this detection channel has good energy resolution (about 10\%) but very moderate angular resolution (about $10^\circ$). The most recent all-sky map of the signal is shown in the top panel of Fig.~\ref{fig:neutrino}. There are no isolated sources visible in the map. The overall signal distribution on the sky is consistent with being isotropic, although some anisotropy aligned with the Galactic Plane direction is detectable at 100~TeV at close to $3\sigma$ level. 

\begin{wrapfigure}{l}{0.4\linewidth}
\includegraphics[width=\linewidth]{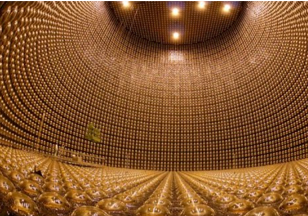}
\caption{Interior of the Super-Kamiokande neutrino detector \cite{superk}.}
\label{fig:sk}
\end{wrapfigure}
The same principle of detection of Cherenkov signal from particles propagating in water is used for detection of much lower energy neutrinos (in the MeV range) by large underground reservoirs, like Super-Kamiokande detector\footnote{http://www-sk.icrr.u-tokyo.ac.jp/sk/index-e.html} shown in Fig.~\ref{fig:sk}. The length of the muon tracks at these low energy is much shorter than in the case of muons detected by IceCube. To catch rate Cherenkov photons a dense network of photomultiplier tubes is installed all over the reservoir walls. The lower energy MeV neutrinos are produced in nuclear reactions and, in particular, in supernova explosions, such as SN1987A which happened in 1987 in the Large Magellanic Cloud \cite{sn1987a}. The sensitivity of current generation low energy neutrino detectors is sufficient for catching the supernova signal from supernovae in the Milky Way (if there will be one) and in the nearby Universe, up to Megaparsec distance \cite{sn_neutrino}.  

\subsection{Gravitational wave astronomy} 

Neutrinos and photons (or electromagnetic waves) are neutral particles which carry information on the physical processes in astronomical sources. They could serve as astronomical "messengers" because they propagate along straight lines from their sources. In a similar way, gravitational waves also carry information along straight lines from their sources, so that measuring the arrival direction of a gravitational wave signal it is possible to know from which source on the sky it comes from. 

Electromagnetic waves are emitted by charged particles moving in different astrophysical environments. Their wavelength is determined by the characteristic distance scales of particle motion (e.g. radius of gyration in magnetic field in the case of synchrotron radiation, motion of electrons confined inside atoms in the case of atomic line emission). 

In a similar way, the wavelength of gravitational waves emitted by astronomical sources is determined by the characteristic distance scales of motion under the fources of gravity. Contrary to the electromagnetic emission, these scales are always macroscopic. The most compact known objects able to emit gravitational waves are stellar mass black holes and neutron stars which have kilometer scale sizes. This suggests the shortest possible wavelength scale $\lambda_{GW}\sim 1$~km for gravitational wave emission from those objects. 

\begin{wrapfigure}{L}{0.5\linewidth}
\includegraphics[width=\linewidth]{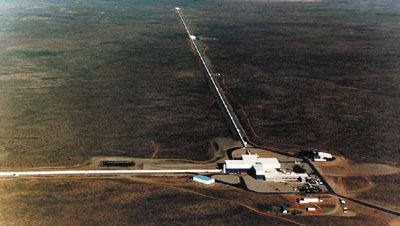}
\caption{LIGO gravitational wave detector, https://www.gw-openscience.org}
\label{fig:ligo}
\end{wrapfigure}

Considering this limitation, the antennae for detection of gravitational waves have kilometer-scale dimensions (see Fig.~\ref{fig:ligo} for an example of LIGO gravitational wave detector). Such detectors are tuned to detect signals from the compact sources. Larger scale gravitational wave sources, such as supermassive black holes in the centers of galaxies could not be observed by the kilometer-scale detectors on the ground, because the characteristic size scale of the supermassive black holes is about one astronomical unit ($10^{13}$~cm), i.e. about the Earth-Sun distance. Detection of the signal with such wavelength requires an antenna of comparable size. This type of antenna will be implemented with LISA space-based gravitational wave detector which will be operate as a constellation of spacecrafts\footnote{http://sci.esa.int/lisa/}. 

First attempts to detect gravitational waves from astronomical sources were performed by Weber in 60th of the last century \cite{gw_weber}, using detectors based on bodies which vibrate resonantly in response to a passing gravitational wave. Modern gravitational wave detectors use a different scheme of antennae using suspended masses in which gravitational wave passage induces a change in distance between the masses. Precision measurement of the distance is done using the laser ranging technique or, more precisely, laser interferometry. Kilometre-Scale interferometers of the Michelson type (Fig.~\ref{fig:ligo}) are used to measure the distance changes in two orthogonal directions, to improve sensitivity and directional reconstruction of the detector. 

The gravitational wave astronomy is the youngest field of the multi-messenger astronomy, which has been born only in 2015 with the discovery of the first astronomical source, a black hole merger \cite{gw}. The two black holes in the binaries are not surrounded by significant amount of matter which could generate electromagnetic emission in conventional ways. In this respect, the black hole mergers are "gravitational-wave-only" sources. Instead, the mergers which involve neutron stars do possess electromagnetic counterpart, as demonstrated by the detection of the neutron star merger event in gravitational and electromagnetic messenger channels \cite{ns_merger}. 

Before being detected directly, the emission of gravitational waves by astronomical sources has been inferred indirectly, via its effect on the parameters of the orbit of a binary pulsar system. Emission of gravitational waves leads to the  energy loss which causes a gradual shrinking of the orbit, the effect which was first observed in the Hulse-Taylor pulsar system \cite{hulse_taylor}. 


\section{Relevant physical processes}
\label{sec:processes}

\subsection{Curvature radiation}

Most of the formulae for the radiative processes involving electrons (synchrotron and curvature radiation, Compton scattering and Bremsstrahlung emission) are different applications of the basic formulae for the dipole radiation of an accelerated charge:
\begin{equation}
\label{eq:curv_power}
I=\frac{2}{3}e^2a_\mu a^\mu\simeq \frac{2}{3}\frac{e^2\gamma^4v^4}{c^3R^2}\simeq 4\times 10^{11}\left[\frac{\gamma}{10^5}\right]^4\left[\frac{R}{10^6\mbox{ cm}}\right]^{-2}\frac{\mbox{eV}}{\mbox{s}}
\end{equation} 
where $a^\mu$ is the 4-vector of particle acceleration, $\gamma$ is the gamma factor of the particles, $v$ is its velocity, $R$ is the curvature radius of the trajectory and $e$ is the particle charge. 

 In the simplest case of circular motion, the spectrum of emission from a non-relativistic particle in a circular orbit with cyclic frequency $\omega_0$  is sharply peaked at the frequency $\omega_0$. The emission has broad angular distribution with significant flux within $4\pi$ solid angle. In the relativistic case, the angular distribution pattern changes due to the Doppler boosting. Most of the flux is emitted within a cone with opening angle 
$\alpha\sim\gamma^{-1}$ with an axis aligned along particle velocity. Beaming within a cone also changes the spectrum of radiation, which appears to the observer as a sequence of short pulses occurring only  during the time $\Delta t_{obs}\sim R/\gamma^3$ when particle velocity is aligned along the line of sight to within the angle $\alpha$. The Fourier transform of the time sequence of pulses detected by the observer has all harmonics up to the frequency $\omega\sim 1/\Delta t_{obs}$, so that the energy of emitted photons is 
\begin{equation}
\label{eq:curv_energy}
\epsilon_{em}=\omega=\frac{\gamma^3}{R}\simeq 2\times 10^{4}\left[\frac{\gamma}{10^5}\right]^3\left[\frac{R}{10^6\mbox{ cm}}\right]^{-1}\mbox{ eV}
\end{equation}
One could understand the above formula in the following way. Relativistic electrons with energies $E_e=\gamma m_e\simeq 100$~GeV confined within a region of the size $R\sim 10$~km, they inevitably emit curvature radiation in the hard X-ray band, at the energies about $\epsilon_{curv}\sim 20$~keV. As an "everyday life" example of such situation one could mention the past times when the LHC accelerator machine in CERN was still electron-positron collider LEP (Large Electron Positron). It was operating at the energies $E\sim 100$~GeV accelerated in the LHC tunnel of the radius $R\sim 10$~km. From Eq. (\ref{eq:curv_energy}) one could find that the electron beam was a source of hard X-rays.  Once injected in a compact region, 100 GeV electrons loose all their energy within the time interval $t\sim E_e/I$ shorter than one second, as one could conclude from Eq. (\ref{eq:curv_power}). All the beam energy was continuously dissipated into the hard X-rays. 

\begin{wrapfigure}{L}{0.4\linewidth}
\begin{center}
\includegraphics[width=\linewidth]{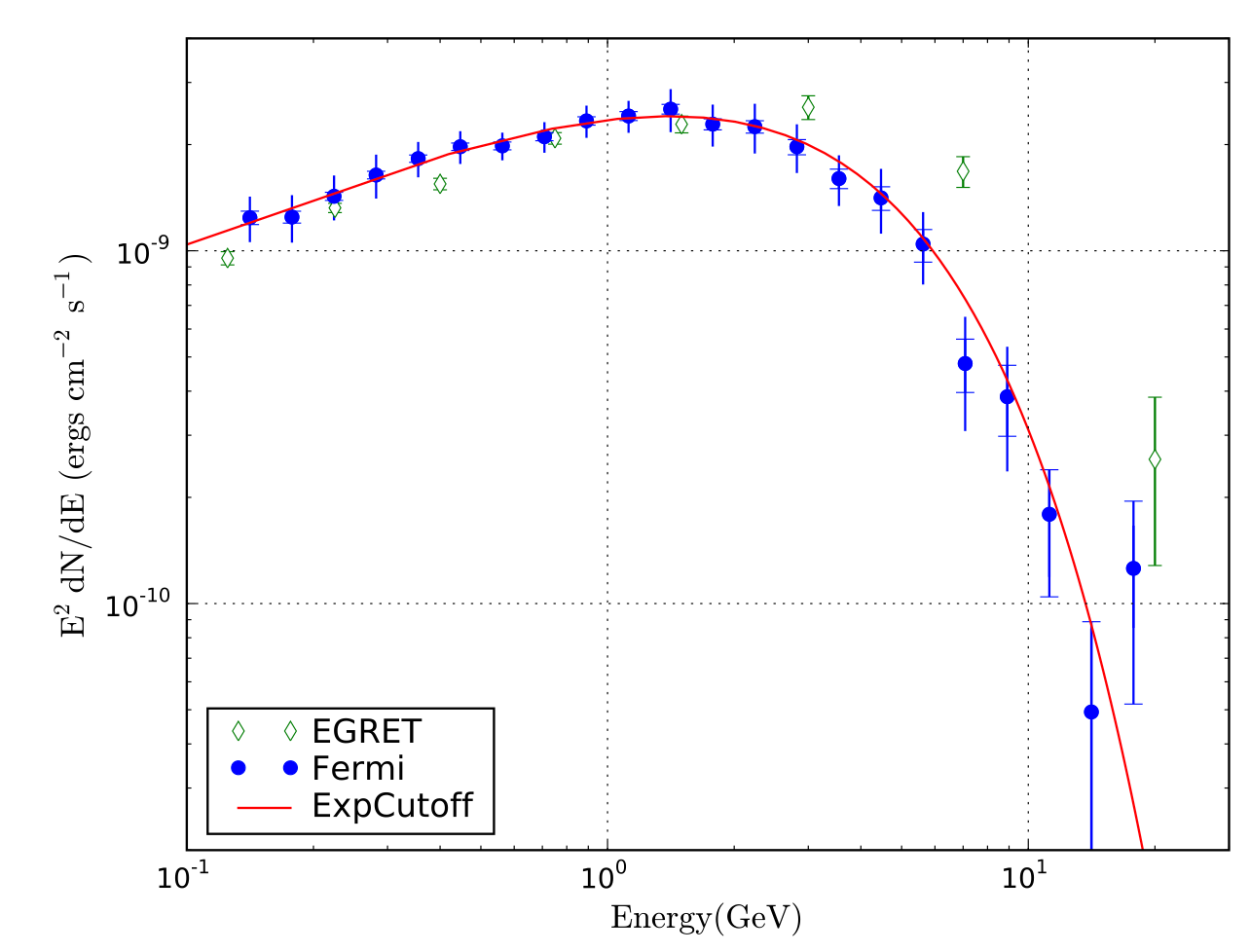}
\end{center}
\caption{Spectrum of pulsed  \gr\ emission from Vela pulsar (the brightest GeV \gr\ point source on the sky). From Ref. \cite{vela_lat}.}
\label{fig:vela_spectrum}
\end{wrapfigure}

The reference example illustrating curvature radiation in astrophysical environments is the currently most often considered model of \gr\ emission from magnetospheres of pulsars. Pulsars are strongly magnetised and fast spinning neutron stars, i.e. compact stars of the size $R_{NS}\sim 10^6$~cm, rotating at frequencies $1-10^3$~Hz and possessing magnetic fields in the range of $B\sim 10^{12}$~G.  Most of the isolated point sources of GeV \gr s in the Galactic Plane, shown in the top panel of Fig.~\ref{fig:fermi} are pulsars. Spectrum of emission from the brightest pulsar on the sky, the Vela pulsar, is shown in Fig.~\ref{fig:vela_spectrum}.

The bright GeV \gr\ emission from the pulsars is pulsed at the period of rotation of the neutron stars ($10^{-3}-1$~s). This implies that the \gr\ photons are produced close to the neutron star, in a region  close to the surface of the neutron star. We adopt a first estimate $R\sim R_{NS}\sim 10^6$~cm. The pulsed emission is detected at the energies exceeding 1~GeV. It is inevitably produced by relativistic particles. 

Relativistic particles confined to a compact spatial region inevitably loose energy at least onto curvature radiation (there might be competing energy loss channels, we will consider them later on). Using Eq. (\ref{eq:curv_energy}) one could estimate the energies of electrons responsible for the observed \gr\ emission, under the assumption that the \gr s are produced via curvature mechanism
\begin{equation}
\label{eq:energy_pul}
E_e\simeq 2\times 10^{12}\left[\frac{\epsilon_{em}}{1\mbox{ GeV}}\right]^{1/3}\left[\frac{R}{R_{NS}}\right]^{1/3}\mbox{ eV}
\end{equation}

\subsection{Synchrotron emission}

Basic relations for the energy of synchrotron photons and power of synchrotron emission could be found directly from the formulae for dipole and, in particular, for curvature radiation simply via substitution of expression for the gyroradius 
\begin{equation}
R_L=\frac{E_e}{eB}\simeq 3\times 10^7\left[\frac{E_e}{10^{10}\mbox{ eV}}\right]\left[\frac{B}{1\mbox{ G}}\right]^{-1}\mbox{ cm}
\end{equation}
Substituting $R_L$ at the palce of $R$ into (\ref{eq:curv_energy}) gives 
\begin{equation}
\label{eq:eps_synch}
\epsilon_{synch}=\frac{\gamma^3}{R_L}=\frac{eB E_e^2}{m_e^3}\simeq 5\left[\frac{B}{1\mbox{ G}}\right]\left[\frac{E_e}{10^{10}\mbox{ eV}}\right]^2\mbox{eV}
\end{equation}
for the energy of synchrotron photons. 

Substituting $R_L$ at the place of $R$ in Eq. (\ref{eq:curv_power}) we find the energy loss rate on the synchrotron emission:
\begin{equation}
\label{eq:synch_loss}
I=-\frac{dE_e}{dt}\simeq \frac{2}{3}\frac{e^4 B^2 E_e^2}{m_e^4}\simeq  4\times 10^5\left[\frac{B}{1\mbox{ G}}\right]^2\left[\frac{E_e}{10^{10}\mbox{ eV}}\right]^2\mbox{ eV/s}
\end{equation}
Electrons emitting synchrotron radiation loose energy on time scale 
\begin{equation}
\label{eq:tsynch}
t_{synch}=\frac{E_e}{-dE_e/dt}=\frac{3m_e^4}{2e^4B^2E_e}\simeq 2\times 10^4\left[\frac{B}{1\mbox{ G}}\right]^{-2}\left[\frac{E_e}{10^{10}\mbox{ eV}}\right]^{-1}\mbox{ s}
\end{equation}

\begin{figure}
\begin{center}
\includegraphics[height=6cm]{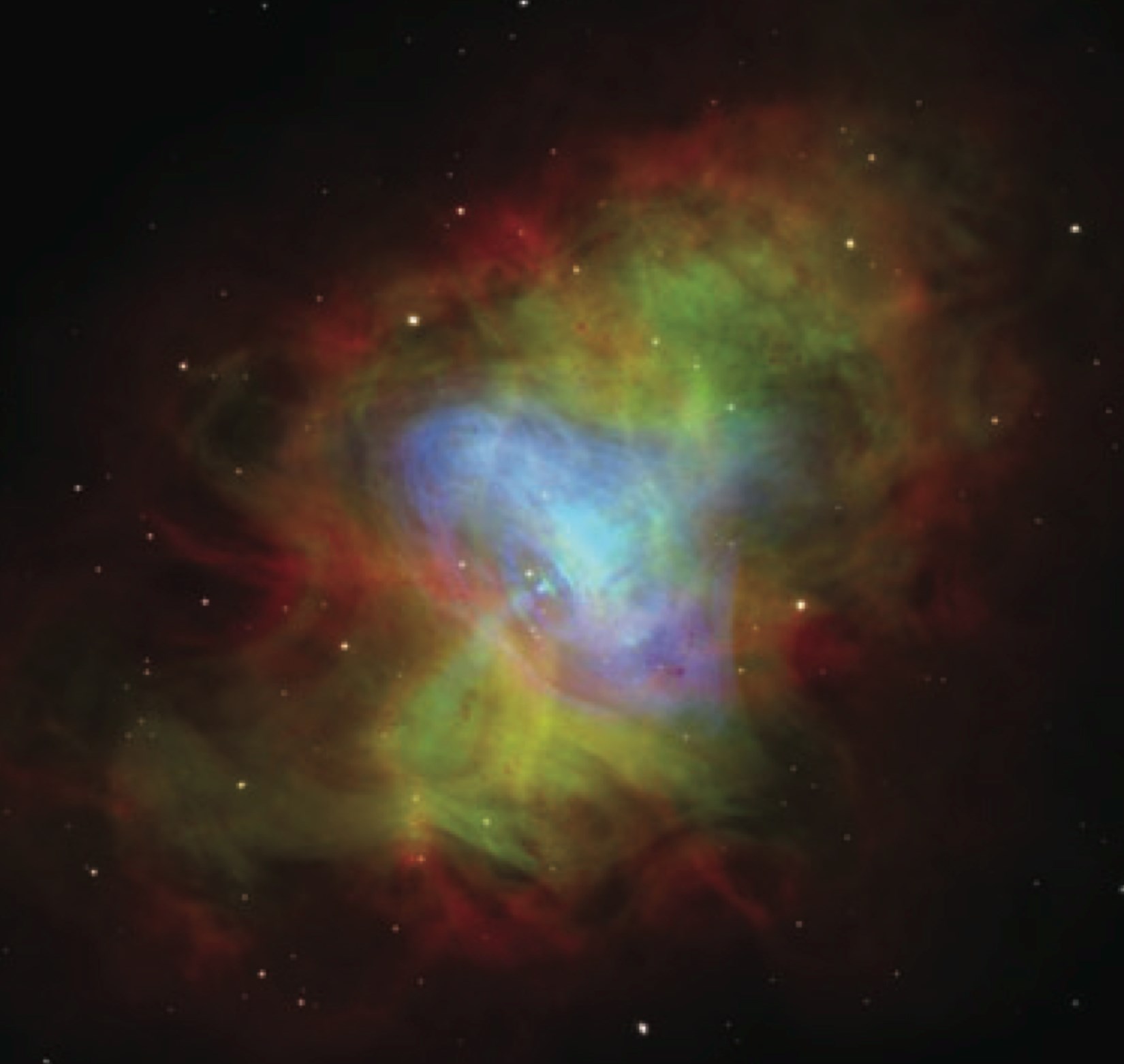}\hspace{5mm}
\includegraphics[height=6cm]{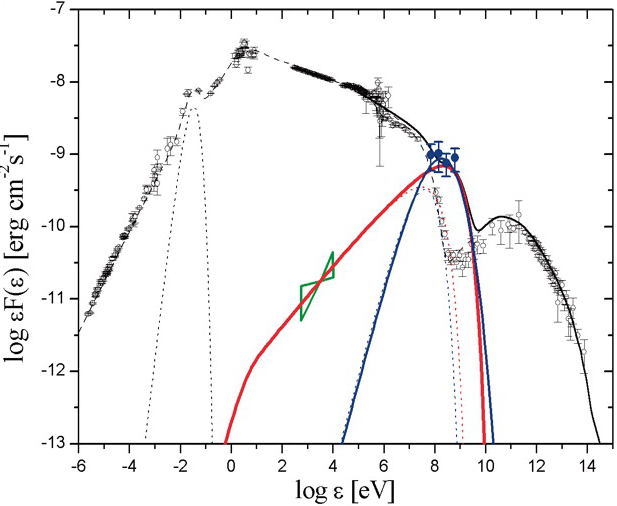}
\caption{Multiwavelength image (left) and spectrum (right) of the Crab pulsar wind nebula. In the image (from \cite{hester08}): blue colour is X-ray, green is in visible light and red is radio emission. In the spectrum (from \cite{tavani11}) black data points show time-averaged spectrum of the source. Blue and violet points show examples of spectra during GeV band flares.} 
\label{fig:crab_nebula}
\end{center}
\end{figure}

A prototypical example of the high-energy source is Crab nebula \cite{crab}, which is a nebular emission around a young ($10^3$ yr old) pulsar. It is one of the brightest \gr\ sources on the sky and serves as a calibration target for most of the \gr\ telescopes. The spectrum of emission from the Crab nebula spans twenty decades in energy, from radio (photon energies $10^{-6}$~eV) up to the Very-High-Energy \gr\ (photon energies up to 100~TeV) band. Large, parsec-scale size and moderate distance (2~kpc) to the nebula enable detailed imaging of the source in a range of energy bands, from radio to X-ray. The composite radio-to-X-ray image of the nebula is shown in Fig.~\ref{fig:crab_nebula}, left panel. Radio to X-ray and up to GeV \gr\ emission from the nebula presents a composition of several broad powerlaw-type continuum spectra, with the photon indices changing from  $\Gamma\simeq 1.3$ in the radio-to-far-infrared band to $\Gamma\simeq 2.2$ in the 1-100~MeV band. Such powerlaw type spectra extending down to the infrared and radio domains are commonly interpreted as being produced by the synchrotron emission mechanism.  From Fig. (\ref{fig:crab_nebula}) one could see that  the spectrum of synchrotron emission has a high-energy cut-off in the 100~MeV range, where is sharply declines. Electrons which produce synchrotron emission in the 100~MeV energy range should have energies $E_{e}\sim 1-10\mbox{ PeV}$ given the magnetic field in the Nebula $B\sim 10^{-4}$~G (see Eq. \ref{eq:eps_synch}). This shows that Crab nebula hosts a remarkably powerful particle accelerator. For comparison, the energies of particles accelerated in the most powerful man-made accelerator machine, the Large Hadron Collider, are in the 10~TeV range, which is three orders of magnitude lower. 

Up to recently, Crab Nebula was believed to be a non-variable source and was conventionally used as a calibration sources for X-ray and \gr\ telescopes, due to its high flux and stability. However, recent observations by {\it Fermi} \cite{crab_flares_fermi} and {\it AGILE}  \cite{crab_flares_agile} \gr\ telescopes have revealed variability of the \gr\ emission from Crab, in the form of short powerful flares, during which the GeV flux of the source rises by an order of magnitude, see Fig.~\ref{fig:crab_nebula}. These flares occur at the highest energy end of the synchrotron spectrum and have durations in the $t_{flare}\sim 1-10$~d$\sim 10^5-10^6$~s range. Comparing the synchrotron cooling times of the 1-10~PeV electrons with the duration of the flares, one finds  that the flares occur in the innermost part of the nebula, in the regions with higher magnetic field ($B\sim 500\ \mu$G), otherwise, long synchrotron cooling time would smooth the flare lightcurve on the time scale $t_{synch}$ and the flare would not have 1~d duration.  The flaring time scale is most probably directly related to the time scale of an (uncertain) acceleration process, which leads to injection of multi-PeV electrons in the nebula. 

\subsection{Compton scattering}

The spectrum of emission from the Crab nebula has apparently two "bumps": one starting in the radio band and ending in the GeV band and the other spanning the 10~GeV-100~TeV needy range with a peak in at $\sim 100$~GeV.  We have interpreted the radio-to-\gr\ bump as being the result of synchrotron emission from high-energy electrons in the source. The synchrotron emission is an "inevitable" radiative loss channel in a wide range of astronomical sources, because most of the known sources possess magnetic fields. There is practically no place in the Universe without magnetic field \cite{durrer}.

Similarly to magnetic fields, the whole Universe is also filled with radiation fields. Radiation was generated by the hot Early Universe. This relic radiation survives till today in the form of Cosmic Microwave Background (CMB). CMB is thermal radiation with temperature $T_{CMB}\simeq 2.7$~K and its present in equal amounts everywhere in the Universe. The spectrum of CMB is the Planck spectrum. Its energy density is 
\begin{equation}
U_{CMB}=\frac{\pi^2}{15}T^4\simeq 0.25\left[\frac{T_{CMB}}{2.7\mbox{ K}}\right]^4\frac{\mbox{ eV}}{\mbox{cm}^3}
\end{equation}
and the number density of CMB photons is
\begin{equation}
n_{CMB}=\frac{2\zeta(3)}{\pi^2}T^3\simeq 4\times 10^2\left[\frac{T_{CMB}}{2.7\mbox{ K}}\right]^3\frac{\mbox{ ph}}{\mbox{cm}^3}
\end{equation}

\begin{wrapfigure}{L}{0.4\linewidth}
\begin{center}
\includegraphics[width=\linewidth]{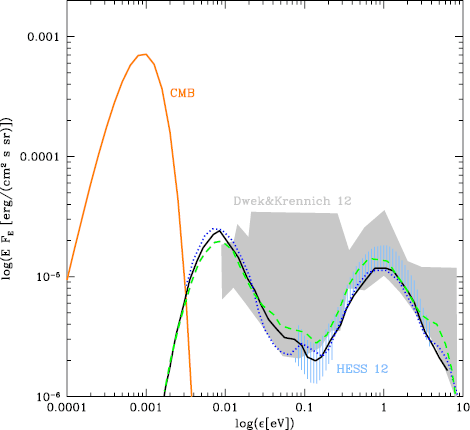}
\end{center}
\caption{The spectrum of Extragalactic Background Light (from \cite{durrer}.  Grey band shows the range of uncertainties of currently existing measurements.} 
\label{fig:EBL_spectrum}
\end{wrapfigure}

Apart from the universal CMB photon background, radiation fields are generated by collective emission from stars and dust in all galaxies over the entire galaxy evolution span. This leads to production of the so-called "Extragalactic Background Light" with a characteristic two-bump spectrum shown in Fig.~\ref{fig:EBL_spectrum} \cite{franceschini}. The bump at the photon energy $\sim 1$~eV is produced by the emission form stars, the bump at $10^{-2}$~eV is due to the scattering of starlight by the dust. Fig.~\ref{fig:EBL_spectrum} shows the level of the starlight and dust emission averaged over the entire Universe. Inside the galaxies, the densities of both the starlight and dust photon fields are enhanced by several orders of magnitude \cite{isrf}.

Still denser photon fields exist inside astronomical sources, e.g. close to the stars or in the nuclei of active galaxies. High-energy particles propagating through the photon backgrounds could occasionally collide with the low energy photons and loose / gain energy in the Compton scattering process. 

The cross-section of scattering of photons by a non-relativistic electron is the Thomson cross-section:
\begin{equation}
\sigma_T=\frac{8\pi e^4}{3m_e^2}\simeq 6.65\times 10^{-25}\mbox{ cm}^2
\end{equation}

Expressing the acceleration of electron by the electric field of electromagnetic wave and substituting it into Eq. \ref{eq:curv_power} one finds the emission power by the moving electron
\begin{equation}
I=\sigma_TU_{rad}
\end{equation}
where $U_{rad}$ is the energy density of the incident radiation.

The most widespread example of Compton scattering in astronomy is the scattering of photons inside stars. The nuclear reactions which power stellar activity proceed most efficiently deep in the stellar cores, where temperatures are significantly higher than at the stellar surface. However, we are not able to observe directly radiation from the nuclear reactions, because the star is "opaque" to the radiation. The process which prevents photons produced deep inside the stars from escaping is the Compton scattering. 

Let us take the Sun as an example. The average density of the Sun is 
\begin{equation}
n_\odot=\frac{M_\odot}{(4\pi/3)R_\odot^3 m_p}\simeq 10^{24}\mbox{ cm}^{-3}\,.
\end{equation}
From the definition of the scattering cross-section, we find that the mean free path of photons with respect to collisions with electrons inside the Sun is just 
\begin{equation}
\lambda=\frac{1}{\sigma_Tn_\odot}\simeq 1\mbox{ cm}\,.
\end{equation}
This mean free path is much shorter than the distance of the order of the size of the Sun $R_\odot\simeq 7\times 10^{10}\mbox{ cm}$ which the photon needs to cross before leaving the surface of the Sun. Since $\lambda\ll R_\odot$, none of the photons produced in the core is able to escape. The source is opaque to photons. The opacity of the source of the size $R$ is often measured in terms of the optical depth, which is, by definition $\tau=R/\lambda\sim 10^{11}$. 

If we assume that Compton scattering results in random changes of the direction of motion of photons, we could describe the process of escape of photons from inside the star as a random walk or diffusion in 3d space. The law of diffusion allows to estimate the time needed for  photons to escape from the core
\begin{equation}
t_{diff}=\frac{R_\odot^2}{\lambda}=\tau \frac{R}{c}\simeq 10^{11}\mbox{ s       for the Sun}
\end{equation} 
Thus, Compton scattering slows down the radiative transfer from the core to the surface.  

Sun-like stars are supported by the balance of gravity force and pressure of the stellar plasma. However, in massive stars with much higher luminosity than that of the Sun, the force due to the radiation pressure competes with the gravity and the equilibrium configuration of the star is supported by the balance of gravity and radiation pressure force
\begin{equation}
\frac{G_NMm_p}{R^2}\simeq \epsilon n \sigma_T
\end{equation}
(we assume that the number of protons and electrons in the star is the same). The strongest radiation pressure force is acting on electrons, stronger gravity force is acting on protons. Expressing the density of radiation through the luminosity $L$
\begin{equation}
\epsilon n=\frac{L}{4\pi R^2}
\end{equation}
and substituting into above equation we find an expression for $L$ through the mass of the star
\begin{equation}
\label{eq:edd}
L_{Edd}=\frac{4\pi G_N M m_p}{\sigma_T}\simeq 10^{38}\left[\frac{M}{M_\odot}\right]\mbox{ erg/s}
\end{equation}
This mass-dependent luminosity, called Eddington luminosity, is, in fact an upper limit on the luminosity of a self-gravitating object. No persistent astronomical source of the mass $M$ could have luminosity higher than $L_{Edd}$, because otherwise the source would be disrupted by the radiation pressure force. 

The effect of Compton scattering of hard X-ray / soft \gr\ photons off electrons in detector material  is used in Compton telescopes described in section \ref{sec:compton}. Photons scattered on  electron at rest  transfer a fraction of their energy to electron. Measuring this energy and the energy of the scattered photon (when it is absorbed in the material provides a possibility to measure the scattering angle of the photon and to finally infer the arrival direction and energy of the photon. 

If electrons are moving, opposite is also possible: electron could transfer a fraction of its energy to photon. In this case Compton scattering works as a radiative energy loss for high-energy electrons. This process is called inverse Compton scattering. 

Scattering of photons by relativistic electrons is called inverse Compton scattering because in this case electrons gives away its energy to photons, rather than absorbs it from the incident electromagnetic wave. Formulae for intensity of radiation and characteristic energy of upscattered photons could be obtained via transformation to the electron rest frame and back to the lab frame. This "double Doppler boost" explains the $\gamma^2$ factor which appears in the expression for the average  energy of the upscattered photon $\epsilon_f$ through the initial photon energy $\epsilon_i$: 
\begin{equation}
\label{eq:compton_energy}
\epsilon_f\simeq \epsilon_i\gamma^2\simeq 3\left[\frac{\epsilon_i}{1\mbox{ eV}}\right]\left[\frac{E_e}{10^{10}\mbox{ eV}}\right]^2\mbox{ GeV}
\end{equation}
Note that this expression is valid as long as $\epsilon_f<E_e$, the regime which is called "Thomson regime of inverse Compton scattering". Otherwise, a relation $\epsilon_f\sim E_e$ holds (the Klein-Nishina regime of inverse Compton scattering). 

Transformations between comoving and lab frames allow also to find the energy loss rate of electron (and of intensity of radiation) from Eq. \ref{eq:curv_power}. In the particular case of electron moving through an isotropic radiation field with energy density $U_{rad}$ this gives 
\begin{equation}
\label{eq:compton_loss}
I=-\frac{dE_e}{dt}=\frac{4}{3}\sigma_TU_{rad}\gamma^2
\end{equation}
Electrons loosing energy through inverse Compton scattering cool in Thomson regime on the time scale
\begin{equation}
t_{IC}=\frac{E_e}{dE_e/dt}=\frac{3}{4}\frac{m_e^2}{\sigma_TU_{rad} E_e}\simeq 3\times 10^7\left[\frac{U_{rad}}{1\mbox{ eV/cm}^3}\right]^{-1}
\left[\frac{E_e}{10^{10}\mbox{ eV}}\right]^{-1}\mbox{ yr}
\end{equation}

At this point we could come back to the example of the broad band spectrum of the Crab Nebula (see Fig.~\ref{fig:crab_nebula}). The synchrotron component of the spectrum is cut-off at GeV energy. Above this energy, one could see a gradually rising new component which reaches maximum power in the 100~GeV energy band. 

This high-energy component is conventionally attributed to the inverse Compton emission from the same electrons which produce synchrotron emission at lower energies. Let us not calculate the properties of inverse Compton emission produced by these electrons. First, we need to understand which low energy photon field provides most of the target photons for inverse Compton scattering. The photon fields present in the Crab nebula include the "universal" soft photon field, the CMB, with the energy density $U_{rad}=0.25$~eV/cm$^3$. Next, since the source is in the Galaxy, it "bathes" in the interstellar radiation field, produced by stars and dust in the Galaxy. Crab is not far for the Sun ($D_{Crab}\simeq 2$~kpc distance) and one could estimate the density of the interstellar radiation field at the location of the Crab  based on the knowledge of the local  interstellar radiation field density which is about $U_{rad}\sim 1$~eV/cm$^3$.

Finally, the synchrotron radiation produced by the high-energy electrons in the Crab Nebula also provides abundant target photon field for the inverse Compton scattering. We could estimate the density of this radiation field from the measured flux of Crab (see Fig.~\ref{fig:crab_nebula}). The flux reaches $F_{crab}\simeq 10^{-7.5}$~erg/cm$^2$s in the visible / IR energy band $\epsilon_{ph}\sim 1$~eV. The size of the innermost part of the Crab Nebula is about $d_{Crab}\simeq 1$~pc. The flux of the photons escaping from the Nebula higher than the flux detected on Earth by a factor $(D_{Crab}/d_{Crab})^2\sim 4\times 10^6$. The flux (measured in erg/cm$^2$s) is related to the energy density of radiation as $F_{crab}=U_{rad}c$, so that the energy density of synchrotron radiation could be estimated as
$U_{rad}=F\left(D_{Crab}/d_{crab}\right)^2\simeq 3\mbox{ eV}/\mbox{cm}^3$
which is somewhat higher than the estimate of the density of the interstellar radiation field and is an order of magnitude higher than the CMB energy density. This means that the main source of the soft photons for inverse Compton scattering in Crab is the synchrotron radiation of the nebula itself. 
 
According to Eq. (\ref{eq:compton_energy}), electrons with the energies about $E_{e,break}$ should upscatter the synchrotron photons with energies $\epsilon_{ph}\simeq 1$~eV up to the energy $E_\gamma\simeq 10^{14}$~eV, which is, obviously, not possible because the energy of \gr\ could not exceed the energy of electron. The upscattered photon energy is equal to electron energy at 
\begin{equation}
E_e\simeq \frac{m_e^2}{\epsilon_i}\simeq 2.5\times 10^{11}\left[\frac{\epsilon_i}{1\mbox{ eV}}\right]^{-1}\mbox{eV}
\end{equation}
This is exactly the energy at which the maximum of the power of inverse Compton emission is reached, see Fig.~\ref{fig:crab_nebula}.

Above this energy the inverse Compton scattering proceeds in the Klein-Nishina regime. 
From Figure \ref{fig:crab_nebula} one could see that the power of inverse Compton energy loss gets suppressed in this regime. In this regime, each scattering event transfers a significant fraction of electron energy to the photon so that $\epsilon_f\simeq E_e$. The inverse Compton energy loss time is then just the time between subsequent collisions of electron with photons, which is the interaction time
\begin{equation}
t_{IC,KN}=\frac{1}{\sigma_{KN} n_{ph}}\simeq 10^6\mbox{ yr}\left[\frac{\sigma_{KN}}{\sigma_T}\right]^{-1}\left[\frac{n_{ph}}{1\mbox{ cm}^{-3}}\right]^{-1}\mbox{ yr}
\end{equation}
It depends on the electron energy, because the cross-section of inverse Compton scattering in this regime $\sigma_{KN}$ is no longer constant. 

The exact expression for the cross-section is derived from quantum mechanical treatment of the scattering process. This could be understood after a transformation to the comoving reference frame of electron. In this reference frame,  the incident photon has an energy higher than the rest energy of electron $\tilde \epsilon_i=xm_e$, $x>1$. Its wavelength $\lambda\simeq \epsilon_i'^{-1}$ is shorter or comparable to the Compton wavelength of electron, $\lambda_C=1/m_e$. This means that electron could not be considered anymore as a classical particle influenced by an incident electromagnetic wave. 

The quantum mechanical expression for the scattering cross-section is \cite{aharonian_book}
\begin{equation}
\sigma_{KN}=\frac{3\sigma_T}{8x}\left\{\left[1-\frac{2(x+1)}{x^2}\right]\ln(2x+1)+\frac{1}{2}+\frac{4}{x}-\frac{1}{2(2x+1)^2}\right\}
\end{equation}
where  $x=\tilde \epsilon_i/m_e$ is the incident photon energy in the comoving frame expressed in units of electron energy.  In the regime $x\gg 1$ the inverse Compton scattering cross section decreases with electron energy as 
\begin{equation}
\sigma_{KN}\sim \frac{\ln(E_e)}{E_e}
\end{equation}
in the limit of large $E_e$. As a result, in the Klein-Nishina regime the inverse Compton cooling time increases with energy as
\begin{equation}
t_{IC,KN}\sim \frac{1}{\sigma_{KN}}\sim \frac{E_e}{\ln(E_e)}
\end{equation}
Thus, the inverse Compton cooling becomes less and less efficient and the power of inverse Compton emission drops at the energies above the Klein-Nishina / Thomson regime transition energy.

\subsection{Bethe-Heitler pair production}

In the electron rest frame the transition between the Thomson and Klein-Nishina regimes of Compton scattering takes place at the photon energy $E_\gamma\simeq m_e$. The  behaviour of Compton scattering cross-section changes from nearly constant value $\sigma\simeq \sigma_T$ to a decreasing function of energy $\sigma\sim E^{-1}\ln(E)$. Photon energy range $E_\gamma\sim m_e$ is remarkable also from another point of view. As soon as the energy of the incident photon reaches $2m_e$, the energy transfer in photon-matter collisions becomes sufficient for production of electron-positron pairs. 

The cross-section of the pair conversion in the Coulomb field is  the Bethe-Heitler cross-section:
\begin{equation}
\label{eq:bh}
\sigma_{N\gamma\rightarrow Ne^+e^-}\simeq \frac{28\alpha^3}{9m_e^2} r_e^2Z(Z+1)
\end{equation}
Being proportional to the third power of the fine structure constant, $\alpha^3$, It is much smaller than the Thomson cross-section $\sigma_T=(8\pi/3)(\alpha/m_e)^2$ in the case of  scattering of photons in the proton Coulomb field. However, in the case of scattering in "high-Z" material with the $Z\gtrsim 1/\sqrt{\alpha}$, the $Z^2$ factor could compensate for the suppression of the cross-section by $\alpha$. The Bethe-Heitler pair production is used in the \gr\ instrumentation, but it is rarely important in the astrophysical environments because the density of matter there is typically much too low to make this process important. 

Instead, a process of pair production in photon-photon collisions, which becomes possible when the energies of the colliding photons, $\epsilon_1,\epsilon_2$ become sufficiently large
\begin{equation}
\epsilon_1\epsilon_2\ge m_e^2
\end{equation}
(this corresponds to the center-of-mass energy sufficient for creation of an electron-positron pair) is more important in astrophysical environments.  A typical example  is of a source emitting high-energy \gr s which, before escaping from the source have to propagate through a soft photon background. Taking the \gr\ energy $E_\gamma$ and the typical soft photon energy $\epsilon$, we find that the source potentially becomes opaque to \gr s with energies higher than
\begin{equation}
\label{eq:thresh}
E_\gamma\simeq 250\left[\frac{\epsilon}{1\mbox{ eV}}\right]^{-1}\mbox{GeV}
\end{equation}

\begin{figure}[ht]
\begin{center}
\includegraphics[width=0.9\linewidth]{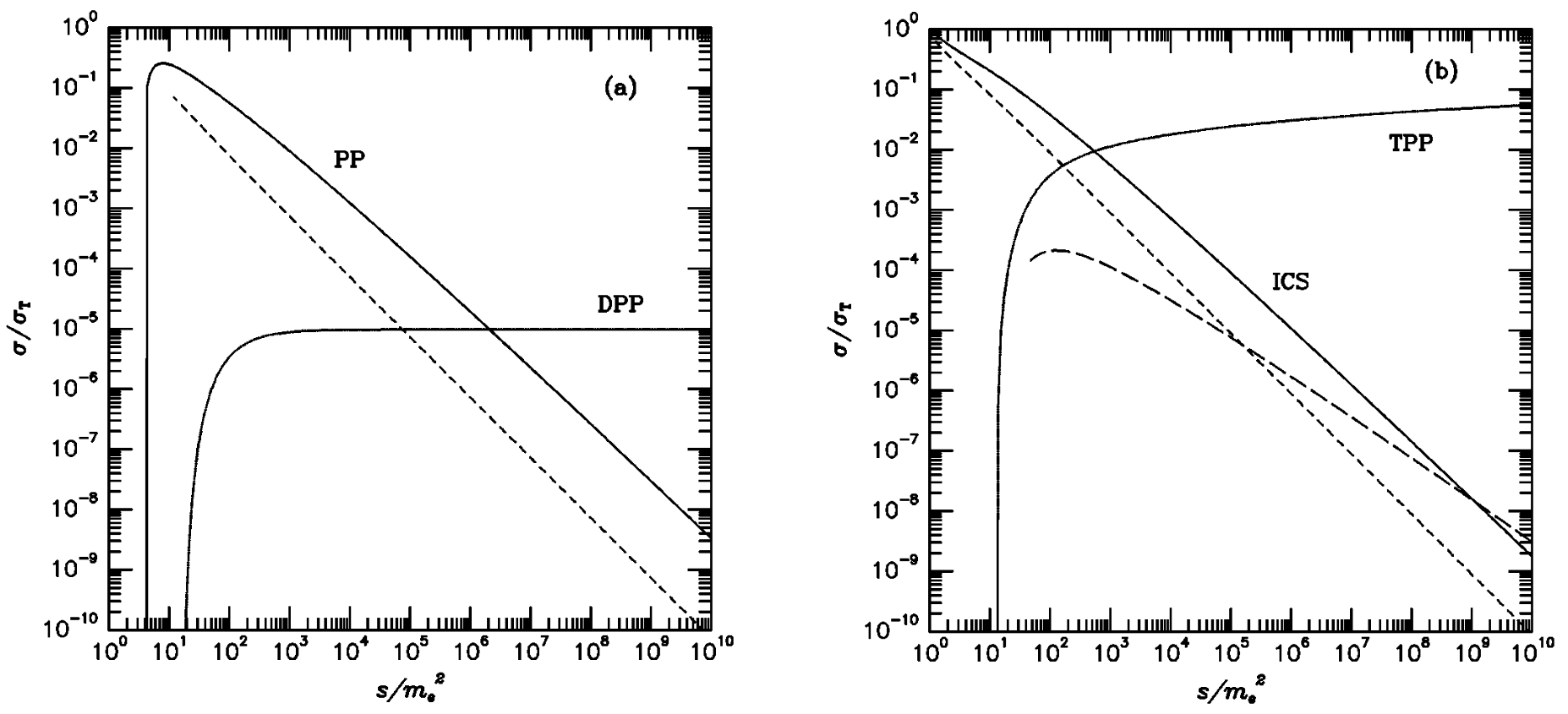}
\caption{Comparison of the cross-sections of inverse Compton scattering (ICS), triplet production (TPP) and $\gamma\gamma$ pair production (PP) as  functions of the square of the centre-of-mass energy $s$. Also shown in the cross-section of double pair production (DPP).  Solid lines show the cross-sections, dashed lines show cross-sections times the elasticity (average fractional energy loss of electron in single collision). From Ref. \cite{lee}. } 
\label{fig:gammagamma}
\end{center}
\end{figure}

The cross-section of gamma-gamma pair production reaches 
\begin{equation}
\sigma_{\gamma\gamma}\simeq \frac{3}{8}\sigma_T\simeq 1.5\times 10^{-25}\mbox{ cm}^2
\end{equation}
at the maximum (Fig.~\ref{fig:gammagamma}). At the energies much larger than the threshold it decreases as 
\begin{equation}
\sigma_{\gamma\gamma}\simeq \frac{3\sigma_T m_e^2}{8E_\gamma\epsilon}\ln\left(\frac{E_\gamma\epsilon}{m_e^2}\right)
\end{equation}
i.e. it decreases as $E\ln E$, similarly to the Compton cross-section in Klein-Nishina regime \cite{aharonian_book}.

A comparison of the energy dependences of the cross-sections of inverse Compton scattering, $\gamma\gamma$ pair production and triplet production is shown in Fig.~\ref{fig:gammagamma}.

The relation between the cross-sections of Compton scattering and pair production has an important implication for the physics of high-energy sources. We remember that many sources (including all the stars) are opaque with respect to the Compton scattering (i.e. the mean free path of photons $\lambda=(\sigma_T n_e)^{-1}$ is much shorter than the size of the source $R$ and the optical depth $\tau=R/\lambda\gg 1$. 

This means automatically that the same sources are opaque also to \gr s with energies in excess of the pair production threshold (\ref{eq:thresh}).  The most famous example of a pair production thick source is given by the Gamma-Ray Bursts (GRB). These are \gr\ sources which occur for short periods of time ($0.1-10^3$~s in stellar explosive events (e.g. supernovae, neutron star mergers). The luminosity of the sources reaches $L_\gamma\simeq 10^{50}$~erg/s, in the soft \gr\ band $E_\gamma\simeq 1$~MeV. The relation of the source to the star with gravitationally collapsing core suggests an estimate of the source size $R\sim 10^6$~cm, of the order of the gravitational radius of the star $R_g=G_N (10 M_\odot)\simeq 10^6$~cm. If we calculate the optical depth of the source with respect to the pair production by the MeV \gr s on themselves ($\epsilon\sim E_\gamma\sim 1$~MeV), we find
\begin{equation}
\tau=\frac{R}{\lambda_{\gamma\gamma}}\simeq \frac{L_\gamma\sigma_{\gamma\gamma}}{4\pi R\epsilon}
\end{equation}
we would find a value in the range of $10^{14}$. This implies that the MeV \gr s should not escape from the source at all.  This pair production opacity problem is encountered in a milder form in a number of other high-energy source types, like e.g. active Galactic Nuclei. It is conventionally resolved by taking into account relativistic motion of the emitting source or its parts. In this case the estimate of the optical depth is modified after the account of Doppler effect. Still even with the account of the Doppler effect, the source could well be optically thick with respect to the pair production, especially if one considers propagation of high-energy \gr s through low energy photon background.  

\begin{wrapfigure}{L}{0.5\linewidth}
\includegraphics[width=\linewidth]{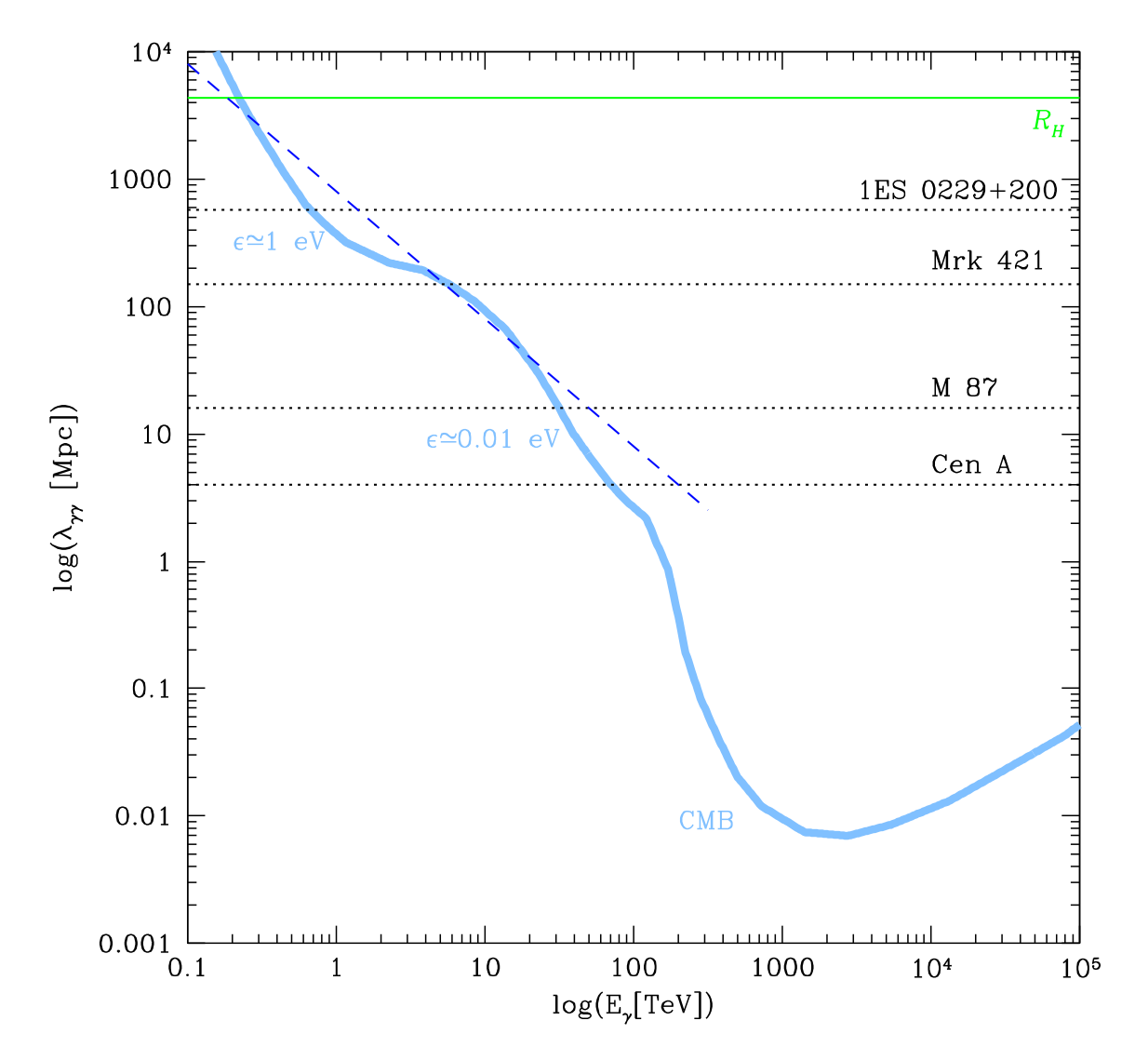}
\caption{Comparison of the energy-dependent mean free path of \gr s with the distances to the known TeV \gr\ sources \cite{durrer}.} 
\label{fig:EBL_abs}
\end{wrapfigure}

Another common example of a source opaque w.r.t. the pair production is the Universe itself. Indeed, the Universe is filled with radiation fields (the CMB, the interstellar radiation field in our Galaxy, the Extragalactic Background Light.  The energy of CMB photons is $\epsilon\simeq 10^{-3}$~eV. This means that \gr s with energies higher than $E_\gamma\simeq 0.3$~PeV (see eq. (\ref{eq:thresh})) could produce pairs in interactions with CMB photons.  The density of the CMB photons is $n_{CMB}\simeq 400$~cm$^{-3}$. The mean free path of the \gr\ s w.r.t. the pair production is, therefore, 
\begin{equation}
\lambda_{\gamma\gamma}=\frac{1}{\sigma_{\gamma\gamma}n_{CMB}}\simeq 8\mbox{ kpc}
\end{equation}
Thus, PeV \gr s are even not able to escape from the host galaxy of the source (typical galaxy sizes are 10-100~kpc). 

Otherwise, lower energy TeV  \gr s could efficiently produce pairs in interactions with the EBL photons of the energies $\epsilon\simeq 1$~eV. The density of the EBL is much lower and the mean free path of photons w.r.t. this process is respectively larger. Still it is shorter than the typical distance to the extragalactic sources of TeV \gr s. Comparison of the \gr\ mean free path with the distances to the known  TeV sources is shown in Fig.~\ref{fig:EBL_abs}.

\subsection{Electromagnetic cascades}
\label{sec:cascade}

The pair production converts high-energy \gr s into electrons and positrons of comparable energies. The inverse Compton scattering process re-generates the high-energy \gr s by transferring the energy of electrons / positrons to the low-energy photons. In this way, a cyclical process of "bouncing" of the energy between electrons / positrons  and \gr s could take place. This process also leads to multiplication of the number of high-energy particles, because each pair production event generates two high-energy particles (electron and positron) of one high-energy \gr. This leads to development of  electromagnetic cascade which is  important in the context of several astrophysical situations, in particular, in the interiors of some high-energy sources. 

The mean free path of photons w.r.t. the pair production is comparable to that of electron / positron w.r.t. the inverse Compton scattering (in Klein-Nishina regime). After propagating approximately one mean free path distance, the \gr\ of energy $E_0$ disappears and transfers energy to electron and positron in roughly equal proportions, so that the energies of both particles are $E_{e,1}\sim E_0/2$. Inverse Compton scattering in the Klein-Nishina regime converts most of the electron /positron energy back into \gr\ energy, so that the energy of the "first generation" \gr s are $E_{\gamma, 1}\simeq E_0/2$. The process  of division of energy between new particles repeats in the "second generation of the cascade. After $n$ generations, the typical energy of the cascade particles is  $E_{e,n}\simeq E_{\gamma,n}\simeq E_0/2^n$
The number of particles in the cascade is $N_n\simeq 2^n$. The process of energy division repeats until the energies of particles decrease below the pair production threshold. 

\subsection{Bremsstrahlung}

Up to now we have concentrated on the radiative losses of electrons due to the interactions with magnetic field (synchrotron) and radiation fields (inverse Compton). High-energy electrons also suffer from energy losses when they propagate through matter and interact with the electrostatic Coulomb field of atomic nuclei. There are two energy loss channels: radiative one, called Bremsstrahlung, and non-radiative one, called ionisation loss. The radiative energy loss is related to the accelerated motion of electron which is deviated by the Coulomb field of an atomic nucleus. 

The Fourier transform of the Larmor formula \ref{eq:curv_power} allows to find the spectrum of emission from this process for a given impact parameter of electron onto the scattering center, $b$:
\begin{equation}
\frac{dI}{d\omega}=
\left\{
\begin{array}{ll}
\frac{\displaystyle 2e^2}{\displaystyle 3\pi}\frac{\displaystyle 4Z^2e^4}{\displaystyle m_e^2 b^2 v^2},&\ \ \omega\ll 2\pi v/b;\\
0, & \ \ \omega\gg 2\pi v/b
\end{array}
\right.
\end{equation}
Here $Ze$ is the charge of atomic nucleus, $v$ is velocity of the electron. 
We notice that the power of radiation scales with the square of the charge of the nucleus. The power is higher for close encounters (smaller $b$) and slow electrons (smaller $v$). 

A typical situation is when electron propagates through a medium. The rate of encounters with different $b$ is determined by the density of the medium. Integration over all possible impact parameters gives 
\begin{equation}
\label{eq:brems_power}
\frac{dI_{tot}}{d\omega\ dt}=\frac{16}{3}\frac{Z^2e^6n}{m_e^2 v}\ln\left(\frac{b_{max}}{b_{min}}\right)
\end{equation}
where $b_{min},b_{max}$ are the minimal and maximal values of the impact parameter, which are difficult to estimate, it is rather tabulated for different materials. The spectrum of Bremsstrahlung emission extends up to the electron energy $\omega\sim E_e=m_ev^2/2$.  Inspecting the Eq. \ref{eq:brems_power} we notice that the power of Bremsstrahlng emission scales with the energy of electron as $dI/dt\simeq \omega dI/d\omega d t\sim v\sim E_e^{1/2}$. The expression for the Bremsstrahlung spectrum in the relativistic case is similar to the non-relativistic expression:
\begin{equation}
\frac{dI_{tot}}{d\omega dt}\simeq \frac{Z^2e^6n}{m_e^2v}\ln\left(\frac{192 v}{Z^{1/3}}\right)
\end{equation}
However, the same formula implies a different scaling of the Bremsstrahlung power with the election energy: $dE_e/dt\simeq \omega dI_{tot}/d\omega d t\simeq Z^2e^6 n E_e/m_e^2\sim E_e$.
This implies that the Bremsstrahlung cooling time
\begin{equation}
t_{brems}=\frac{E_e}{dE_e/dt}\simeq \frac{m_e^2}{Z^2e^6 n}=\frac{1}{\sigma_{brems}n}
\end{equation}
is roughly energy independent for relativistic particles. We have introduced in the last equation the cross-section $\sigma_{brems}\simeq Z^2e^6 / m_e^2$ One could notice that this cross-section is  $\sigma_{brems}\sim 10^{-2}Z^2\sigma_T$
similarly to the cross-section of Bethe-Heitler pair production. 

\begin{wrapfigure}{L}{0.4\linewidth}
\includegraphics[width=\linewidth]{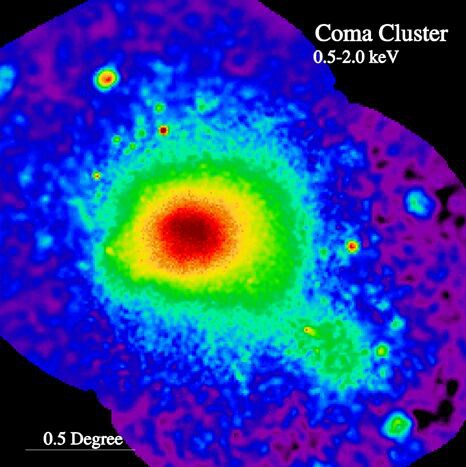}
\caption{X-ray image of Coma galaxy cluster, from Ref. \cite{coma}.}
\vskip-0.2cm
\label{fig:coma}
\end{wrapfigure}

The Bremsstrahlung emission is responsible for the X-ray flux from galaxy clusters. Gas which falls in the gravitational potential well of the clusters with mass $M\sim 10^{15}M_\odot$ heats liberates its gravitational potential energy which is converted in the energy of the random motions of particles or, in other words into heat. The temperature of the  gas could be estimated from the virial theorem
\begin{equation}
T\sim  \frac{mv^2}{2}\sim \frac{GMm_p}{R}\sim 10\left[\frac{M}{10^{15}M_\odot}\right]\left[\frac{R}{1\mbox{ Mpc}}\right]^{-1}~\mbox{ keV}
\end{equation}
Gas heated to this temperature emits Bremsstrahlung photons with energies comparable to the energies of electrons in the gas, see Fig.~\ref{fig:coma} for an example of X-ray Bremsstrahlung emission from Coma galaxy cluster \cite{coma}. 

Density of the gas in the cluster about $n\sim 10^{-3}$~cm$^{-3}$. In such conditions  the Bremsstrahlung cooling time 
\[
t_{brems}\sim 10^{10}\left[n/10^{-3}\mbox{ cm}^{-3}\right]^{-1}\left[T/10\mbox{ keV}\right]^{1/2}\mbox{ yr}
\] 
is comparable to the age of the Universe, so that the hot gas residing in the clusters  is about to cool down at the present epoch of evolution of the Universe. Gas cooling is expected to produce a "cooling flow" with decreasing temperature and increasing density in the center of the clusters \cite{boehringer}. The increase of the density and decrease of temperature speed up the cooling so that the cold gas should quickly accumulate in the cluster core.  This cooling flow process is most of the time counteracted by the activity of the central galaxy of the cluster which produces relativistic outflows displacing the cooling flow and heating the intracluster medium. 

\subsection{Ionisation losses}

The energy loss of electron scattering on an atomic nucleus is not limited to the radiative Bremsstrahlung loss. Another energy loss channel is the kinetic energy of nucleus recoil. An electron propagating through a medium interacts with many nuclei at different shooting parameters $b$. The overall energy transferred to the nuclei per path length $dx$ is  given by the Bethe-Bloch formula
\begin{equation}
\frac{dE_e}{dx}=\frac{4\pi Z^2e^4n}{m_ev^2}\left[\ln\left(\frac{2\gamma^2m_ev^2}{I}\right)-v^2\right]
\end{equation}
where $I$ is the ionisation energy of the atoms composing the medium. Taking a medium with the density water, $n\simeq 10^{24}$~cm$^{-3}$, we could estimate the energy loss as  $dE_e/dx\simeq (several)\mbox{ MeV/cm}$
for a mildly relativistic electron ($v\sim 1,\ \ \gamma\sim1$).  The energy dependence of ionisation loss has a pronounced minimum at $v\gamma\simeq 3$. Particles of this energy are called "minimum ionising particles" in particle physics. 

The ionisation energy loss rate of relativistic particles grows only logarithmically with energy. The ionisation loss time is then  $t_{ion}=E_e/(dE_e/dt)\sim E_e/\mbox{ln}(E_e)$ becomes longer with the increasing energy. Comparing the ionisation and Bremsstrahlung energy loss times we find
\begin{equation}
\frac{t_{ion}}{t_{brems}}\simeq \frac{Z^2e^6nE_e}{m_e^2 v}\frac{m_ev^2}{Z^2 e^4 n}\sim \frac{e^2 E_e}{m_e}
\end{equation}
As soon as the gamma-factor of electron $E_e/m_e$ reaches $1/e^2\sim 100$, the Bremsstrahlung energy loss starts to dominate over the ionisation energy loss. 

Bremsstrahlung, ionisation and Bethe-Heitler pair production processes are involved in the EAS physics, which is key for the operation of ground-based \gr\ telescopes and neutrino detectors. 

\subsection{Interactions of high-energy protons}

Classical radiative energy losses of protons are much smaller than those of electrons. This applies for all the main classical radiative losses, including curvature, synchrotron, inverse Compton and Bremsstrahlung radiation.  Energy losses for protons are more important for the quantum processes of production of new particles. 

The first example is given by the process of pair production in  interactions of high-energy protons with energy $E_p$  with low energy photon background composed of photons with energy $\epsilon$ \cite{pair_production}. This process is essentially the same as the Bethe-Heitler pair production by \gr s propagating through a medium. Its cross-section could be found from Eq. (\ref{eq:bh}):  $\sigma_{p\gamma\rightarrow pe^+e^-}\simeq 4\times 10^{-27}\mbox{ cm}^2$. This process is possible above an energy threshold 
\begin{equation}
E_{p,thr}=\frac{m_pm_e}{\epsilon}\simeq 5\times 10^{14}\left[\frac{\epsilon}{1\mbox{ eV}}\right]^{-1}\mbox{ eV}
\end{equation}
In spite of the sizeable cross-section, the efficiency of the pair production as proton energy loss in astrophysical conditions is usually very low. This is because of the small "inelasticity" of the process. Proton looses only a small fraction of its energy in each pair production event. In the reference system comoving with the proton, both the proton and the newly produced electron and positron are almost at rest. Their energies are $m_p$ and $m_e$, respectively. Transforming to the lab frame, one finds that the electron and positron carry away only  a fraction $\kappa=2m_e/m_p\simeq 10^{-3}$ of the proton energy. 

A reference example of marginally important pair production by protons is given by the effect of interactions of high-energy cosmic ray protons with the CMB photons . The threshold energy for this process is $E_{p,thr}\simeq 10^{18}$~eV. The density of CMB is $n_{CMB}\simeq 400$~ph/cm$^3$. Proton mean free path w.r.t. the pair production is
$\lambda_{p\gamma\rightarrow pe^+e^-}\simeq  10^{24}\mbox{ cm}
\simeq 0.3\mbox{ Mpc}$, while the energy loss distance is much larger, $l_{p\gamma\rightarrow pe^+e^-}=\kappa^{-1}\lambda_{p\gamma\rightarrow pe^+e^-}\simeq 300\mbox{ Mpc}$ This corresponds to the energy loss time $t=l_{p\gamma\rightarrow pe^+e^-}\simeq 1\mbox{ Gyr}$, just an order of magnitude below the age of the Universe. 

A more efficient energy loss of protons is via production of heavier particles, e.g. pions. Pions are two-quark particles with masses in the $m_\pi\simeq 100$~MeV range, i.e. two orders of magnitude higher than electron. Repeating the calculation which led to the estimate inelasticity, we find that in the case of the pion production the inelasticity $\kappa\simeq m_\pi/m_p\simeq 0.1$ is much higher. This means that proton looses a significant fraction of its energy in just several collisions. 

The threshold for this reaction could be found from the kinematics considerations
\begin{equation}
E_{p,thr}=\frac{m_pm_\pi(1+m_\pi/(2m_p))}{2\epsilon}\simeq 10^{17}\left[\frac{\epsilon}{1\mbox{ eV}}\right]^{-1}\mbox{ eV}
\end{equation}

The cross-section of this process is determined by the physics of strong interactions. Close to the threshold is is as large as $\sigma_{p\gamma}\simeq 6\times 10^{-28}$~cm and drops to $\simeq 10^{-28}$~cm much above the threshold.

If we consider again the example of cosmic rays interacting with the CMB photons, we find that the mean free path and the energy loss with respect to the pion production reaction are, respectively, longer and shorter, compared to the pair production
$\lambda_{p\gamma}= 1/(\sigma_{p\gamma}n_{CMB})\simeq  1-10\mbox{ Mpc}$
and the energy loss distance $l_{p\gamma}=\kappa^{-1}\lambda_{p\gamma}\simeq 10-100\mbox{ Mpc}$. 
The pion production loss has, therefore, stronger effect on the cosmic rays spectrum, significantly suppressing the flux of cosmic rays with energies $E_p>E_{p,thr}\simeq 10^{20}\left[\epsilon/10^{-3}\mbox{ eV}\right]^{-1}$~eV. This is the domain of Ultra-High-Energy Cosmic Rays (UHECR), the highest energy particles ever detected. The interactions with the CMB suppress the flux of cosmic rays above the threshold of the pair production. This effect is know Greisen-Zatsepin-Kuzmin (GZK) cut-off \cite{greisen,zk,gzk_auger}.

Similar pair and pion production effects take place also in proton-proton collisions. Kinematics of the reaction allows to calculate the threshold
\begin{equation}
E_{p, thr}=\frac{m_\pi^2+4m_\pi m_p}{2m_p}\simeq 280\mbox{ MeV}
\end{equation}
The cross-section of this reaction is also determined by the strong interactions and is about the geometrical cross-section of the proton, $\sigma_{pp}\simeq 4\times 10^{-26}\mbox{ cm}$ to $10^{-25}\mbox{ cm}$, depending  on (growing with) the proton energy.

Pion production affects  propagation of cosmic ray protons in the interstellar medium. Typical density of the interstellar medium around us is $n_{ISM}\sim 1$~cm$^{-3}$. The mean free path of the proton is $\lambda_{pp}\simeq 1/\sigma_{pp}n_{ISM}\simeq 3-10\mbox{ Mpc} $.
Inelasticity of the reaction in the case of proton-proton collisions is quite high, $\kappa\simeq 0.5$, so that single collision takes away a sizeable fraction of the proton energy. The  cooling time due to the pion production process is about $(1-3)\times 10^8$~yr. This is somewhat longer than the residence time of cosmic rays in the Galaxy, but still, a fraction of cosmic rays interacts in the Galactic Disk before escaping from it. 

Neutral  and charged pions $\pi^0,\pi^\pm$ are unstable particles which decay into \gr s, $\pi^0\rightarrow 2\gamma$, neutrinos and  muons $\pi^\pm\rightarrow\mu^\pm+\nu_\mu$. Muons, in turn , are also unstable and decay into electrons and neutrinos $\mu^\pm\rightarrow e^\pm+\nu_e+\nu_\mu$. Thus, pion production in $pp$ collisions results in production of \gr s, neutrinos and high-energy electron / positrons.  Gamma-ray emission induced by interactions of cosmic rays with the protons from the interstellar medium is the main source of high-energy \gr s from the Milky Way galaxy \cite{fermi_diffuse,hess_diffuse_gp}.

\section{Multi-Messenger sources} 
\label{sec:sources}

\subsection{Gravitational collapse at the end of life of massive stars}

Multi-messenger sources residing in our Galaxy which and operating particle accelerators originate from evolution of massive stars. 
Stellar evolution proceeds through the synthesis of heavier elements,  from the lighter ones (starting from the primordial H, He). At the end of each stage of evolution (e.g. synthesis of He from H) the energy output of nuclear reaction diminishes. This leads to the decrease of radiation pressure and to contraction of the part of the star not supported by the radiation anymore. Contraction reheats the star and starts next stage of nuclear synthesis. The process could repeat up to the point at which Fe nuclei are produced in nucleosynthesis.  At the end of evolution, the star is composed of several layers of different elements.

The nucleus of the evolved star is composed of iron, the nuclear reactions stop because the synthesis of still heavier elements is not possible energetically (the nuclear binding energy of heaver elements is smaller than that of iron). The iron core is in hydrostatic equilibrium with pressure $P$ at distance $r$ balancing the gravity force on a unit volume element with density $\rho$:
\begin{equation}
\frac{dP}{dr}=\frac{GM\rho}{r^2}
\end{equation}
The main contribution to the pressure is that of degenerate electron gas. Electrons are fermions and the relation between the pressure and density of electron gas could be derived directly from the Heisenberg uncertainty relation  $\Delta p\Delta x\sim 1$ in the following way. The mean distance between particles in the gas of density $n$ is  $\Delta x\simeq n^{-1/3}$. The momentum of non-relativistic particles is   $\Delta p=m v$ where $m$ is the particle mass and $v$ is its velocity. Assuming $\Delta p\sim p\sim (\Delta x)^{-1}\sim n^{1/3}$ one finds that the pressure of a gas of non-relativistic fermions is
$P\sim nvp\sim n_e^{5/3}/m_e=Y_en_N^{5/3}/m_e$ where $Y_e$ is the number of electrons per nucleon and $n_N\sim M/(R^3m_p)$ is the density of nucleons.
 Averaging the hydrostatic equilibrium equation over the volume of the iron core one finds its size 
\begin{eqnarray}
\label{core}
R\sim \frac{Y_e^{5/3}}{Gm_p^{5/3}m_eM^{1/3}}
\end{eqnarray}
Accumulation of the iron "nuclear waste" in the stellar core which leads to the growth of the core mass $M$ also leads to the decrease of the core size $R$.  The contraction, in turn, leads to the increase of the density of degenerate electron gas. The increase of the density of electron gas leads to the increase of velocities of electrons, $v\sim n_e^{1/3}/m_e$.  At some point of the stellar evolution, typical velocities of electrons in the stellar core become comparable to the speed of light. 

At this moment the equation of state of degenerate electron gas changes from non-relativistic, $P\sim n^{5/3}$ to the relativistic one, $P\sim n^{4/3}$.  An immediate consequence of the change of equation of state is modification of the conditions of hydrostatic equilibrium to 
\begin{equation}
\frac{P}{R}\sim \frac{Y_e^{4/3}M_*^{4/3}}{m_p^{4/3}R^5}\sim \frac{G_NM_*^2}{R^5}
\end{equation}
which is valid only if
\begin{equation}
M_{Ch}\sim \frac{Y_e^2 M_{Pl}^3}{m_p^2}\simeq  1.4 M_\odot
\end{equation}
 This fundamental mass scale is named Chandrasekhar mass \cite{chandrasekhar}.  

As soon as the stellar core reaches the mass $M_{Ch}$, any further accumulation of the "nuclear waste" in the core would lead to an instability, since the hydrostatic equilibrium equation does not have solutions with $M$ exceeding $M_{Ch}$. At this moment the only possible further step of the stellar evolution is the gravitational collapse of the core \cite{janka}.
 
The size of the stellar core just before the onset of collapse is about $10^9$~cm (i.e. about the size of the Earth).
Matter density at the onset of the gravitational collapse is $\rho\sim 10^7$~g/cm$^3$. This density is much higher than the density of matter which one could find in conventional laboratory conditions (compare with e.g. the density of water  $\sim 1$~g/cm$^3$). At the same time, it is much lower than the density in the interior of atomic nuclei. It is also much lower than the characteristic density which one could ascribe to protons or neutrons. Dividing the proton mass by the cube of its radius $r_p\sim 10^{-13}$~cm one finds $\rho_N\sim m_p/r_p^3\simeq 10^{15}\mbox{ g/cm}^3$.
Contraction of matter distribution in the process of gravitational collapse should lead to increase of the density,  up to the atomic nuclei density scales. 

An obstacle to a prompt gravitational collapse of an unstable configuration is heating of the collapsing matter by the released gravitational energy. Applying the virial theorem to the collapsing core one finds that the temperature should reach some
\begin{equation}
\label{virial}
T\sim \left<\frac{m_pv^2}{2}\right>\sim \frac{GMm_p}{2R}\sim 25\left[\frac{R}{10^6\mbox{ cm}}\right]^{-1}\left[\frac{M}{M_\odot}\right]\mbox{ MeV}
\end{equation}
as soon as the entire core shrinks to the size of about 10 km. 

This temperature is higher than the difference of mass between proton and neutron $m_p-m_n\simeq 1.3$~MeV and than the typical binding energies of atomic nuclei. This has two consequences. First, occasional collisions of atomic nuclei in the collapsing core result in their breakdown, so that the process of gravitational collapse "undoes" all the nucleosynthesis work done at the previous stages of stellar evolution. Next, collisions between free protons and neutrons in the collapsing matter could convert protons into neutrons and vice versa.

\subsection{Neutron star and neutrino emission}

Neutrons could be produced via the "electron capture" reaction $p+e^-\rightarrow n+\nu_e$.
The electron participating in this reaction is taken from the Fermi "sea" and has the energy close to Fermi energy which, for the marginally unstable configuration is about the rest energy of electron. The neutrino produced in electron capture freely escapes from the core and takes away energy. The collapsing core of a massive star should, therefore, be a strong source of neutrinos (see below).

At the same time, the neutron decay reaction $n\rightarrow p+e^++\overline{\nu}_e$
could not proceed in the same way as it does in the low density environment, because the electron which should be produced in this reaction has to return to the degenerate Fermi gas, in which all the states up to the Fermi energy $E_F\simeq m_ec^2$ are already occupied. In this way, the neutron decay is suppressed and the entire matter of the core gets gradually saturated with the neutrons which could not decay, the process called neutronization leading to formation of neutron star \cite{janka}. 

Nuclear reactions, in particular, massive production of neutrons, transform the state of matter in the collapsing body so that at some stage of this transformation the equation of state of matter will change in such a way that the pressure of matter would again support the configuration from further collapse.  The size of this configuration could be estimated by replacing $m_e$ with $m_p$ in expression (\ref{core})  for $R_{Ch}$. This gives $R_{NS}\sim 10^6$~cm. The gravitational energy released in the collapse is
\begin{equation}
E_{grav}=\frac{G_NM_{Ch}^2}{R_{NS}}\sim 10^{53}\mbox{ erg}
\end{equation}

This energy could be released only through neutrino emission, because the time of diffusion of photons to the surface of the star is very long. The density of the neutron star is so high that even neutrinos diffuse out of its interior, rather than escape freely.  Typical energies of  neutrinos are estimated from Eq. (\ref{virial}). 

The rate of the core collapse events of massive stars in our Galaxy is roughly 1 per 100~yr. This means that nearby Galactic core collapse events which might provide the strongest neutrino signal are rare. Different types of neutrino detectors with sensitivity sufficient for detection of neutrino flux from Galactic core collapse events exist only since some $\sim 30-40$~yr. 

The only convincing case of detection of neutrino emission from an astronomical source other than the Sun was the detection of neutrinos from a core-collapse supernova SN 1987A \cite{sn1987a} which was situated in a satellite galaxy of the Milky Way, called Large Magellanic Cloud (LMC). 

\begin{wrapfigure}{L}{0.5\linewidth}
\includegraphics[width=\linewidth]{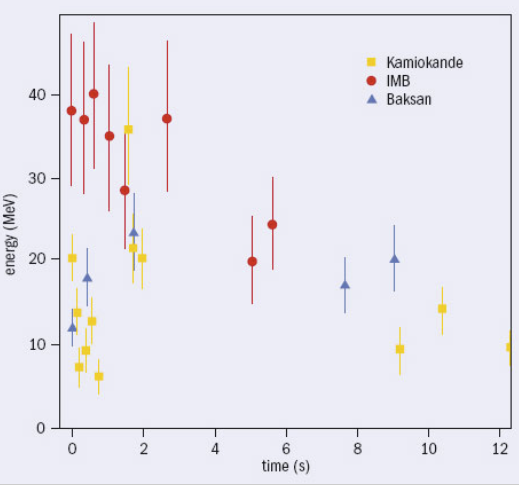}
\caption{Compillaiton of neutrino signal from SN 1987A, from \cite{sn1987_nu}.}
\label{fig:sn1987a}
\end{wrapfigure}



The distance to the LMC is $D_{LMC}\simeq 50$~kpc. Estimate the typical neutrino fluence from the process of proto-neutron star cooling gives the neutrino flux 
\begin{equation}
F_\nu\sim \frac{0.1 E_{grav}}{4\pi D_{LMC}^2}\sim 3\times 10^4\frac{\mbox{erg}}{\mbox{cm}^2}
\end{equation}
so that the total number of neutrinos which passed through each square centimeter was about  $N_\nu\sim F_\nu/E_\nu\sim 10^{9}\mbox{ cm}^{-2}$.

At the time of SN 1987A event the existing neutrino detectors had physical size of the order of $R_{det}\sim 10$~m, so that the total number of neutrinos which passed through the detectors was about $N_\nu R_{det}^2\sim 10^{15}$. The cross-section of neutrino interactions with matter is extremely small, so that only a small fraction of these huge neutrino number could have interacted inside the detectors. Typical density of the detector matter is 
$\rho_{det}\sim 1 $~g/cm$^3$. The mean free path of neutrinos through the detector matter is $\lambda=(\sigma_W\rho_{det}/m_p)^{-1}\sim 10^{16}$~cm, assuming the cross-section of weak interactions $\sigma\sim 10^{-40}$~cm$^2$ in the relevant energy range $E_\nu\sim 10-100$~MeV. This means that the probability for a neutrino to interact inside the detector was $p\sim R_{det}/\lambda\sim 10^{-13}$. Thus, out of $\sim 10^{15}$ neutrinos passing through the detector only about $\sim 10^2$ have interacted there. Assuming that the efficiency of detector is some 10\%  (i.e. only a fraction of the neutrino interaction events in the detector would be noticed) one finds that the 10 m size neutrino detectors could have found some ten(s) of neutrinos from SN1987A, as indeed was observed simultaneously  by several neutrino detectors, Kamiokande in Japan \cite{kamiokande}, IMB \cite{imb} in In US and Baksan in Soviet Union \cite{baksan}. All in all these three detectors found some 25 neutrinos with energies in the range 10-40 MeV. The arrival times of these neutrinos  and their energies are shown in Fig.~\ref{fig:sn1987a}.

Detection of the neutrino flux from a known core collapse event, with the time, energy and overall fluence consistent with the theoretical expectations provides a strong support for the core collapse scenario. The techniques of neutrino detection have significantly evolved since the SN 1987A event. With the next supernova explosion in our Galaxy or in a Milky Way satellite galaxy, detailed time and energy evolution of the neutrino signal will be studied, because the expected statistics of the neutrino events will be orders of magnitude higher than in the SN 1987A event \cite{sn_neutrino}. 

\subsection{Supernova and supernova remnants}

\begin{figure}
\includegraphics[width=\linewidth]{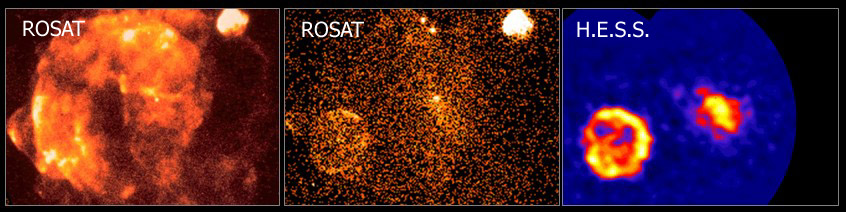}
\caption{X-ray and TeV gamma-ray images of a region of sky containing Vela (larger bubble, 5 degrees diameter), Vela Junior (smaller bubble on the left) and Puppis A (bubble on the right top) supernova remnants \cite{vela_rosat,hess_velajr,hess_puppis}. (image credit: https://www.mpi-hd.mpg.de/hfm/HESS/pages/home/som/2006/05/) }
\label{fig:vela}
\end{figure}

Following the gravitational collapse, a significant fraction of massive stars produces supernova explosions in which an outer part of the stellar envelope is ejected carrying overall kinetic energy  $E_{SN}\sim 10^{51}\mbox{ erg}\sim 10^{-2}E_\nu$
which constitutes only a small fraction of the energy carried away by neutrinos. This kinetic energy is estimated from the observations of expanding supernova remnant shells like those shown in Fig.~\ref{fig:vela}  \cite{vela_rosat,hess_velajr,hess_puppis}. In this figure one could see three shells superimposed on each other, with the biggest one being the shell around Vela supernova remnant, which is visible in the "soft" band image of ROSAT X-ray telescope (left), Smaller shell is "Vela Junior" supernova remnant visible in the hard band of ROSAT (middle) and also in the TeV energy band (right). Still smaller shell is that of Puppis A supernova. The expansion velocity of the supernova ejecta reaches $V_{ej}\sim 10^9$~cm/s, for the ejecta mass $M_{ej}\sim M_\odot$, which provides an estimate $E_{SN}=M_{ej}V_{ej}^2/2\sim 10^{51}$~erg.

The details of the mechanism of ejection of the matter from the envelope are not clear \cite{janka,supernovae}. The most popular idea is that of "neutrino driven supernova explosion". In this scenario the ejection of a part of the stellar envelope happens because of the excess pressure produced by neutrinos escaping from the neutron star. Indeed, the energy flux of neutrinos traversing the stellar envelope is so high that a small fraction of this energy flux absorbed in the envelope would already deposit sufficient energy to eject the envelope matter. However, the simplest versions of "neutrino-driven supernova explosion" scenario do not work and it is perhaps a combination of neutrino pressure plus the large scale 3d motions of matter in the collapsing star (different hydrodynamical instabilities) which produce the explosions. 

\subsection{Neutron star properties}

Neutron stars, formed in result of gravitational collapse of massive stars, are the one of the major classes of high-energy sources.  They are  observed as X-ray emitting "compact central objects" inside the remnants of supernova explosions \cite{cco}.  Isolated neutron stars power activity of radio / gamma-ray pulsars \cite{pulsars}. They are also responsible for the observed X-ray emission from a large fraction of X-ray binaries \cite{binaries}. Before moving into the discussion of observed astrophysical phenomena produced by the neutron stars, we review in this section the existing understanding of the physical properties of these objects.

\begin{wrapfigure}{R}{0.5\linewidth}
    \includegraphics[width=\linewidth]{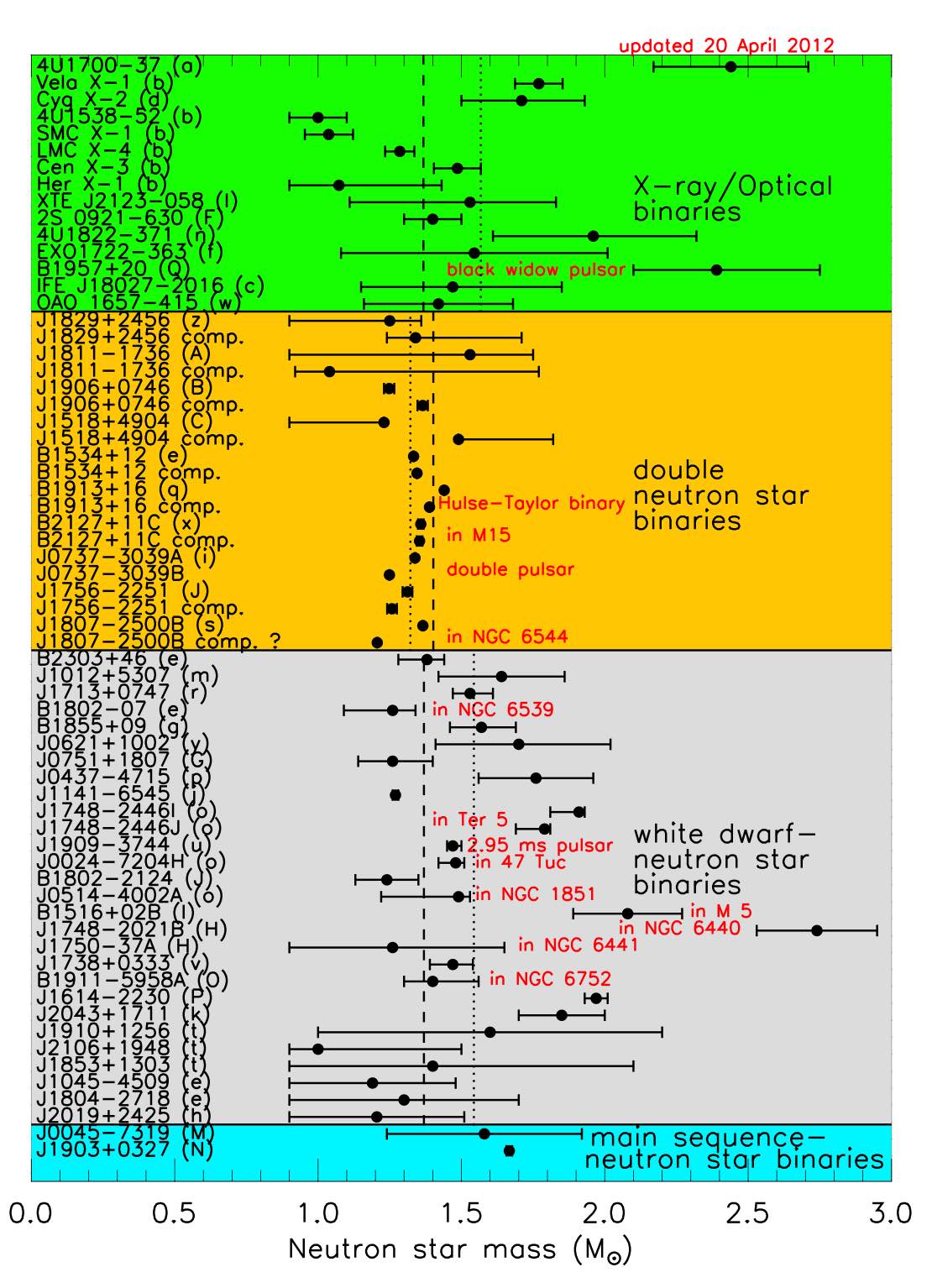}
     \caption{Measurements of masses of the neutron stars \cite{ns_masses}. }
      \label{fig:NS_masses}
\end{wrapfigure}

Basic estimates of the properties of neutron stars derived from the consideration of the gravitational collapse already give order-of-magnitude estimates for the mass, size and density of these objects, in the range $M_{NS}\gtrsim 1.5 M_\odot$, $R_{NS}\sim 10$~km, $\rho_{NS}\sim 10^{15}$~g/cm$^3$. Several more parameters which were not discussed in the previous section are the spin and magnetic field. First "naive" estimates of these parameters at the moment of the birth of the neutron star could be obtained from the laws of conservation of angular momentum and magnetic flux
\begin{eqnarray}
L&=&MR^2/P\sim const\nonumber\\
\Phi&=&BR^2\sim const
\end{eqnarray} 
where $P$ is the rotation period, $B$ is the magnetic field. Taking the Sun as the (best studied) example of a non-collapsed star, we use the estimates $B_\odot\sim 1-100$~G, $R_\odot\sim 10^{11}$~cm for the magnetic field and size and get 
\begin{equation}
B_{NS}\sim  B_\odot \frac{R_\odot^2}{R_{NS}^2}\sim 10^{12}\mbox{ G}
\end{equation}
as an estimate of the magnetic field. Re-scaling of the rotation period $P_\odot \sim 2\times 10^6$~s (26 days) gives
\begin{equation}
P_{NS}\sim P_\odot \frac{R_\odot^2}{R_{NS}^2}\sim 10^{-3}\mbox{ s}
\end{equation}
for the initial spin period of the neutron star.

Similarly to the conventional stars, the hydrostatic equilibrium of the neutron stars is supported by the balance between the gravity force and matter pressure. The matter pressure is provided by the degenerate gas of neutrons, with additions from the pressure of other matter forms present in the star (protons, atomic nuclei). 

Measurements of masses and sizes of the neutron stars are potentially important in the general context of fundamental physics, because they would provide tests of the theoretical model calculations involving strong gravity theory (presumably General Relativity) and the theory of strong interactions \cite{ns_masses}. The physical conditions under which the tests of these theories would be achieved are impossible to create in laboratory conditions here on Earth.

Measurements of masses of neutron stars are most conveniently done via observations of dynamics of these objects which are parts of binary systems. The set of neutron star masses estimated in this way is shown in Fig.~\ref{fig:NS_masses}. Measurements of the radii of the neutron stars are possible through  the measurement of their temperatures $T_{NS}$ and luminosities $L$. This allows to derive the size $R_{NS}\sim L/T_{NS}^4$. Observations of thermal X-ray emission from neutron stars are  challenging because of the low brightness of the sources and contamination from other X-ray emission processes. An additional uncertainty stems from the uncertainty of distance measurements. 

A way to bypass the distance measurement uncertainty (which prevents an estimate of luminosity based on flux measurement) is to consider neutron stars which exhibit accretion-driven outbursts during which the luminosity reaches the Eddington level (which depends only on the mass of the star) \cite{ns_radii}. 

A new direct method of measurement of the neutron star masses and radii based on the properties of the gravitational wave signal has become possible with the discovery of neutron star merger events \cite{ns_radii_ligo}. 

\subsection{Pulsars and pulsar wind nebulae}

\begin{wrapfigure}{L}{0.35\linewidth}
    \includegraphics[width=\linewidth]{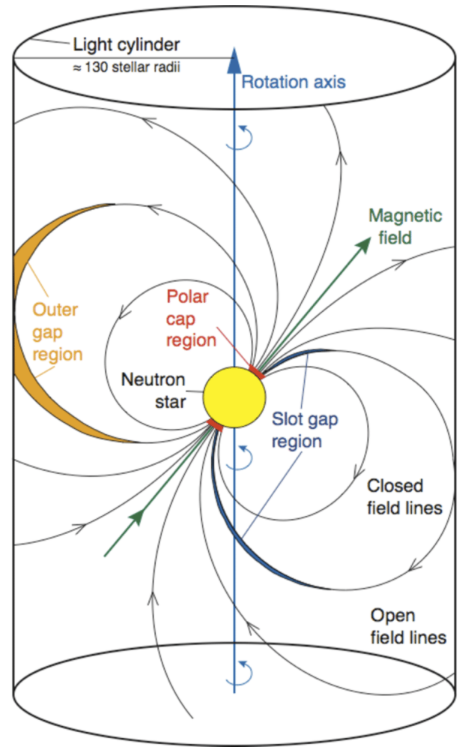}
     \caption{Geometry of pulsar magnetosphere and possible locations of particle acceleration regions ("vacuum gaps") in pulsar magnetosphere (figure fro Ref. \cite{magnetosphere}). }
      \label{fig:gaps}
\end{wrapfigure}

Young neutron stars reveal themselves in a spectacular phenomenon of pulsars. Pulsars are astronomical sources of highly regular pulsed emission with periods in the range between $1$~ms and $\sim 10$~s powered by fast-rotating magnetised neutron stars. The most famous example of pulsar and pulsar wind nebulae is the Crab Nebula shown in Fig.~\ref{fig:crab_nebula} \cite{crab}. The center of the nebula hosts Crab pulsar which has the period $P=33$~ms. Its spin-down power is $\dot E_{\rm rot}\sim 10^{38}$~erg/s. The system is situated at the distance $d\simeq 2$~kpc. It is a remnant of a supernova explosion which was observed by Chinese astronomers in AD 1054, so the system is just about $10^3$~yr old.

As it is discussed above, Crab pulsar works as a very efficient particle accelerator. In fact, in pulsars and pulsar wind nebulae  two different types of accelerators operate. 

First, non-zero electric field around the magnetized neutron star is created in result of the star rotation. To understand this effect one has to consider the surface layer of the neutron star which in which free protons and electrons (and a certain amount of heavier atomic nuclei). Electrons and protons corotate with the neutron star with velocity 
$v_\phi=\Omega R\sin\theta$ where $\theta$ is the angle between the particle radius vector and rotation axis and $\Omega=2\pi/P$ is the angular velocity of the neutron star. Electrons and protons are separated by the action of Lorentz force, with different charges driven toward equatorial or polar regions. 

 This leads to charge separation which, in turn, leads to the creation of electrostatic field. Electric field component parallel to the neutron star surface could be found from the condition of exact compensation of the initial Lorentz force which acts in different direction on them.  Charge separation creates a potential difference between the poles  of the star and infinity:
 \begin{equation}
\label{emax}
U=B_0R^2/\Omega/3\simeq 2\times 10^{19}\left[\frac{B}{10^{12}\mbox{ eV}}\right]\left[\frac{P}{1\mbox{ ms}}\right]^{-1}\mbox{ V}
\end{equation}
where $\Omega$ is the angular velocity of the neutron star.  The electrostatic field produced by the charge redistribution induced by rotation of the neutron star could accelerate particles in the vicinity of the pulsar to extremely large energies. Particle acceleration initiates development of electromagnetic cascades in the space around the neutron star. This process deposits electron-positron pairs which form, together with electromagnetic field, a "magnetosphere",whose structure is presented in Fig.~\ref{fig:gaps}.

The force-free magnetosphere could be maintained only if  $e^+e^-$ pairs are continuously supplied to the space around the pulsar. The problem is that plasma which could freely move along the magnetic field lines is continuously "washed out" from the magnetosphere (together with electromagnetic field) by the centrifugal force. Continuous supply of the plasma could be provided only by the pair production process. The pair production process could work only if there are high-energy particles producing sufficiently high energy gamma quanta. This means that high-energy particles have to be continuously produced at some (as for now, uncertain) locations in the magnetosphere. These particle acceleration sites are called "vacuum gaps" (in the sense that particle acceleration in the gaps is similar to the particle acceleration in the vacuum case considered above). Possible locations of the magnetospheric vacuum gaps are shown in Fig.~\ref{fig:gaps}. Different possible gap locations imply slightly different pair production thresholds for the curvature $\gamma$ quanta.

Rotation of the  magnetosphere produces a centrifugal effect, so that particles are continuously lost from the magnetosphere into a "pulsar wind" which starts at the distance of the "light cylinder" (radius at which the co-rotation velocity reaches the speed of light). Beyond this distance electrons and positrons could not anymore co-rotate because their velocities could not exceed the speed of light. The pulsar wind, composed of relativistic electrons / positrons spreads over large distances finally terminates in interaction with material of the interstellar medium or of the supernova remnant in which the pulsar was born. This produces the phenomenon of pulsar wind nebulae, like the Crab nebula shown in Fig.~\ref{fig:crab_nebula} \cite{crab}.  In some cases, the pulsars dissipating their rotation energy on particle acceleration could be found in binary stellar systems. In this case, interaction of the pulsar wind with the stellar wind of companion star results in a "compactified" pulsar wind nebula which exhibits periodically variable activity like e.g. in the PSR B1259-63 system \cite{chernyakova}.

\subsection{Stellar mass black holes}

\begin{wrapfigure}{L}{0.4\linewidth}
\includegraphics[width=\linewidth]{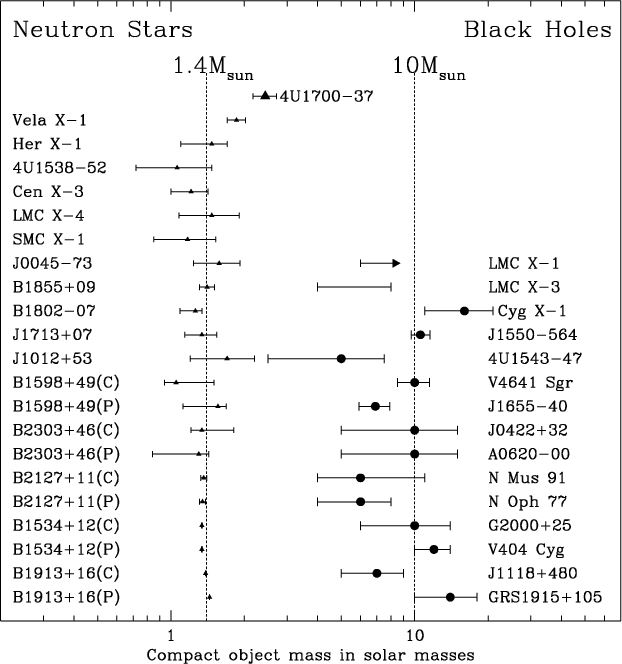}
\caption{Masses of compact objects in X-ray binary systems. Objects of the masses above $3M_\odot$ are black holes. From Ref. \cite{bh_masses}.}
\label{fig:BH_masses}
\end{wrapfigure}

The hydrostatic equilibrium of the neutron star is supported by the pressure  of neutron matter. Similarly to the degenerate electron gas in the iron core of a massive star, the pressure of the degenerate neutron gas could counteract gravity only up to a certain limit. Similarly to the iron cores of massive stars, neutron stars loose stability at about $2.7M_\odot$ \cite{ns_radii}. If the neutron star accumulates more than this mass  during the gravitational collapse of the parent  massive star, the collapse process will not end at the neutron star configuration, but instead will proceed toward a more compact configuration, most probably a black hole. 

First "traces" of the stellar mass black holes  were revealed with the start of the X-ray astronomy. An unexpected discovery of the first X-ray telescopes (initially constructed to study extremely weak X-ray emission from normal stars like the Sun) was the detection of bright X-ray sources with luminosities reaching $\sim 10^5L_\odot$.  Soon after the discovery, it became clear that the X-ray emission originates from stellar binary systems, with only one of the companion stars visible in the optical band, while the other emitting in the X-rays. The X-ray luminosity of the optically invisible star,  is at the level of Eddington luminosity (\ref{eq:edd}). The temperature of the system accreting at the Eddington rate  is related to its size: $L_{Edd}\sim R^2 T^4$. The fact that the temperature is in X-ray range implies that the object is very compact:
\begin{equation}
\label{tedd}
T\sim \frac{L_{Edd}^{1/4}}{R^{1/2}}\simeq 5\left[\frac{M}{M_\odot}\right]^{1/4}\left[\frac{R}{10\mbox{ km}}\right]^{-1/2}\mbox{ keV}
\end{equation}
Thus, the X-ray emitting component of the binary system has to be a compact star with mass about the solar mass and size about the neutron star size.  Measurements of the periodic variations of the radial velocity of the visible component of the X-ray binary allow to determine the dynamical parametes of the system and measure the mass of the compact object. Fig.~\ref{fig:BH_masses} \cite{bh_masses} shows the results of such measurements in a number of X-ray binaries. One could see that the binaries are divided onto two types. In a significant fraction of the binaries the mass of the compact object is close to the Chandrasekhar mass, which indicates that the compact object is most probably a neutron star. At the same time, in some binaries the mass of the compact object is larger than $\sim 3M_\odot$, a theoretical limit of stability of neutron stars. The compact objects in these binaries are most probably black holes. 

Dissipation of liberated gravitational energy of matter falling into gravitational potential well of a black hole  is one of the most efficient mechanisms of extraction of energy from matter. Indeed, the total energy potentially extractable from a particle of the mass $m$ is its rest energy, $E=m$. Consider a particle orbiting a black hole on a Keplerian orbit at a distance $r$. The gravitational energy which needs to be released by the particle to settle at this Kelperian orbit is $U=GM_{BH}m/r$, where $M_{BH}$ is the black hole mass. This energy could be released radiatively, when the accretion flow compresses and heats up while approaching the black hole. By the time when particles reach the distance of the innermost stable circular orbit situated at $6\times$ the gravitational radius, 
\begin{equation}
\label{rg}
R_{isco}=6R_g\sim GM_{BH}\simeq 10^2\left[\frac{M_{BH}}{10M_\odot}\right]\mbox{ km}
\end{equation}
the released gravitational energy per particle reaches the value $U\sim GM_{BH}m_p/ R_{isco}\sim 0.1m_p$, i.e. it becomes a sizeable fraction of the total rest energy of the particle. Detailed calculations based on General Relativity show that the efficiency of extraction of the rest energy of accretion flow depends on the rotation moment of the black hole. Matter approaching the innermost stable circular orbit around  non-rotating black holes could convert $\simeq 10\%$ of the rest energy into radiation. The innermost stable circular orbit around  maximally rotating black holes (those for which the horizon surface rotates with the speed of light) approach closer to $R_g$ and the liberated gravitational energy constitutes 40\% of the rest energy of the accreting matter. 


\begin{figure}[h]
\includegraphics[width=0.55\linewidth]{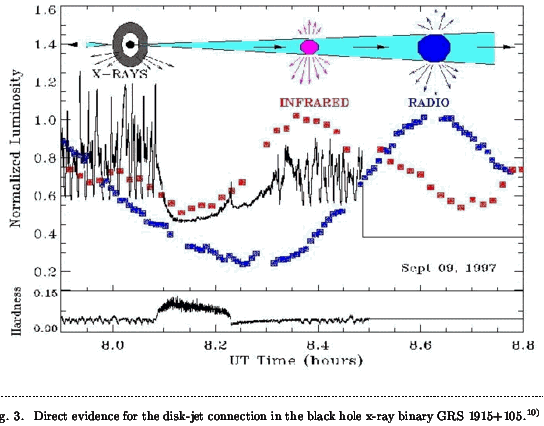}\hspace{2pc}%
\begin{minipage}[b]{0.4\linewidth}\caption{\label{fig:jet}Evolution of the X-ray, infrared and radio fluxes at the moments of matter ejection into the jet of X-ray binary GRS 1915+105 (from Ref.~\cite{mirabel}).}
\end{minipage}
\end{figure}


The companion "donor" star in the binary system could have mass smaller or larger than $\simeq 8M_\odot$. These two cases correspond to different observational appearance of the X-ray binary: low-mass and high-mass X-ray binary. The low- and high-mass binary populations differ in several aspects. First, the massive companion stars in the high-mass X-ray binary are "short-lived": they end their life in supernova explosions after 1-100 Myr. Because of these, the high-mass X-ray binaries could not travel far from their birth places in the Galactic disk. They all appear aligned with the Galactic Plane on the sky. In particular, the hard X-ray sky image of INTEGRAL shown in Fig.~\ref{fig:coded_mask} has a high-mass X-ray binary component along the Galactic Plane. To the contrary, the low-mass companion stars (solar like) have much longer life times and the low-mass X-ray binary systems could travel far. The population of higher Galactic latitude sources in Fig.~\ref{fig:coded_mask} is dominated by the low-mass binaries \cite{arash}.

Otherwise, the high-mass companion stars experience mass loss through the "line-driven" winds (the radiation pressure on ions is so high that they are expelled and pull stellar material with them). Because of this, the black holes orbiting the massive star "bath" in the stellar wind material and pick it up as they go along the orbit. This accretion regime is called "accretion from the wind". In the low-mass binaries, the companion star does not produce strong wind and the main channel of accretion is via "Roche lobe overflow": if the companion star size is large enough, part of the star could find itself in the gravitational potential well of the black hole).  The accretion from the wind is often associated with "outbursts", because of the clumpy structure of the wind \cite{bh_binaries_review}.

Black holes in X-ray binaries produce episodically (via an uncertain mechanism) jet-like outflows \cite{bh_binaries_review} (Fig.~\ref{fig:jet}). During the jet-ejection episodes, blobs of plasma filled with high-energy electrons are ejected in one and the same direction (presumably along the black hole rotation axis). Synchrotron emission produced by the high-energy electrons could be detected from the jets up to parsec-scale distances (distance scale some 10 orders of magnitude larger than the size of the black hole), like in Cyg X-1 system \cite{cygx1}. The jet activity is observable across energy bands, up to the \gr s \cite{cygx3}. 

\subsection{Gravitational wave bursts}

\begin{wrapfigure}{l}{0.4\linewidth}
\includegraphics[width=\linewidth]{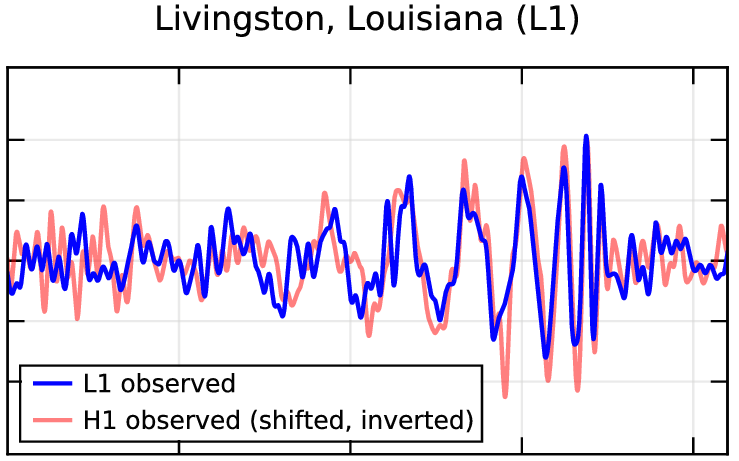}
\includegraphics[width=\linewidth]{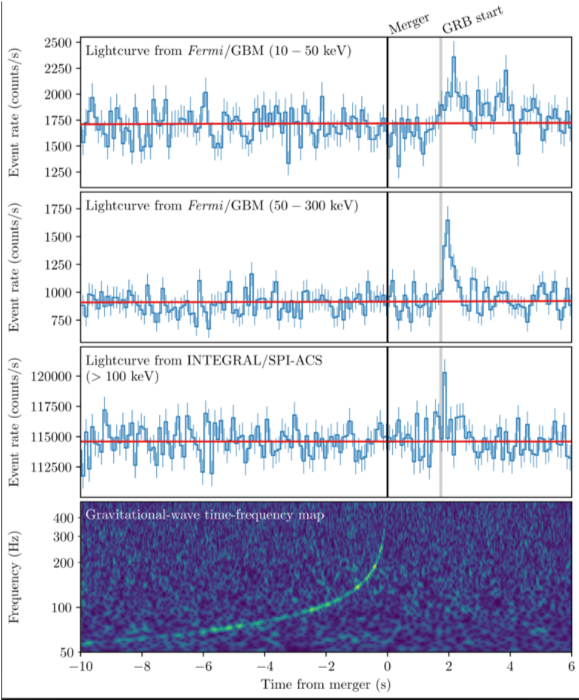}
\caption{Top: gravitational wave signal from a binary black hole merger GW150914 \cite{gw}.  Bottom: gamma-ray and gravitational wave signal from the neutron star merger GW170817 \cite{ns_merger}.}
\label{fig:gw}
\end{wrapfigure}

Binary systems with two massive stars are short-lived (each star leaves some 1-100~Myr, much shorter than the e.g. the Sun). Each star of the binary finally ends its life in a core collapse resulting in formation of either black hole or a neutron star. A small fraction of those systems in which one star has already exploded, but the other has not yet, are observed as High Mass X-ray binary systems. The fate of such systems is formation of binary neutron star, black hole + neutron star and binary black hole systems. Binary black hole systems are remarkably difficult to observe because there is no matter left around the black holes and these systems could not be powered by accretion. First explicit observation of a binary black hole system was done in 2016 with the help of a gravitational wave detector Advanced LIGO, which has detected an event of merger of two black holes in a binary system. The gravitational wave signal of this merger event, GW150914, is shown in Fig.~\ref{fig:gw}. 

Within the General Relativity description of relativistic gravity, the gravitational radiation is quadrupole, with luminosity 
\begin{equation}
L_{GW}=\frac{1}{5}G_N\left<\dddot{I}_{jk}\dddot{I}_{jk}\right>
\end{equation}
proportional to the third time derivative of the quadrupole moment of the mass distribution. For a binary system with companion object masses $M_1\simeq M_2\simeq M$ and binary separation $r$, $I\sim Mr^2$ and its third time derivative is proportional to the orbital angular frequency
\begin{equation}
\label{eq:gw}
\Omega=\sqrt{\frac{GM}{r^3}}
\end{equation}
in the third power: $\dddot I\sim \Omega^3MR^2$. 
This gives
\begin{eqnarray}
&&L_{GW}\simeq G\Omega^6M^2r^4\simeq \frac{G^4M^5}{r^5}\simeq \nonumber\\ &&10^{51}\left[\frac{r}{10^8\mbox{ cm}}\right]^{-5}\left[\frac{M}{10^{34}\mbox{ g}}\right]^5\mbox{ erg/s}
\end{eqnarray}
At the same time, total gravitational energy which could be released by a merger of companions is a fraction of the rest energy of the two bodies:
\begin{equation}
E_{grav}=\frac{GM^2}{r}\lesssim M \sim 10^{55}\left[\frac{M}{10^{34}\mbox{ g}}\right]\mbox{ erg}
\end{equation}
A close binary composed of two stellar mass black holes with  separation just about 10-100 times the size of the black hole horizon would loose all its energy onto gravitational radiation on an hour  time scale. A binary with the orbital separation $10^3$ times larger looses energy via gravitational radiation on the time scale comparable to the age of the Universe. 

The final stage of the merger occurred when the binary separation was about the gravitational radius of the black holes, $r\sim GM$. The characteristic time scale of this final stage is 
\begin{equation}
t\sim \frac{E_{grav}}{L_{GW}}\simeq \frac{r^4}{G^3M^3}\gtrsim 1\mbox{ ms}
\end{equation}
This is the time scale of the fastest and strongest amplitude oscillations of the signal in Fig.~\ref{fig:gw}. 

Contrary to the black holes, neutron star mergers are generically expected to possess electromagnetic counterpart, and they do possess it as was demonstrated by the detection of the neutron star merger GW170817 simultaneously in gravitational waves and \gr s (with a delay of 2 s), see bottom panel of Fig.~\ref{fig:gw}.

\subsection{Active galactic nuclei.}

Compact sources of radio-to-$\gamma$-ray emission are found in the centers of certain types of galaxies. These sources are called radio-loud  AGN. Their luminosities reach the level of $>10^{45}$~erg/s, i.e. larger than typical luminosity of the Milky Way type galaxies. 

\begin{wrapfigure}{l}{0.5\linewidth}
\includegraphics[width=\linewidth]{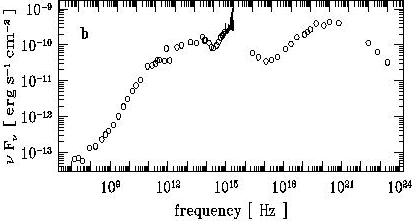}
\caption{Broad band spectrum of a quasar 3C 273 \cite{3c273}.}
\label{fig:3c273}
\end{wrapfigure}

Study of variability of emission from AGN reveals variability time scales down to days, hours and sometimes even minutes and seconds. This implies that the variable emission comes from a compact region of the size $R\le ct_{var}\sim 1\left[t_{var}/10\mbox{ min}\right]$~AU. Such enormous energy output, equivalent to $10^{12}$ Suns, could not be produced by a compact stellar cluster confined to the region of the size of the order of the distance from the Earth to the Sun. Similarly to the case of X-ray sources in binary systems, the only reasonable energy reservoir which could power AGN is thought to be gravitational energy of matter accreting on a compact region.

Re-applying the formula for the Eddington luminosity (\ref{eq:edd}) to AGN one finds that the mass of the object onto which matter accretes should be at least  $10^{7}M_\odot$ to provide the power at the level of $10^{45}$~erg/s. The estimate of the mass of the compact object in the AGN "central engine" is often obtained also from the study of statistics of AGN observations. This study shows that the observed number of AGNs could be explained if the lifetime of these sources in the nuclei of galaxies is $T_{AGN}\sim 10^7 - 10^8$~yr. If the source produces luminosity at the level of $L_{AGN}\sim 10^{45}$~erg/s over this  time span, the total energy output of the source is $E_{AGN}\sim L_{AGN}T_{AGN}\sim 10^{60}$~erg. 

\begin{figure}[ht]
\begin{center}
\includegraphics[width=0.8\linewidth]{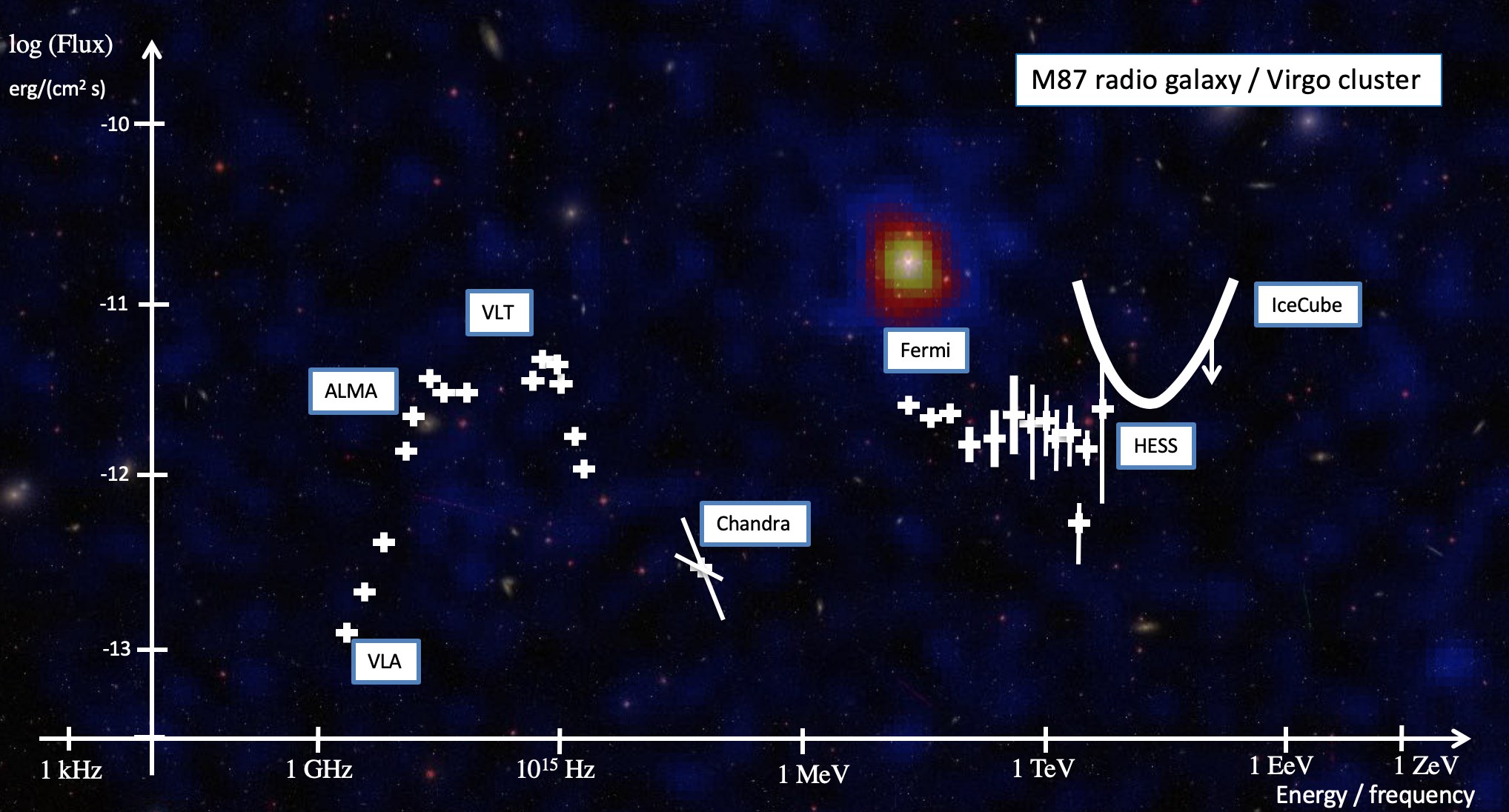}
\caption{Broad band "multi-messenger" spectrum of M8	7 radio galaxy. Background shows the Fermi/LAT image of M87.}
\label{fig:M87_spectrum}
\end{center}
\end{figure}


Assuming the best possible efficiency of conversion of the energy of the rest energy of the "fuel" which powers the AGN into radiation, $\sim 10\%$, we could find that the mass of the "waste" which should be left by the AGN activity is  $M\ge 10E_{AGN}/c^2\sim 10^7 M_\odot$. This mass should reside somewhere in the source, so it is presumably confined within the region of the size $\sim 1$~AU, found from the variability time scale. Comparing this size to the gravitational radius of a $10^7M_\odot$ body (\ref{rg}), we find that all the mass is, in fact contained in the region of the size comparable to the size of a $10^7M_\odot$ black hole. Based on this observation, the most common hypothesis is that the objects powering the AGN are "supermassive" black holes. 

The range of phenomena associated to the supermassive black holes is similar to the phenomena observed in the stellar mass black holes (the term "microquasars" is sometimes used for the black hole powered X-ray binary stellar systems). Estimating the temperature of accretion flow onto the supermassive black hole from Eq. (\ref{tedd}) one finds that for the supermassive black holes 
\begin{equation}
T\simeq \frac{L_{Edd}^{1/4}}{R_g^{1/2}}\sim 100\left[\frac{M_{BH}}{10^7M_\odot}\right]^{-1/4}\mbox{ eV}
\end{equation}
is expected to be in the UV range. such (quasi)thermal radiation form the accretion flow is observed in the form of "Big Blue Bump" in the spectra of quasars (see Fig.~\ref{fig:3c273}, the Big Blue Bump is visible in the lower panel of the figure, in the frequency range $\sim 10^{15}-10^{16}$~eV). 

\begin{wrapfigure}{R}{0.8\linewidth}
\includegraphics[width=0.5\linewidth]{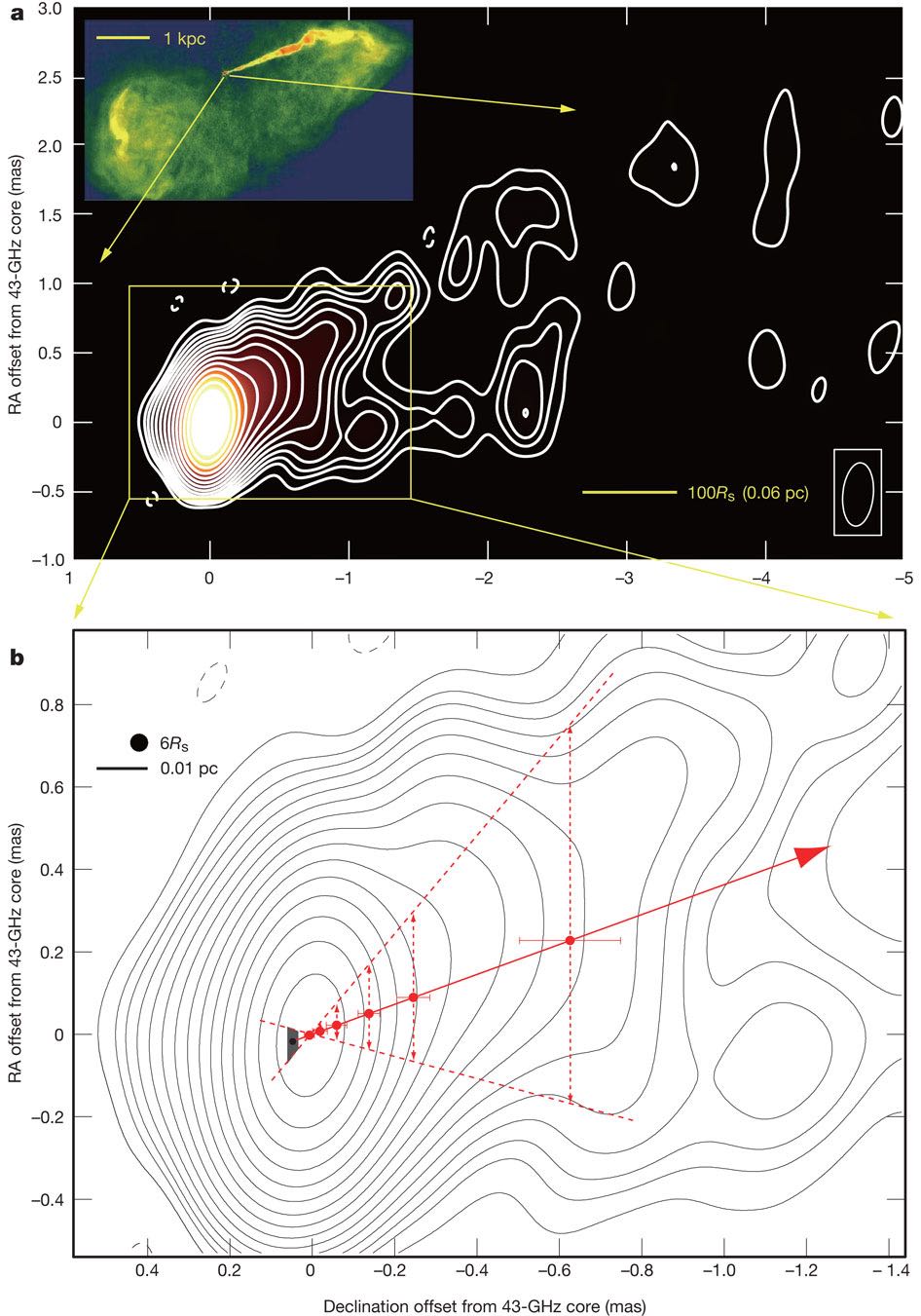}
\includegraphics[width=0.49\linewidth]{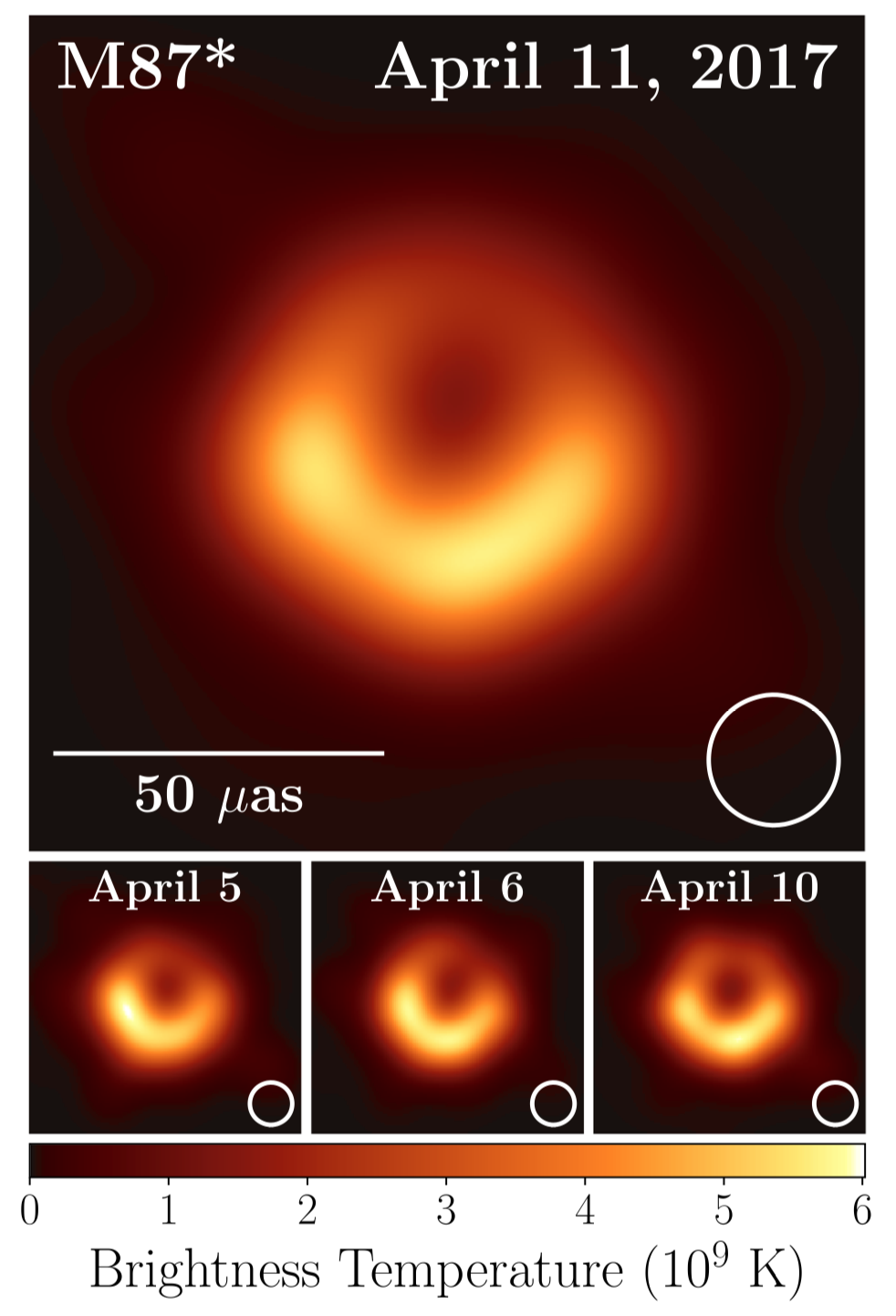}
\caption{Left: radio jet of the galaxy M87 on different distance scales \cite{m87_vlbi}. Image credit: National Radio Astronomy Observatory. Right: high-resolution image of the "central engine" of M87 jet, powered by supermassive black hole. Figure from Ref. \cite{eht}.}
\label{fig:M87}
\end{wrapfigure}

Most of the AGN (about 90\%) are "radio-quiet", in the sense that they they are not strong radio wave emitters. Radio emission is conventionally associated with synchrotron emission from relativistic particles and "radio-quiet" term implies that those AGN do not operate particle accelerators. Their observational signature is the accretion flow "big blue bump", presumably produced by an accretion disk onto the black hole,  plus an X-ray emission from the "hot corona" which surrounds the accretion disk. The origin of "corona" could be understood if we start from a simple observation that single particle falling into the black hole "from infinity" is expected to acquire relativistic velocity close to the black hole (e.g. based on virial theorem: $mv^2\sim G_NMm/R$, with $R\sim G_NM$). Thus, one naively expects to find plasma heated to relativistic temperatures ($\sim 10^9-10^{12}$~K) close to the black hole, which is not observed.
 Instead, radiative processes manage to "cool down" the plasma  to 100~eV. This cooling is, however, not equally efficient everywhere in the accretion flow. Parts of the flow which are not completely cooled form a "corona" whose temperature could be much higher, so that the emission is in the X-ray and hard X-ray band. 
Emission from the hot coronae of AGN is supposed to be responsible for the bulk of hard X-ray emission from extragalactic sky, forming the X-ray background \cite{xrb}. 

Radio-loud AGN,  the remaining 10\% of the AGN population, produce jet-like outflows which, contrary to the jets of the stellar mass black holes are generated steadily, rather than episodically, by the AGN central engines. The jets extend up to the 100~kpc -- Mpc distance scales (again, ten orders of magnitude larger than the size of the black hole) and originate directly from the supermassive black hole \cite{jets_review}. Large size of the supermassive black holes enables direct imaging observations of the jets  at the distance scales as short as the gravitational radius of the black hole in the case of the nearest and largest supermassive black holes. 

The best studied example is the jet of the nearby "radio-loud" AGN M87, which is found in the nucleus of a giant elliptical galaxy in the center of the nearest-to-us galaxy cluster Virgo. The radio image of the M87 jet is shown in Fig.~\ref{fig:M87}  \cite{m87_vlbi}, Due to its proximity, the optical emission from the jet is visible already with moderate power telescopes. The M87 jet was discovered in the visible light back in 1919, almost hundred years ago.  The black hole mass in M87 is in the range $\sim 6\times 10^9M_\odot$, so that the size of the black hole horizon is $\sim 10^{15}$~cm. At the distance $16$~Mpc, the angular size of the supermassive black hole in the nucleus of M87 is $\theta=10^{15}\mbox{ cm}/16\times 3\times 10^{24}$~cm$\sim 4\ \mu$as. To resolve the region of the size about the size of the black hole horizon, a telescope with angular resolution in the micro-arcsecond range is required. Contours in Fig.~\ref{fig:M87} show the image of the core of M87 taken with the Very Large Baseline Array (VLBA) radio telescope, which is a network of radio antennae working as a single telescope with aperture equal to the distance between the telescopes. This allows to achieve angular resolution in the range $\theta_{resolution}\sim \lambda/D\sim 0.1-1$~mas, where $\lambda$ is the wavelength of the radio signal  and $D$ is the distance between the telescopes. From Fig.~\ref{fig:M87} one could see that the M87 jet originates from the region of the size not larger than  $\simeq 10$ the black hole horizon size. 

The jet launch point has recently been resolved with Very-Long Baseline Interferometry technique in mm wavelength range \cite{eht}. This allows to achieve angular resolution sufficient for imaging of the immediate vicinity of the black hole horizon, see right panel of Fig. \ref{fig:M87}. 

Systematic re-observation of AGN jets on the time scales of years shows that the jets are dynamic structures, with bright "blobs" moving away from the AGN "central engine". In fact, the measured speeds of the jet blobs often exceed the speed of light. This effect is not related to the real superluminal motion, but is an artifact of projection and Doppler effects. The Doppler effect is also responsible for the one-sided appearance of the M87 jet evident from Fig.~\ref{fig:M87}. The supermassive black hole of M87 galaxy is powered by radiatively inefficient accretion flow, which does not possess a disk, but is rather wrapping the black hole around. In this case the brightest emission is expected from the direct vicinity of the horizon, so that one naively would expect to observe a "glowing sphere" in the high-resolution image, possibly with well identifiable asymmetry related to the jet launch locations (north and south pole). This is not the impression one could get from the image in the right panel of Fig. \ref{fig:M87}. A prominent feature on the image is a ring-like structure which is the result of extreme gravitational lensing by the black hole. The light from the direct vicinity of the horizon is strongly deflected so that not only light from the geometrically visible side of the black hole horizon could reach the telescope on Earth, but also the light from the "back side" of the black hole. This light is directed along the  ring of the angular radius 
\begin{equation}
\theta=\frac{\sqrt{27}GM}{D}\simeq 20\left[\frac{M}{6\times 10^9M_\odot}\right] \left[\frac{D}{17\mbox{ Mpc}}\right]^{-1}\mu\mbox{as}
\end{equation}
for the reference parameters of the M87 black hole: the mass $M\simeq 6.2\times 10^9M_\odot$, distance $D\simeq 16.9$~Mpc \cite{eht1}.

\subsection{Supermassive black hole in the Milky Way.}

\begin{figure}[ht]
\begin{center}
\includegraphics[height=6.5cm]{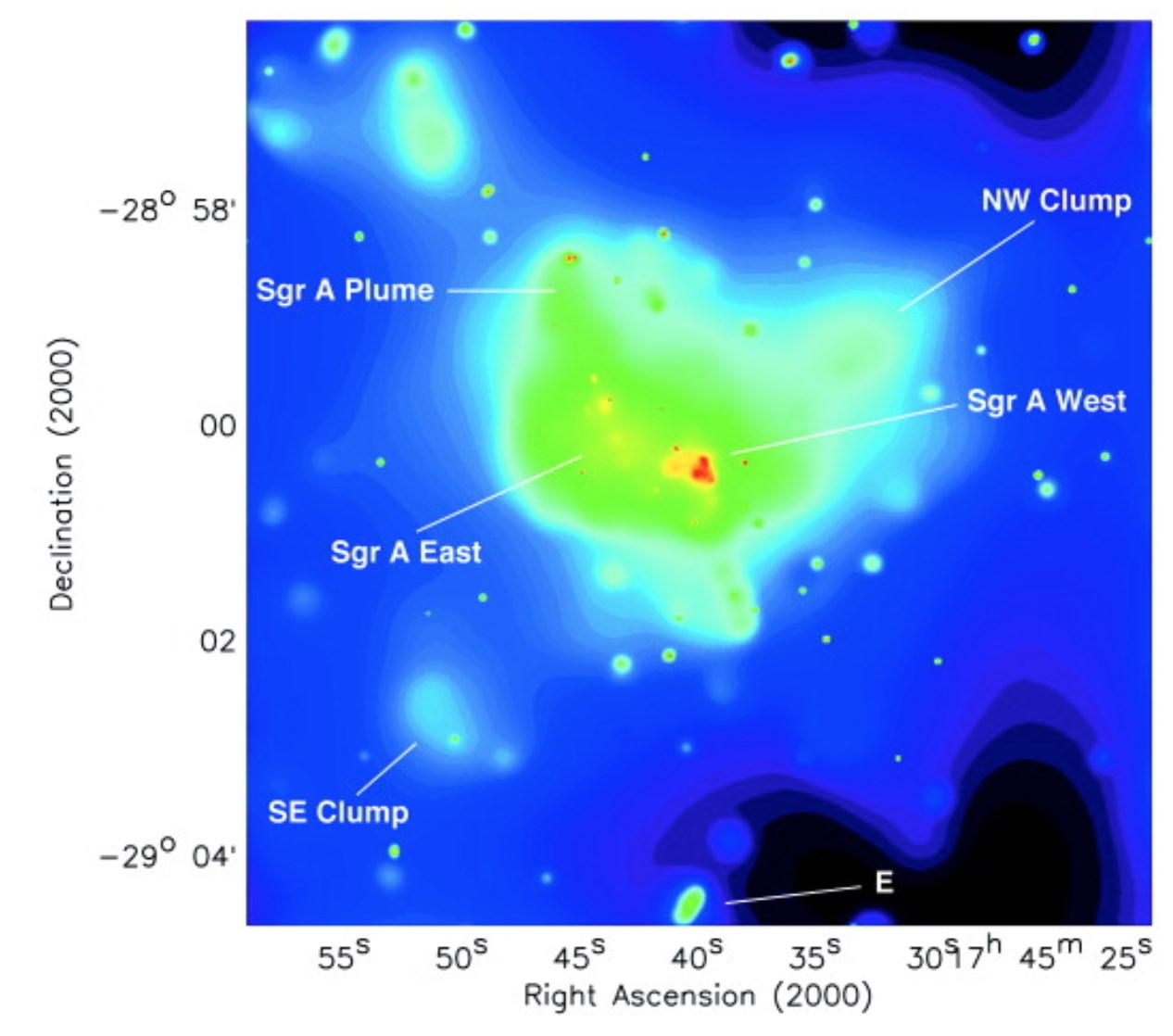}\hspace{5mm}
\includegraphics[height=6.5cm]{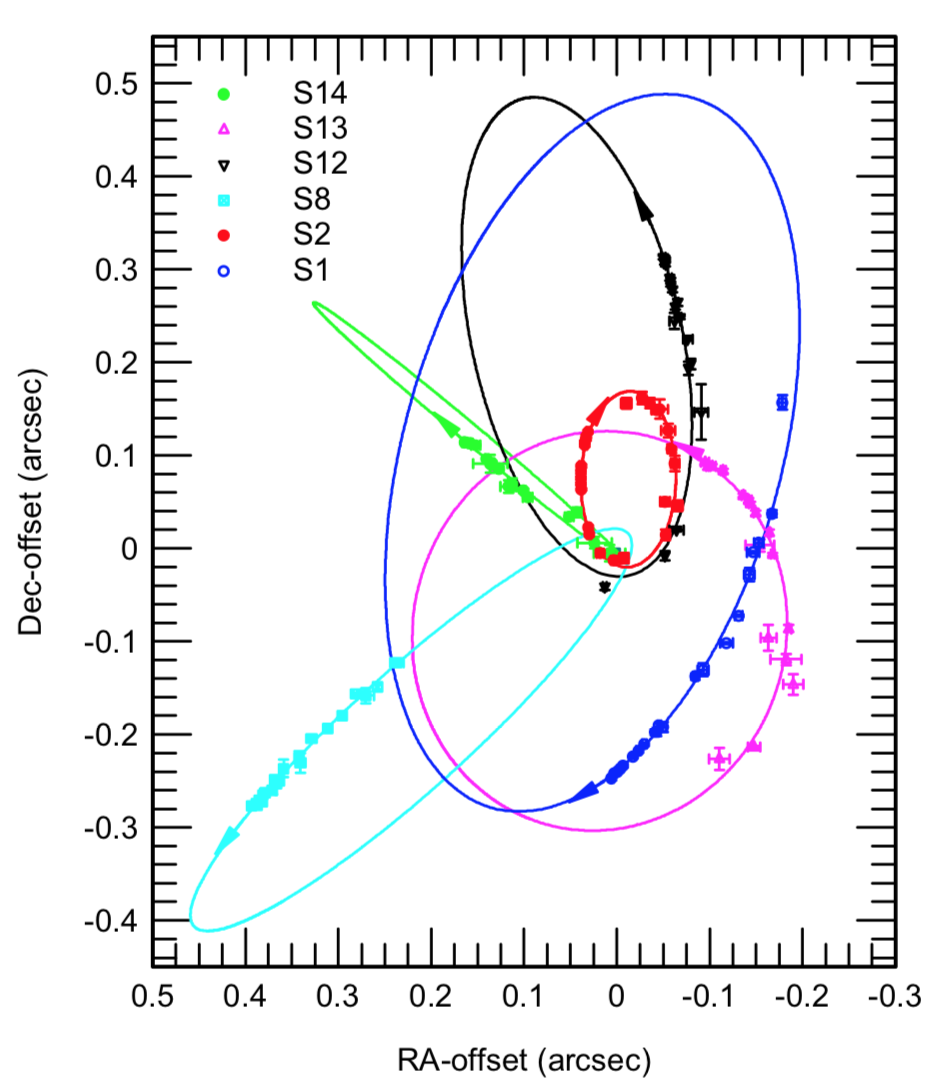}
\caption{X-ray image of the source Sgr A region hosting the supermassive black hole of the Milky Way galaxy \cite{sgra_chandra}. Blue bubble-like structure around the source is Sgr A shell of the size $\sim 10$~pc. Right: Stellar orbits around Sgr A* \cite{orbits}.}
\label{fig:sgra}
\end{center}
\end{figure}

The best studied supermassive black hole in galactic nucleus is the Milky Way supermassive black hole. It is situated in the source in the constellation of Sagittarius, called Sgr A* \cite{gc}. Image of this source in the X-ray band is shown in Fig.~\ref{fig:sgra}. 

The center of the Milky Way is situated at the distance $\simeq 8$~kpc away from the Sun. In the visible band it obscured by the dust filling the Galactic Disk. However, stars in the Galactic Center region are visible in the infrared band. The source Sgr A* itself is most of the time not visible in the infrared, except for the flaring periods of increased activity. Observations of motion of the stars in the region around Sgr A* show that  they move along elliptical orbits (Fig.~\ref{fig:sgra}) \cite{orbits}. The periastron of the closest approaching orbit is just $\sim 10^{15}$~cm (100 AU) away from Sgr A*. Dynamics of the orbits is consistent with motion around a central point object of the mass $M_{BH}=4\times 10^6M_\odot$. The most straightforward model for the compact central object is that it is a supermassive black hole. 

The angular size of the supermassive black hole in the Milky Way center could be calculated similarly to the size of the M87 black hole. This gives some $\sim 20\ \mu$as. The required angular resolution was, in fact, already reached by the radio telescopes using  the technique of very-long-baseline interferometry. First constraints on the size of the radio source in Sgr A* obtained with this technique indicate that the radio emission originates directly from the black hole horizon.


\section*{References}
\bibliography{em}

\end{document}